\documentclass[physics]{ejs_author}
\usepackage{upgreek}
\usepackage{color}
\usepackage{amsmath,amssymb,amsfonts,bbm,graphicx,graphics,times}
\usepackage[english]{babel}
\usepackage{bbm}

\usepackage{bm}
\usepackage{epsfig}
\usepackage{hyperref}

\title{Hybrid optomechanics for Quantum Technologies}

\articletype{Review Article}

\author{B. Rogers\inst{1}, N. Lo Gullo\inst{2},  G. De Chiara\inst{1}, G. M. Palma\inst{3}, and M. Paternostro\inst{1}}
\institute{\inst{1}Centre for Theoretical Atomic, Molecular and Optical Physics, School of Mathematics and Physics, Queen's University, Belfast BT7 1NN, United Kingdom
\inst{2} Dipartimento di Fisica e Astronomia Galileo Galilei and CNISM, Universit\`a di Padova, Via Marzolo 8, 35122 Padova, Italy
\inst{3}NEST Istituto Nanoscienze-CNR and Dipartimento di Fisica e Chimica, Universit\'a degli Studi di Palermo, via Archirafi 36, I-90123 Palermo, Italy}
\abr{QMTR}
\journal{Quantum Measurements and Quantum Metrology}

\abstract{We review the physics of hybrid optomechanical systems consisting of a mechanical oscillator interacting with both a radiation mode and an additional matter-like system. We concentrate on the cases embodied by either a single or a multi-atom system (a Bose-Einstein condensate, in particular) and discuss a wide range of physical effects, from passive mechanical cooling to the set-up of multipartite entanglement, from optomechanical non-locality to the achievement of non-classical states of a single mechanical mode. The reviewed material showcases the viability of hybridised cavity optomechanical systems as basic building blocks for quantum communication networks and quantum state-engineering devices, possibly empowered by the use of quantum and optimal control techniques. The results that we discuss are instrumental to the promotion of hybrid optomechanical devices as promising experimental platforms for the study of non-classicality at the genuine mesoscopic level.}

\newcommand{\ket}[1]{\left\vert#1\right\rangle}
\newcommand{\bra}[1]{\left\langle#1\right\vert}

\newcommand{\nbar}{\overline{n}}
\newcommand{\Nbar}{\overline{N}}
\newcommand{\modulo}[1]{\vert#1\vert}

\newcommand{\funcx}[4]{#1_{#3}^{#4}({\bf #2})}
\newcommand{\opx}[4]{\hat{#1}_{#3}^{#4}({\bf #2})}
\newcommand{\opt}[4]{\hat{#1}_{#3}^{#4}(#2)}
\newcommand{\op}[3]{\hat{#1}_{#2}^{#3}}
\newcommand{\mean}[2]{\langle#1\rangle_{#2}}

\newcommand{\braketM}[3]{\langle#1\vert#2\vert#3\rangle}
\newcommand{\braket}[2]{\langle#1\vert#2\rangle}
\newcommand{\ketbra}[2]{\vert#1\rangle\langle#2\vert}


\begin{document}
\firstpage{1}

\maketitle

The interest in delivering winning architectures for quantum technologies has now extended well beyond the academic domain that is more closely linked to modern quantum mechanics to involve the industrial and policy-making sectors~\cite{dwave,insane}. Major financial investments aimed at the realisation of fully functioning prototypes of quantum-empowered devices have or are about to be made to boost the steps made so far in this area and catalyse the paradigm shift that the implementation of a disruptive quantum technological platform promises to embody. The major obstacle in this respect is very well summarised by the following question: {\it Are we exploring the correct experimental scenarios for the delivery of quantum technologies?} This is a sensible question that aims at understanding if the current formulation of quantum information processing (QIP) and the physical candidates to the implementation of a fully function quantum information processing device are fit for the task. In a way, it is well possible that something similar to what triggered the birth of the ``second generation" of (classical) computers would be needed: the discovery of a new platform (semi-conductor transistors, in the case of classical machines) able to turn the whole technological paradigm completely and boost miniaturisation, scalability, and performance efficiency.

In contrast with the so-far dominating QIP architecture that makes use of homogeneous information carriers (ions, neutral atoms, semi- or super-conducting chips, photonic circuits, among the most prominent examples~\cite{NielsenChuang}), recently the idea of hybridisation has started becoming increasingly popular. Indeed, the community interested in QIP architectures and their implementation is realising that the combination of information carriers and processors of heterogeneous nature might be a winning strategy. By putting together the strengths of different individual technological platforms at the price of engineering controllable interfaces, hybrid devices would be more flexible, adaptive and performant than their homogeneous counterparts~\cite{Wallquist}. 

This is precisely the context in which this review will  be set: we will illustrate the factual benefits for QIP capabilities coming from the hybridisation of cavity-optomechanics setups~\cite{reviews,AspelmeyerRMP}, which are currently raising considerable attention in light of the possibilities that they offer for quantum state engineering, quantum control, and investigations on the foundations of quantum mechanics and its potential modifications~\cite{Bassi}. In particular, we will review some of the advantages for state preparation, manipulation and diagnostics that are provided by the cooperation established between optically driven mechanical systems (operating at the quantum level) and simple atomic-like systems embedded into an optical cavity. We shall showcase the rich range of relevant effects emerging from such an admixture of different physical information carriers (light, atomic systems, and mechanical ones). We will pinpointing the opportunities opened by such hybrid structures for improved quantum control and revelation, focusing in particular on the achievement of non-classical features at the full mechanical level, a goal that is currently at the centre of much of the experimental endeavours in the area of cavity optomechanics and that, as we will argue, might be significantly ``aided" by the adoption of hybrid setups~\cite{arcizetNV, tip, hunger,sillanpaa}. 

In detail, the structure of this paper is as follows. In Sec.~\ref{Sottraggopaper} we address a protocol for quantum state engineering of a mechanical mode operated by an optical photo-subtraction mechanism. After discussing the general features of this scheme, we briefly sketch how such an operation can be performed by using the hybridisation methods based on the use of atomic-like systems coupled to the field of an optomechanical cavity. besides illustrating a non-trivial example of mechanical quantum-state engineering through all-optical means, this scenario provides an interesting motivation for studying in details the opportunities offered by such hybrid models for control, sensing and processing at the genuine quantum level.

With such motivations at hand, in Sec.~\ref{BECpaper} we introduce and discuss a hybrid optomechanical system consisting of a BEC interacting with the field of an optomechanical cavity it is trapped in. We address the quantum back-action induced on the mechanical device by the coupling of the field with the collective state of the atoms in the BEC. We show that such setup offers a very rich set of possibilities for both quantum control and entanglement distribution. Indeed, as addressed in Sec.~\ref{Gabrielepaper}, a very interesting structure of entanglement sharing is set among the parties at hand, including the possibility for observing genuine tripartite entanglement set among continuous variable systems. The analysis illustrated in Secs.~\ref{BECpaper} and \ref{Gabrielepaper} will be conducted at the steady-state reached by the system after a sufficiently long interaction time. However, as Sec.~\ref{Benpaper} reveals, very important features can be gathered from a time-resolved analysis of the dynamics of the hybrid system at hand. In fact, we will see how the short-time dynamics is characterised by entanglement set between genuinely mesoscopic degrees of freedom (both atomic and mechanical) that is prevented at the stationary state. By borrowing techniques that are typical of the emerging field of optimal control, we will show that a considerable improvement of the degree of mesoscopic entanglement shared by mechanical mode and the BEC can be achieved if one implements a rather simple form of driving modulation, thus demonstrating the advantage of mixing up strategies for quantum control theory and the flexibility of a hybridised setup. The usefulness of a BEC-hybridized optomechanical setup will be epitomised by the study performed in Sec.~\ref{Nicolapaper}, where we address the problem of the revelation of quantum coherence in the state of a single-clamped cantilever by mapping its state onto the magnetic behaviour of a spinor BEC. In Sec.~\ref{Giovannipaper} we change perspective completely and address a different form of hybridisation, this time based on the use of a single spin system. In particular, we exploit a three-level atom, trapped within an optomechanical cavity, to demonstrate a dynamical regime that is able to generate a state quite close to a Schr\"odinger cat state. The incorporation of a simple postselection stage allows us to prepare the mechanical system in a highly non-classical state, as shown by a criterion based on the negativity of the Wigner function. 
Finally, Sec.~\ref{conclusions} allows us to draw our conclusions.


\section{Quantum state engineering through photo-subtraction}
\label{Sottraggopaper}
In order to introduce the formalism that will be used across a large part of the manuscript without the complications of dealing immediately with a multipartite system, in this Section we concentrate on the case of a purely optomechanical system and discuss a protocol for quantum state engineering of a massive mechanical mode based on the combination of radiation-pressure coupling and photon subtraction from a light field~\cite{grangier,kim2}. We show a dynamical regime where non-classical states of a mechanical oscillator can be in principle achieved under non-demanding conditions: cooling of the oscillator down to its ground-state energy is not required as the scheme prepares non-classical states for operating temperatures in the range of $1$ K and inefficiencies at the photon-subtraction stage do not hinder the effectiveness of the method. As we will discuss in the last part of this Section, the required photon subtraction step can indeed be realised by using ancillary elements, therefore embodying a genuine case of hybrid optomechanics. Therefore, this study is instrumental to illustrate an interesting instance of non-classical state-engineering protocol that makes use of the advantages provided by hybridisation and also to provide the necessary mathematical tools that will be used explicitly when addressing the BEC-hybridised configurations presented in Secs.~\ref{BECpaper}, \ref{Gabrielepaper} and \ref{Benpaper}. 

We consider the prototypical optomechanical setting consisting of a cavity of length ${\cal L}$ driven through its steady input mirror by an intense light field of frequency $\omega_L$ and endowed with a highly reflecting end-mirror that can oscillate along the cavity axis around an equilibrium position.  The vibrating mirror, which is modelled as a mechanical harmonic oscillator at frequency $\omega_m$, is in contact with a background of phononic modes in equilibrium at temperature $T$. We write the Hamiltonian of the system made out of the cavity field, the movable mirror and the BEC as
\begin{equation}
\label{modellobase}
\hat{\cal H}=\hat{\cal H}_M+\hat{\cal H}_C+\hat{\cal H}_{MC}
\end{equation}
where the mirror and cavity Hamiltonians are 
\begin{eqnarray}
\hat{\cal H}_M&=&m \omega_m^2 \hat{q}^2/2 +{\hat{p}^2}/{(2m)},
\\
 \hat{\cal H}_C&=&\hbar (\omega_C{-}\omega_L)\hat{a}^\dagger\hat{a}{-}i \hbar \eta (\hat{a}-\hat{a}^\dagger),
\end{eqnarray}
respectively. Here $\hat{q}$ ($\hat{p}$) is the mirror position (momentum), $m$ is its effective mass, $\omega_C$ is the cavity frequency and $\hat{a}$ ($\hat{a}^\dag$) is the corresponding annihilation (creation) operator. We have included a {\it cavity pumping} term $-i\hbar\eta(\hat{a}-\hat{a}^\dag)$ with coupling parameter $\eta=\sqrt{2\kappa{\cal R}/\hbar\omega_L}$ (${\cal R}$ is the laser power and $\kappa$ is the cavity decay rate). For small mirror displacements and large cavity free spectral range  with respect to $\omega_m$ (which allows us to neglect scattering of photons into other mechanical modes), the mirror-cavity interaction can be written as 
 \begin{equation}
{\hat{\cal H}_{MC}=-\hbar \chi \hat{q}\hat{a}^\dagger\hat{a}}
\end{equation}
with $\chi=\omega_C/{\cal L}$ the optomechanical coupling coefficient. 
Before addressing the full-fetched configuration for photon subtraction-aided quantum optomechanics, it is worth gathering some insight into the features of the system itself. This analysis will justify {\it a posteriori} some of the conclusions that will be reached later on.


The properties of the field-oscillator state are well characterized using these covariance matrix ${\cal V}_{MC}$ having elements $({\cal V}_{MC})_{ij}{=}\langle\hat{\bm q}_i\hat{\bm q}_j+\hat{\bm q}_j\hat{\bm q}_i\rangle/2~(i,j=1,..,4)$ with $\hat{\bm q}{=}(\hat q,\hat p,\hat x,\hat y)$ and the dimensionless field quadratures $\hat x=(\hat a{+}\hat a^\dag)/\sqrt{2}$, $\hat y=i(\hat a^\dag-\hat a)/\sqrt{2}$. 
In such an ordered operator basis, the covariance matrix of the simple optomechanical system is written as
\begin{equation}
\label{cvmatrix}
{\cal V}_{MC}=
\begin{pmatrix}
{\bm M}&{\bm R}\\
{\bm R}^T&{\bm C}
\end{pmatrix}
\end{equation}
with ${\bm M}{=}\textrm{Diag}[m_{11},m_{22}]$ a diagonal matrix encompassing the local properties of the mechanical mode and 
${\bm J}={\bm C}, \bm{R}$ with elements $({\bm J})_{ik}=j_{ik}$ accounting for either the field's properties or the correlations between the two subsystems, respectively. 
In general, the dynamics encompassed by $\hat{\cal H}$ is made difficult by the non-linearity inherent in $\hat{\cal H}_{MC}$. However, for an intense pump laser, the problem can be linearised by introducing quantum fluctuation operators as $\hat{O}{\rightarrow}{\cal O}_s{+}\delta\hat{O}$ with $\hat{\cal O}$ any of the operators entering into $\hat{\cal H}_{MC}$, ${\cal O}_s$ the corresponding mean value and $\delta\hat{O}$  the associated first-order quantum fluctuation operator~\cite{Vitali2007}. The explicit form of the elements of ${\cal V}_{MC}$  is found using the solutions of the Langevin-like equations regulating the open-system dynamics undergone by the mechanical and optical fluctuation operators. These can be written in the compact form
\begin{equation}
\label{eq:dotphi}
\partial_t{\hat{\bm \phi} }_{MC}={\cal K}_{MC}\hat{\bm \phi}{+}\hat{\cal {\bm N}}_{MC},
\end{equation}
where we have introduced the vector of fluctuation operators $\hat{\bm \phi}^T_{MC}=(\delta\hat{x}~\delta\hat{y}~\delta\hat{q}~\delta\hat{p})$,  the noise vector $\hat{\cal\bm N}^T_{MC}=(\!\sqrt{\kappa}(\delta\hat{a}^\dag_{in}{+}\hat{a}_{in})~i\sqrt{\kappa}\delta(\hat{a}^\dag_{in}{-}\hat{a}_{in})~0~\hat{\xi})$ and the dynamical coupling matrix ${\cal K}_{MC}$ given, in the chosen basis, by
\begin{equation}
{\cal K}_{MC}=
\begin{pmatrix}
0&\omega_m&0&0\\
-\omega_m&-\gamma&G&0\\
0&0&-\kappa&\Delta\\
G&0&-\Delta&-\kappa
\end{pmatrix}.
\end{equation}
In these expressions $G=\chi\sqrt{\hbar/(2m\omega_m)}$ is the effective optomechanical coupling rate and $\gamma$ is the energy decay rate of the mechanical oscillator. Moreover, $\hat \xi$ is a Langevin operator that describes the Brownian motion of the mechanical mode (induced by environmental phonon modes) at temperature $T$. The statistical properties of this operator will be addressed in some detail in Sec.~\ref{BECpaper}. Here it is sufficient to mention that in the limit of small mechanical damping and large enough temperature, $\hat \xi$ describes a delta-correlated noise of strength $\gamma(\nbar+1)$ with $\nbar$ the mean phonon number of the mechanical system. Eq.~\eqref{eq:dotphi} can be straightforwardly transformed into the dynamical equation for ${\cal V}_{MC}$~\cite{Vitali2007}
\begin{equation}
\label{Lyapunovdue}
\dot{\cal V}_{MC}={\cal K}_{MC}{\cal V}_{MC}+{\cal V}_{MC}{\cal K}^T+{\cal D}_{MC}
\end{equation}
with ${\cal D}_{MC}{=}\text{diag}[0,\gamma(2\overline{n}{+}1),\kappa,\kappa]$ accounting for the noise affecting the bipartite system at hand here. We will consider initial conditions such that the mechanical mode is prepared in a thermal state at temperature $T$ and the cavity field in a coherent state of amplitude determined by the intensity of the pumping field.

  
 \begin{figure*}[t]
{\bf (a)}\hskip6cm{\bf (b)}\hskip6cm{\bf (c)}
\includegraphics[width=0.35\textwidth]{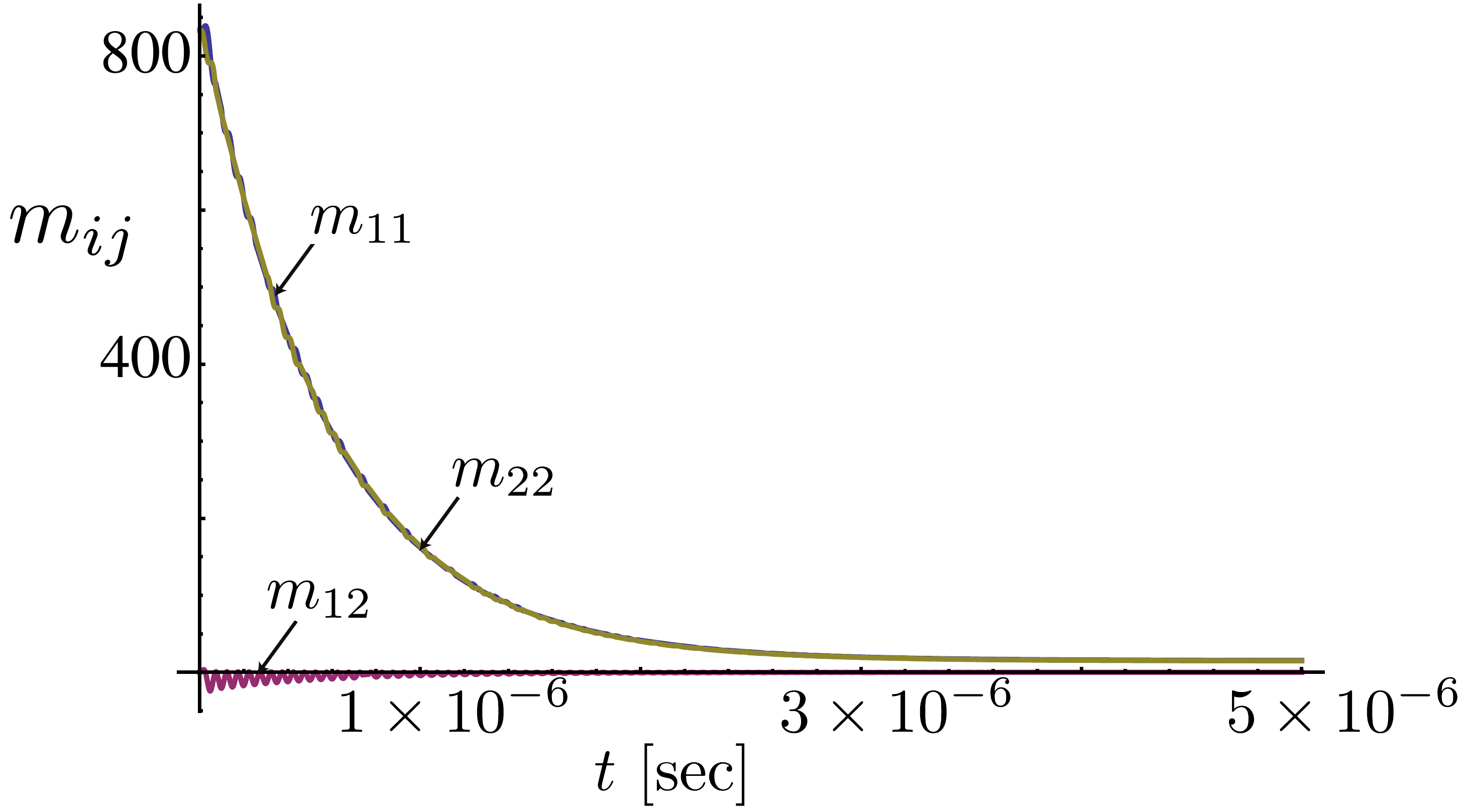}\includegraphics[width=0.35\textwidth]{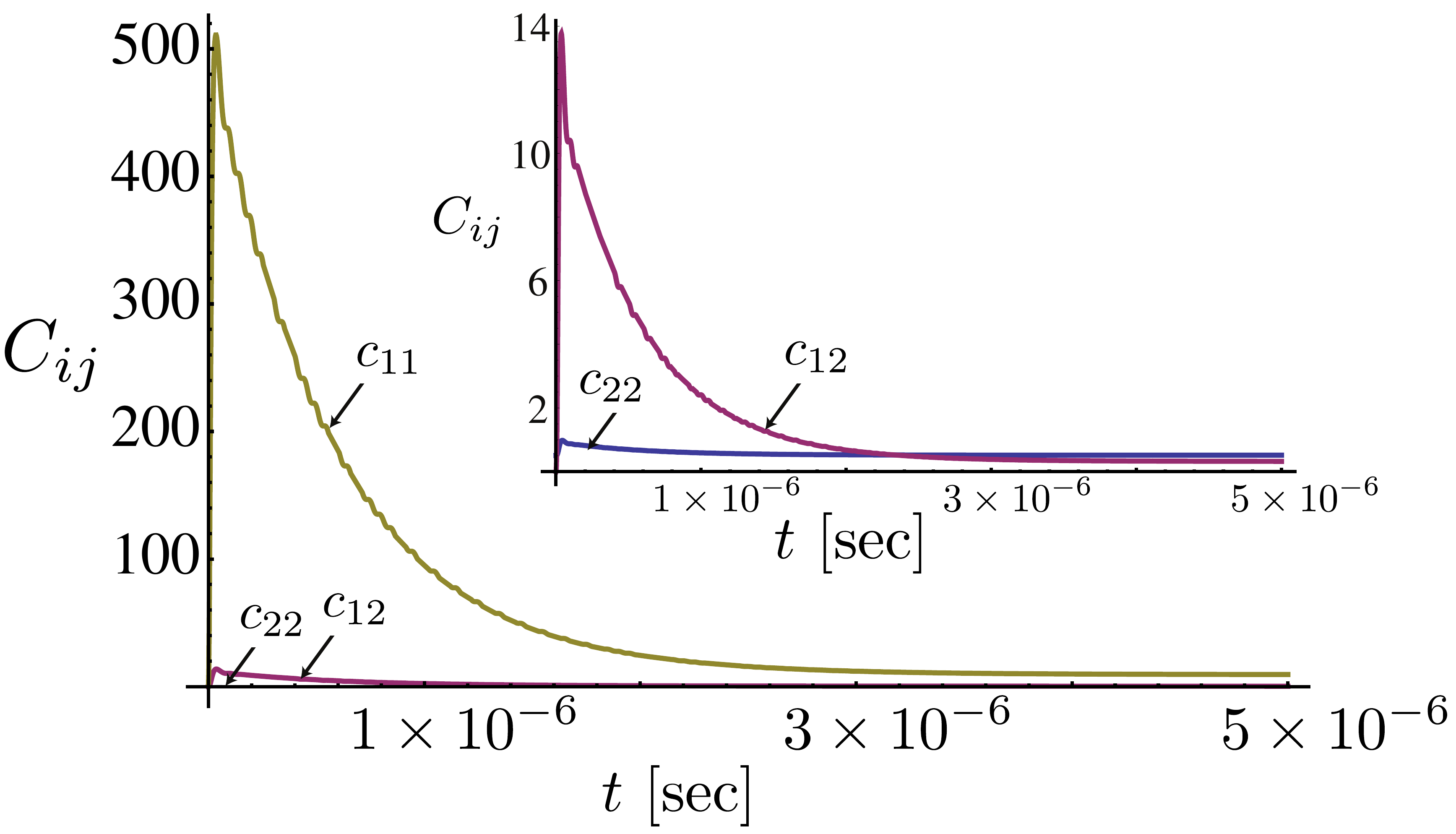}\includegraphics[width=0.35\textwidth]{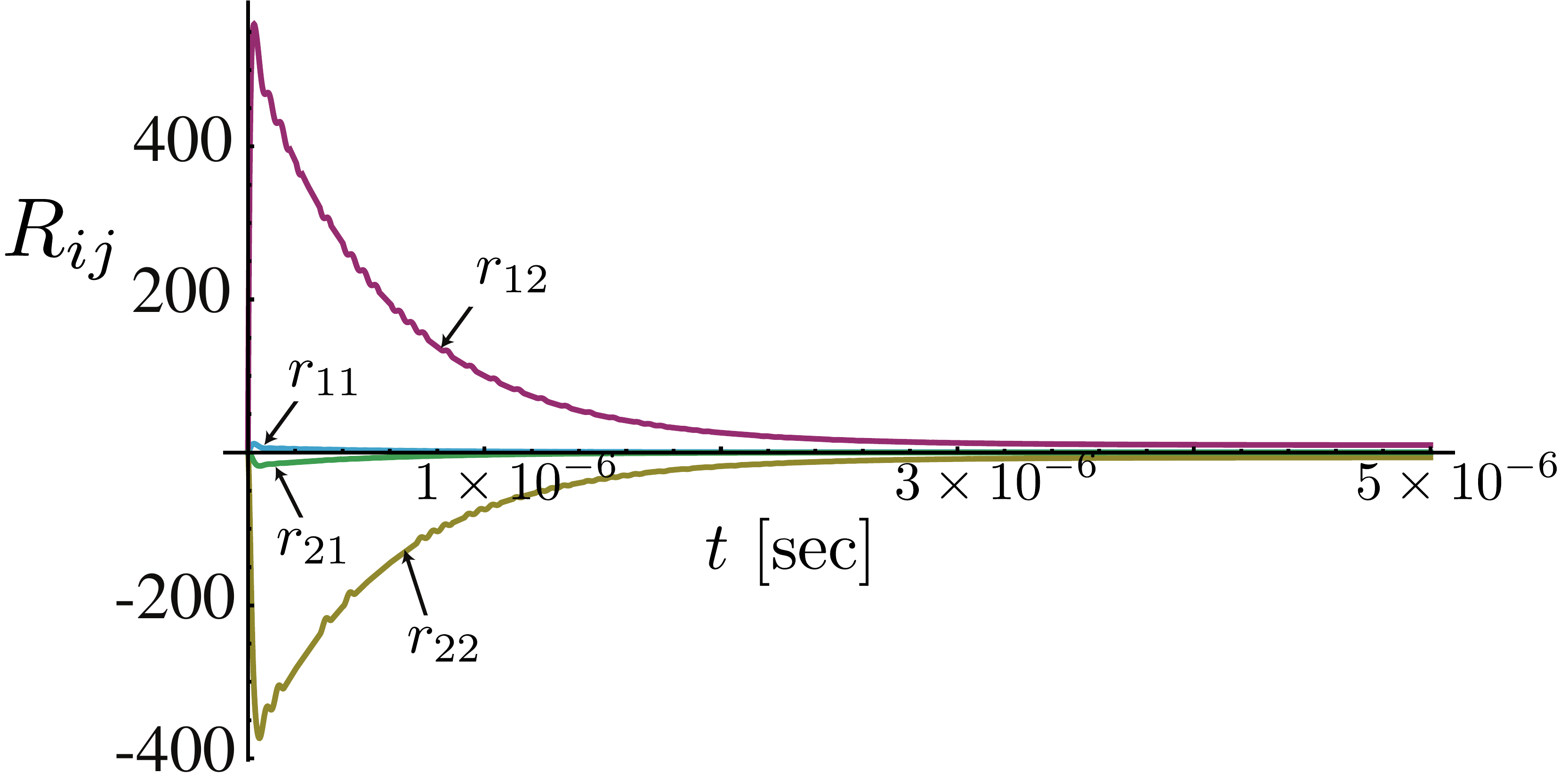}
\caption{Elements of the covariance matrix ${\cal V}_{MC}$ plotted against the interaction time $t$ for $\Delta/\omega_m=0.05$, $T=0.4$K and $m=5\times10^{-12}$Kg. We have taken a cavity of length ${\cal L}=1$mm, cavity frequency $\omega_c/2\pi\simeq4\times10^{14}$Hz and finesse $10^4$, pumped with $20$mW. The mechanical damping rate is as small as $\sim10$Hz. In panel {\bf (a)} we show the elements of the block ${\bm M}$ pertaining to the mechanical mode (notice that elements $m_{11}$ and $m_{22}$ are almost indistinguishable, while the steady-state form of such block is diagonal). Panel {\bf (b)} shows the elements of the field's block ${\bm C}$ (the inset shows a magnification of the plot for values of the matrix entries $[0,14]$. This allows to appreciate the elements that are not clearly visible from the main panel). Panel {\bf (c)} is for the elements of the correlation block ${\bm R}$.}
\label{ele}
\end{figure*} 

 \begin{figure}[b]
\includegraphics[width=0.75\linewidth]{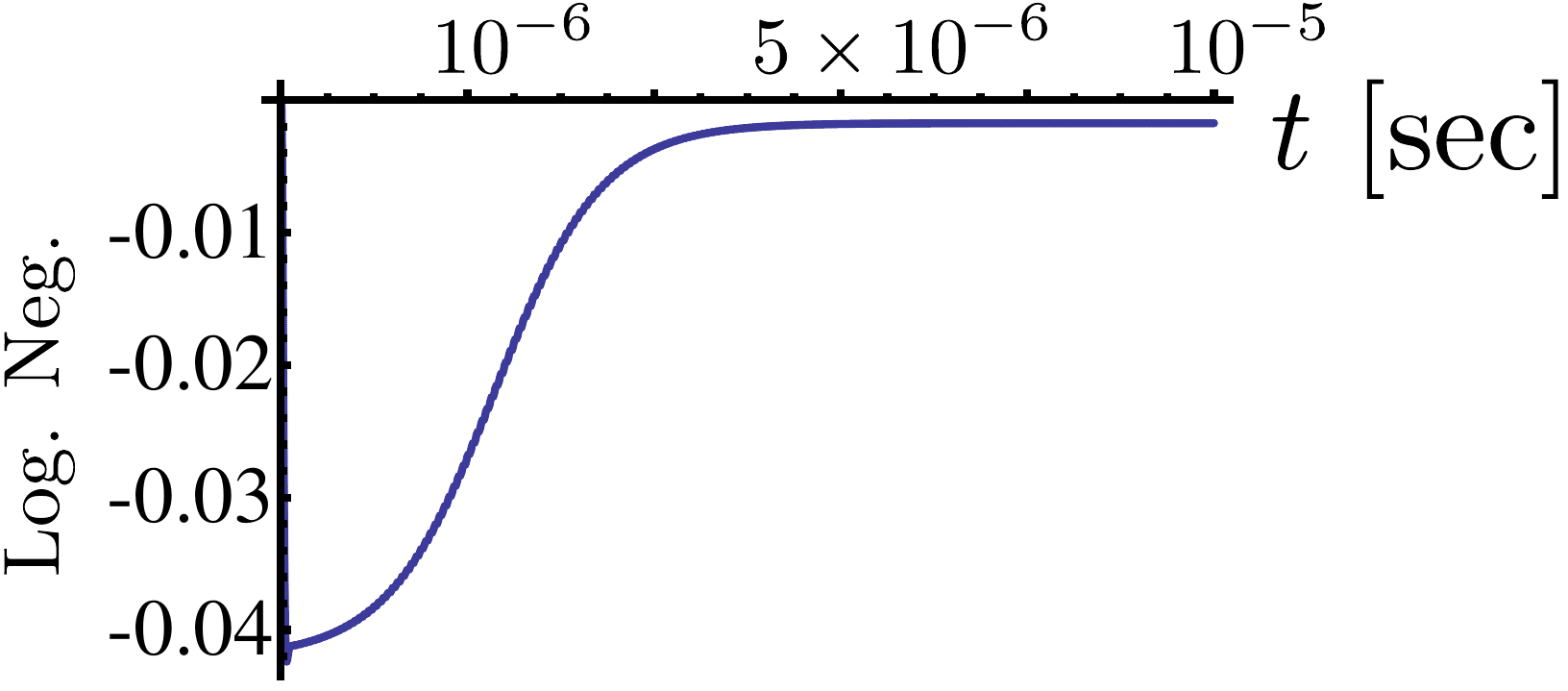}
\caption{Logarithmic negativity of the opto-mechanical state plotted against the interaction time $t$ for $\Delta/\omega_m=0.05$, $T{=}0.4$K and $m=5\times10^{-12}$Kg. We have taken a cavity of length ${\cal L}=1$mm, cavity frequency $\omega_c/2\pi\simeq4\times10^{14}$Hz and finesse $10^4$, pumped with $20$mW. The mechanical damping rate is as small as $\sim10$Hz.}
\label{noent}
\end{figure} 
A typical example of the time-behaviour of the matrix elements of ${\cal V}_{MC}$ is displayed in Fig.~\ref{ele} for a choice of the relevant set of parameters in our problem. The stability of the dynamical equations is guaranteed throughout the whole evolution. The system reaches its steady-state on a timescale roughly dictated by $\omega^{-1}_m$. In such long-time conditions, the behaviour of the system is fully captured by the Lyapunov equation 
\begin{equation}
\label{Lya}
{\cal K}_{MC}{\cal V}_{MC}+{\cal V}_{MC}{\cal K}^T_{MC}=-{\cal D}_{MC}
\end{equation}
whose explicit solution has been given in the Supplementary Material accompanying Ref.~\cite{mauro2}. As the corresponding expressions are too lengthy to be informative, we do not report them here. 

With these tools at hand, it is straightforward to evaluate the entanglement within the opto-mechanical device as quantified by the logarithmic negativity ${E}_{MC}=\max[0,{-}\ln 2\nu]$~\cite{logneg}. Here, $\nu$ is the smallest symplectic eigenvalue of the the matrix ${\cal V}'_{MC}=P{\cal V}_{MC}P$, where $P=\text{diag}(1,1,1,-1)$ performs momentum-inversion in phase-space. 
The results are shown in Fig.~\ref{noent}, where only the quantity $-\ln 2\nu$ is plotted to show that no entanglement is found in the optomechanical system, nor dynamically neither at the steady state. Although Fig.~\ref{noent} shows just an instance of this, our extensive numerical exploration confirmed this result for a large range of the key parameters in our problem. As we will see later on, however, the absence of entanglement does not hinder the validity of the state engineering scheme.  

\begin{figure}[b]
\centerline{\includegraphics[width=0.75\linewidth]{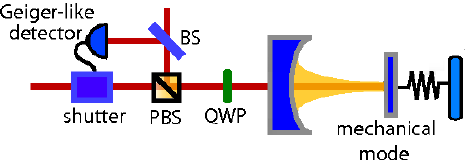}}
\caption{Sketch of the thought experiment. A pulsed laser field (with a set polarization) enters a cavity and drives the oscillations of an end mirror embodying a mechanical mode. Steady-state of the opto-mechanical system is reached on a timescale of a few multiples of $\omega_m$. The field is then photon-subtracted by a high-transmittivity beam splitter (BS) and a Geiger-like photo-detector. A click at the latter triggers a shutter (such as an electrically driven half wave-plate) that blocks the pumping process. Also shown are the symbols for a polarizing beam splitter (PBS) and a quarter wave-plate (QWP) used to direct the field to the cavity or the photo-subtraction unit.}
\label{schema}
\end{figure}

We now pass to the description of the scheme discussed in this Section. The field reflected by the mechanical mirror undergoes a single photon-subtraction process (that correspondingly stops the cavity-pumping process). A sketch of the proposed system is given in Fig.~\ref{schema}. The idea behind such proposal is that the correlations (not necessarily quantum) set between the mechanical oscillator and the field are enough to ``transfer" the non-classicality induced in the conditional state of the field by the photon-subtraction process to the state of the mechanical mode. In this respect, this proposal is along the lines of the scheme by Dakna {\it et al.}~\cite{dakna}, where a photon-number measurement on one arm of an entangled two-mode state projects the other one into a highly non-classical state. However, we should remark again that no assumption on an initially entangled optomechanical state will be necessary. While here we are interested in the formal aspects of the mechanism behind our proposal, a physical protocol will be addressed later on. 

Given the covariance matrix of the bipartite state of the system, we calculate the Weyl characteristic function as $\chi_W(\eta,\lambda){=}{\text{e}}^{-\frac{1}{2}\tilde{\bm q}{\bm\sigma}\tilde{\bm q}^T}$ with $\eta{=}\eta_r{+}i\eta_i$, $\lambda{=}\lambda_r{+}i\lambda_i$ and $\tilde{\bm q}{=}(\eta_r,\eta_i,\lambda_r,\lambda_i)$ the vector of complex phase-space variables. With this, the density matrix of the joint field-oscillator system can be written as $\varrho_{MC}={\pi^{-2}}\int\!\chi_W(\eta,\lambda)\hat{D}^\dag_M(\eta){\otimes}\hat{D}^\dag_C(\lambda){d}^2\eta\,{d}^2\lambda$~\cite{glauber}. Here, $\hat D_j(\alpha){=}\exp[\alpha\hat{a}^\dag_j-\alpha^*\hat{a}_j]$ is the displacement operator of mode $j{=}m,f$ of amplitude $\alpha{\in}\mathbb{C}$. 
The mechanical state resulting from the subtraction of a single quantum from the field is then described by
\begin{equation}
\label{ridotta}
\varrho_M=\frac{\cal N}{\pi^2}\int\!\chi_W(\eta,\lambda)\hat{D}^\dag_M(\eta)\,{\text{tr}}[\hat{a}\hat{D}^\dag_{C}(\lambda)\hat{a}^\dag]{d}^2\eta\,{d}^2\lambda
\end{equation}
with ${\cal N}$ a normalization constant. Eq.~(\ref{ridotta}) can be manipulated to get rid of the degrees of freedom of the cavity field by using the transformation rule of $\hat{a}$ induced by $\hat{D}^\dag_{C}(\lambda)$ and the closure relation $\pi^{-1}\int{d}^2\alpha\ket{\alpha}\!\bra{\alpha}_C{=}\hat\openone_C$, where $\ket{\alpha}$ is a coherent state~\cite{glauber}. After some algebra, one gets
\begin{equation}
\label{strumento}
{\text{tr}}[\hat{a}\hat{D}_C^\dag(\lambda)\hat{a}^\dag]{=}\!\!\int\frac{(|\alpha|^2{-}|\lambda|^2{+}2i\text{Im}[\lambda^*\alpha]{+}1)e^{-\frac{|\lambda|^2+2i\text{Im}[\lambda^*\alpha]}{2}}}{\pi}{d}^2\alpha.
\end{equation}
We now first perform the integration over $\lambda$, introduce the function 
\begin{equation}
{\cal C}(\alpha,\eta,\lambda){=}\chi_W(\eta,\lambda)(|\alpha|^2{-}|\lambda|^2{+}2i\text{Im}[\lambda^*\alpha]{+}1)e^{-\frac12|\lambda|^2}
\end{equation}
and cast the state of the mechanical mode as 
\begin{equation}
\varrho_M=\frac{\cal N}{\pi^3}\int\hat{D}^\dag_{M}(\eta){\cal F}[{{\cal C}(\alpha,\eta,\lambda)}]{d}^2\eta{d}^2\alpha
\end{equation}
with ${\cal F}[{{\cal C}(\alpha,\eta,\lambda)}]$ the Fourier transform of ${\cal C}(\alpha,\eta,\lambda)$. Such function encompasses any effects that the photon subtraction might have on the state of the mechanical system. As discussed before, the idea behind our proposal is that the correlations (not necessarily of a quantum nature) shared by the field and the mechanical mode are sufficient for the latter to experience the effects of the de-Gaussification induced by the photon subtraction. In what follows, we show that this is indeed the case. 

\begin{figure}[b]
\centerline{\includegraphics[width=0.45\textwidth]{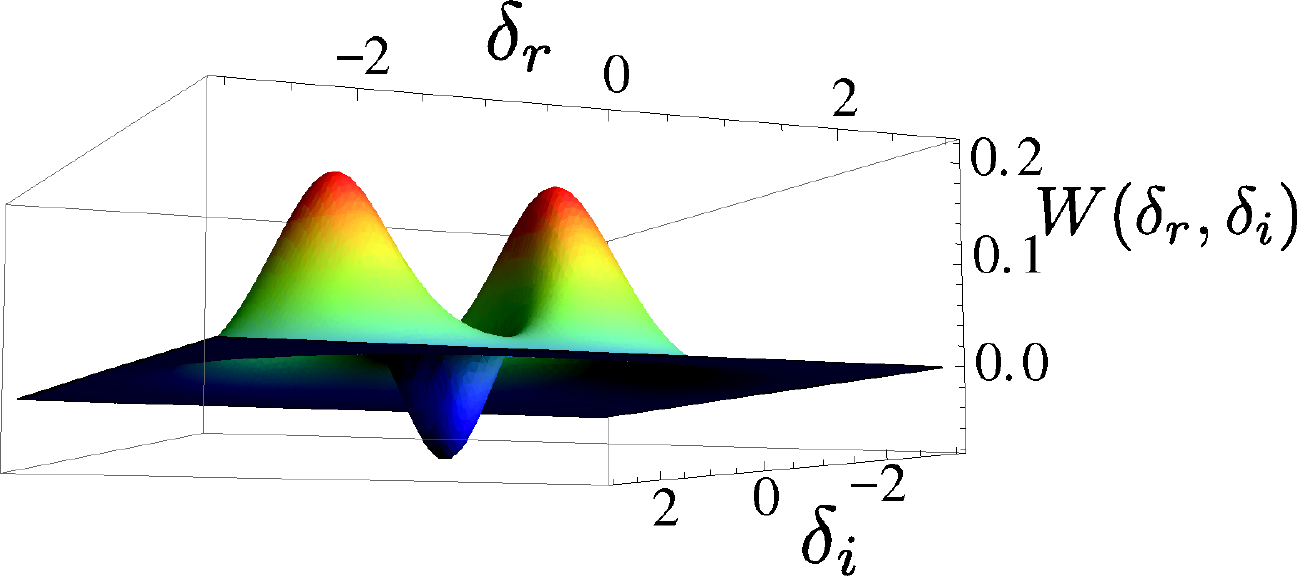}}
\caption{Wigner function of the mechanical mode for $\omega_m/2\pi=10$~MHz, $\Delta/\omega_m=0.05$, $T{=}0.4$~K and $m=5\times10^{-12}$~Kg. We have taken a cavity $1$ mm long,  frequency $\omega_c/2\pi\simeq4\times10^{14}$ Hz and finesse $10^4$, pumped with $20$ mW. In line with current experimental values, the mechanical damping rate is as small as $\sim10$ Hz. }
\label{esempio}
\end{figure}

In order to determine the features of $\varrho_M$, we address its Wigner function 
\begin{equation}
W(\delta_r,\delta_i){=}\frac{1}{\pi^2}\int\Xi(\gamma)\text{e}^{\gamma^*\delta-\gamma\delta^*}{d}^2\gamma~~~(\delta{=}\delta_r{+}i\delta_i),
\end{equation}
which is calculated using the characteristic function $\Xi(\mu){=}\text{tr}[\hat{D}_M(\mu)\varrho_M]$ evaluated at the phase-space point $\mu\in\mathbb{C}$. A lengthy yet straightforward calculation leads to
\begin{equation}
\label{wignerformale}
W(\delta_r,\delta_i)=\frac{2\pi{\cal A}}{(\textrm{det}\,{\bm M})^{5/2}(c_{22}{+}c_{11}{-}2)}\,e^{-2\left(\frac{\delta^2_i}{m_{11}}+\frac{\delta^2_r}{m_{22}}\right)}
\end{equation}
with 
\begin{equation}
\begin{aligned}
{\cal A}&=m^2_{22}[(c_{11}+c_{22}{-}2)m^2_{11}+(4\delta^2_i-m_{11})(r^2_{11}+r^2_{12})]\\
&+m^2_{11}(4\delta^2_r-m_{22})(r^2_{22}+r^2_{21})\\
&-8m_{11}m_{22}(r_{11}r_{21}+r_{12}r_{22})\delta_{r}\delta_{i}.
\end{aligned}
\end{equation}
The polynomial dependence of ${\cal A}$ on $\delta$ entails the non-Gaussian nature of the reduced mechanical state. We now seek evidences of non-classicality. A rather stringent criterion for deviations from classicality is the negativity of the Wigner function associated with a given state. This embodies the failure to interpret it as a classical probability distribution, which is instead possible whenever the Wigner function is positive~\cite{kenfack}. Building on the so-called Hudson theorem~\cite{hudson}, which proves that only multi-mode coherent and squeezed-vacuum states have non-negative Wigner functions, measures of non-classicality based on the negativity of the Wigner function have been formulated~\cite{kenfack}. More recently, operational criteria for inferring quantumness through the negative regions in the Wigner function have been proposed~\cite{mariWigner}. 
By inspection, we find that Eq.~(\ref{wignerformale}) can indeed be non-positive and achieves its most negative value for $\delta_{r,i}=0$. Assuming that none of the variances of the mechanical oscillator and the field are squeezed below the vacuum limit 
we have $W(0,0){<}0$ for
\begin{equation}
\label{forneg}
\frac{m_{11}}{m_{22}}>\frac{({c_{11}+c_{22}-2})m_{11}{-}({r^2_{11}+r^2_{12}})}{({r^2_{22}+r^2_{21}})},
\end{equation}
which is quite an interesting finding. First it shows that the non-classicality of the mechanical mode depends on its initial {degree of squeezing} given by the ratio $m_{11}/m_{22}$~\cite{commentsq}. 
Second, we remark the ``plug$\&$play" nature of Eq.~(\ref{forneg}): by determining the matrix ${\cal V}_{MC}$ of the two modes under scrutiny, which can be performed as described in~\cite{Vitali2007,MauroPRL,mari}, one can determine the amplitude of the negative peak of $W(\delta_r,\delta_i)$ without the necessity of reconstructing the full Wigner function. Clearly, this is a major practical advantage that allows us to bypass the demanding needs for a tomographically complete set of data. In addition, as the entries of a covariance matrix are determined with a rather good precision~\cite{fabre}, we expect a covariance matrix-based criterion for non-classicality to be less prone to artifacts (such as large error bars) that would mask negativity and thus erroneously make the Wigner function consistent with a classical probability distribution (the reconstruction of ${\cal V}_{MC}$ can be performed as described in~\cite{Vitali2007,MauroPRL,mari} via all-optical procedures enjoying a rather good precision and accuracy~\cite{fabre}). Finally, Eq.~(\ref{forneg}) is handy to gauge the quality of the parameters of a given experiment with respect to the achievement of a non-classical mechanical state. 
Fig.~\ref{esempio} shows that $W(\delta_r,\delta_i)$ can take significantly negative values for proper choices of the parameters and  quite a large temperature.
\begin{figure}[t]
{\bf (a)}\hskip3.5cm{\bf (b)}
\includegraphics[width=0.5\textwidth]{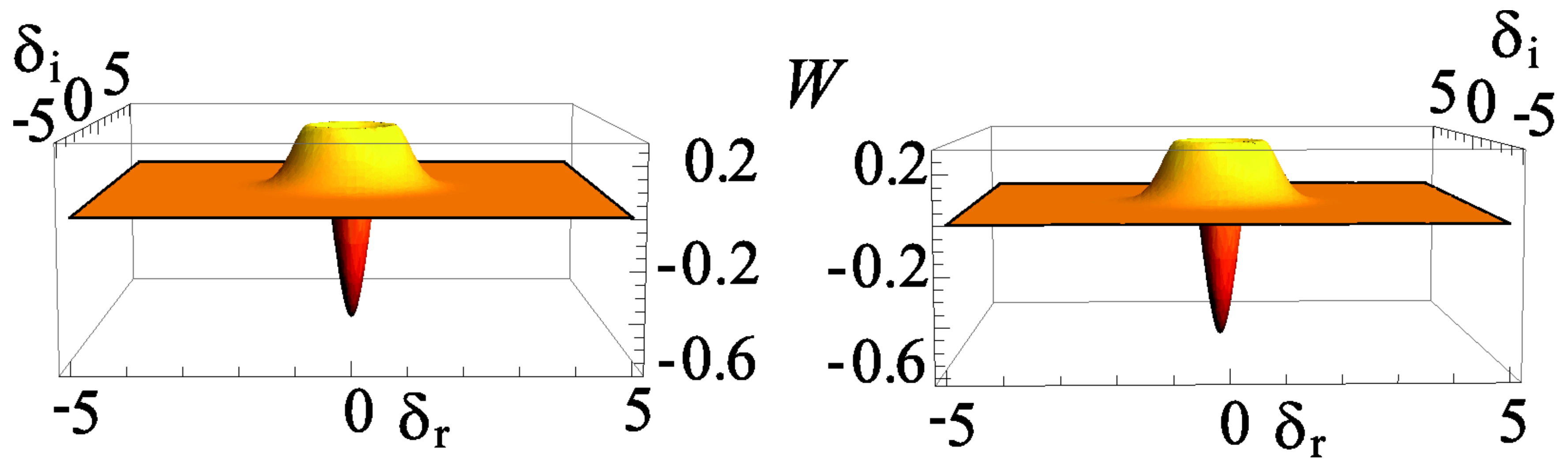}
\caption{{\bf (a)} Conditional Wigner function of the mechanical mode after photon subtraction for $T=4\times10^{-3}$~K and $m=5\times10^{-12}$~Kg. {\bf (b)} Same as panel {\bf (a)} but assuming that modes $M$ and $C$ are initially in a pure two-mode squeezed vacuum state of squeezing factor $\zeta=0.4$.} 
\label{paragone}
\end{figure}

\begin{figure}[b]
\includegraphics[width=0.5\textwidth]{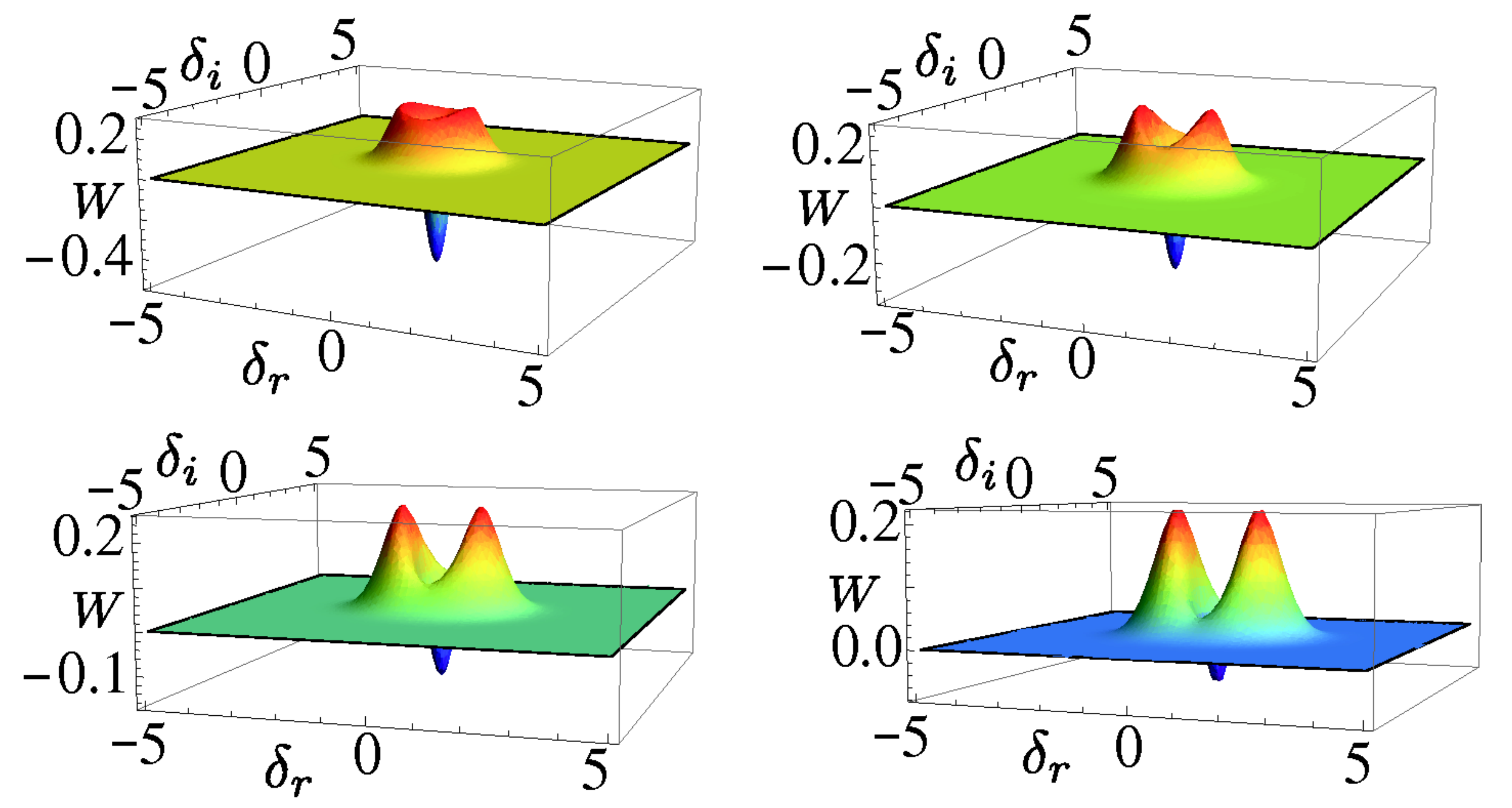}
\caption{Snap-shot of the phase-space dynamics of the Wigner function of the mechanical mode for increasing values of the temperature. We have taken $T=0.1,0.2,0.3,0.4$ K in going from left to right, top to bottom panel. Other parameters as in Fig.~\ref{esempio}.}
\label{sequenza}
\end{figure}

\begin{figure}[b]
\centerline{
\includegraphics[width=0.45\textwidth]{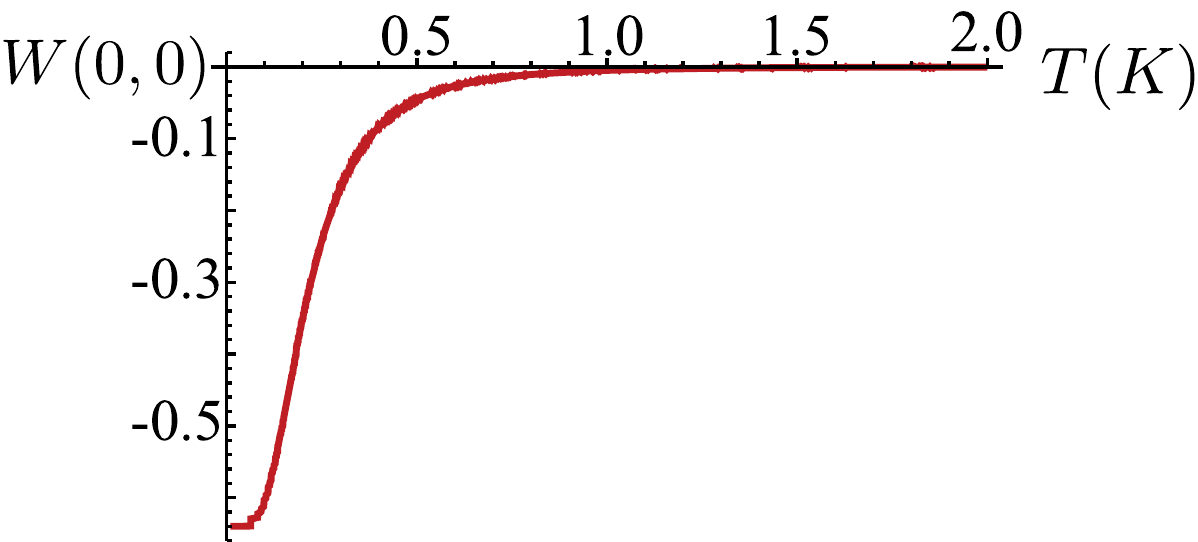}~~
}
\caption{Negativity of $W(0,0)$ against temperature for $\Delta/\omega_m=0.05$. Other parameters as in Fig.~\ref{esempio}. }
\label{effetti}
\end{figure}

Depending on the parameters being used, optomechanical entanglement can persist up to temperatures of about $20$~K~\cite{Vitali2007,MauroPRL}. We now wonder whether the conditional state-engineering scheme proposed here enjoys this very same feature. First, we notice that by subtracting a single photon from mode $2$ of a two-mode squeezed vacuum $\ket{\zeta}=(\cosh\zeta)^{-1}\sum^\infty_{n{=}0}\lambda^n\ket{n,n}_{12}$ with squeezing factor $\zeta{<}1$ and $\lambda{=}\tanh\zeta$, the Wigner function of the unmeasured mode $1$ is basically identical to $W(\delta_{r},\delta_i)$ in the limit of small temperature ($T{\sim}{1}$~mK). This is quantitatively illustrated in Fig.~\ref{paragone} {\bf (a)} and {\bf (b)}, where the Wigner function of mode $1$ for $\zeta{=}0.4$ is shown to be indistinguishable from the analogous function of mode $m$ after the application of our scheme. Such a similarity is understood as follows: The high-quality mechanical mode, large-finesse cavity and low-temperature limit used here make a unitary approach to the time evolution of the optomechanical system quite appropriate. The dynamics, in such case, involve two-mode squeezing of modes $M$ and $C$~\cite{MauroJPB}, which explains the similarity seen in Fig.~\ref{paragone} and discussed here. Such an analogy is illuminating as it is straightforward to see that the effects experienced by mode $1$ in the (unnormalized) unilaterally photon subtracted state $\hat{a}_2\ket{\zeta}\bra{\zeta}_{12}\hat a^\dag_2$ can be interpreted as the addition of a photon, which is the origin for non-classicality of the resulting state (as signaled by the negativity of its Wigner function). When $T$ is increased, however, the rotational invariance of $W(\delta_r,\delta_i)$ is progressively lost. As a result of the loss of coherence, $W(\delta_r,\delta_i)$ splits into two localized peaks, which become progressively Gaussian-shaped as the temperature grows and represent the thermal average of displaced states in the phase-space. This effect is clearly illustrated in Fig.~\ref{sequenza}, where a snap-shot of the phase-space dynamics against $T$ is shown. As anticipated above, for the parameters chosen in our analysis, quite large negative values are observed in $W(\delta_r,\delta_i)$ for $T{\ll}{1}$~K and the Wigner function remains negative up to ${1}$~K [see Fig.~\ref{effetti}]. In Sec.~\ref{robustness} we will estimate the life-time of the enforced non-classicality.  

\subsection{Thought experimental scheme}

Our approach so far was to consider photon subtraction at a formal level. Although, as we will demonstrate shortly, the accuracy of the quantitative results achieved in this way is excellent we now go beyond such an abstract description and assess a close-to-reality version of our proposal. In a real experiment, the non-Hermitian operation of subtracting a photon is realized by superimposing, at a high transmittivity beam splitter (BS), mode $C$ to an ancilla $\aleph$ prepared in the vacuum state~\cite{grangier,grangier1}. This makes ours a three-body system characterized by the variance matrix  ${\cal V}_{MC\aleph}=(\openone_m{\oplus}{\bf B}^T_{C\aleph})({\cal V}_{MC}{\oplus}\openone_\aleph)(\openone{\oplus}{\bf B}_{C\aleph})$, where we have introduced the symplectic BS transformation ${\bf B}_{C\aleph}{=}\openone{\otimes}(\tau\openone_\aleph)-i\sigma_y{\otimes}(r\openone_\aleph)$. 
Here $\tau$ is the transmittance of the BS ($r^2{+}\tau^2{=}1$). The characteristic function of such correlated three-mode state is $\tilde\chi_W(\eta,\lambda,\xi)=\exp[{-\tilde{\bm q}{\cal V}_{MC\aleph}\tilde{\bm q}^T}/2]$, where $\tilde{\bm q}{=}(\eta_r,\eta_i,\lambda_r,\lambda_i,\xi_r,\xi_i)$ is the vector of phase-space variables of the three modes and $\xi{=}\xi_r{+}i\xi_i$. The corresponding density matrix is thus given by 
\begin{equation}
\tilde\varrho=\frac{1}{\pi^3}\int\tilde\chi_W(\eta,\lambda,\xi)\hat{D}^\dag_M(\eta)\hat{D}^\dag_C(\lambda)\hat{D}^\dag_\aleph(\xi){d}^2\eta{d}^2\xi{d}^2\lambda.
\end{equation}
We post-select the event where a single click is obtained at a photo-resolving detector measuring the state of mode $\aleph$, thus projecting its state onto $\ket{1}_\aleph$. This gives the conditional state  \begin{equation}
\tilde\varrho_{M}=\frac{\tilde{\cal N}}{\pi^2}\int\tilde\chi_W(\eta,0,\xi)\hat{D}^\dag_{M}(\eta)(1-|\xi|^2)\text{e}^{-\frac{1}{2}|\xi|^2}{d}^2\eta\, d^2\xi
\end{equation} 
with $\tilde{\cal N}$ the normalization factor and where the formula ${}^{}_{\aleph}\!\bra{1}\hat D^\dag_\aleph(\xi)\ket{1}_\aleph{=}\textrm{e}^{-|\xi|^2/2}(1{-}|\xi|^2)$ has been used~\cite{glauber}. The calculation of the Wigner function $\tilde{W}(\delta_r,\delta_i)$ of the mechanical mode then proceeds along the lines sketched above. The resulting analytic expression is however very involved and can only be managed numerically. A thorough analysis shows that already at $\tau^2{=}{0.8}$, $\tilde{W}(\delta_r,\delta_i)$ reproduces very accurately the behaviour of $W(\delta_r,\delta_i)$. For instance, at the value of $\Delta$ used to produce the figures in this paper, we get $|\tilde{W}(0,0){-}W(0,0)|{\simeq}{10^{-8}}$. Clearly, the quality of the agreement depends crucially on the BS transmittivity. It is enough to take $\tau{\sim}0.9$ to get full agreement between $\tilde{W}(\delta_{r},\delta_i)$ and its formal counterpart over the range $\Delta\in[0,0.1]\omega_m$, where non-classicality is observed.

\subsection{Role of imperfections}

For the sake of a practical implementation, it is important to assess the role that imperfections play in the performance of our scheme. The most relevant one for our tasks is the inability of discriminating the number of photons impinging on the detector used to subtract a single photon from $f$. We thus consider a finite-efficiency Geiger-like detector modelled by the positive operator valued measurement $\{\hat{\Pi}^\text{nc}_\aleph,\openone_\aleph{-}\hat{\Pi}^\text{nc}_\aleph\}$ with $\hat{\Pi}^\text{nc}_\aleph=\sum^\infty_{j=0}(1-\epsilon)^j|j\rangle\langle{j}|_\aleph$ the projection operator accounting for ``no-click'' at the detector. Due to the finite efficiency $\epsilon\in[0,1]$, a photonic state with $j$ photons has a probability $(1{-}\epsilon)^j$ to be missed. It is straightforward to see that the Wigner function corresponding to the state of mode $M$ is then given by ${\cal W}(\delta_r,\delta_i){=}\pi^{-2}{\cal F}[{\Xi(\mu,\epsilon)}]$ with
\begin{equation}
\begin{aligned}
\Xi(\mu,\epsilon)\propto\tilde{\chi}(\mu,0,0)&-\sum^\infty_{j=0}\frac{(1{-}\epsilon)^j}{\pi}\\
&\times\int\tilde{\chi}(\mu,0,\xi)e^{-\frac{|\xi|^2}{2}}{\cal L}_{j}(|\xi|^2)d^2\xi 
\end{aligned}
\end{equation}
and ${\cal L}_j(|\xi|^2)$ the Laguerre polynomial of order $j$. By 
inverting the order of sum and integration and using the generating function of Laguerre polynomials~\cite{grad},
 we have 
 \begin{equation}
 \sum^\infty_{j=0}(1{-}\epsilon)^j{\cal L}_{j}(|\xi|^2){=}\frac{e^{-\frac{2-\epsilon}{2\epsilon}|\xi|^2}}{\epsilon}, 
 \end{equation}
 so that $\Xi(\mu,\epsilon)\propto\tilde{\chi}(\mu,0,0)-\Phi(\mu,\epsilon)$ with
\begin{equation}
\label{WeylIneff}
\Phi(\mu,\epsilon)=-\frac{1}{\pi\epsilon}\int\tilde{\chi}(\mu,0,\xi)e^{-\frac{2-\epsilon}{2\epsilon}|\xi|^2}d^2\xi.
\end{equation}
The effects of detection inefficiency are thus quantified by considering that $\Phi(\mu,\epsilon)$ is the only term that depends on $\epsilon$ in $\Xi(\mu,\epsilon)$. Therefore, $|\Phi(\mu,1){-}\Phi(\mu,\epsilon)|$ provides a quantitative estimate of the differences due to a non-ideal detector. Numerically, for $\epsilon\ge0.7$ we have found negligible values of this quantity (${\sim}{10}^{-2}$), almost uniformly with respect to $\tau$: Fig.~\ref{effetti} is reproduced without noticeable differences. Moreover, the performance of our scheme is not affected by even smaller detection efficiency. This is in line with the analysis conducted on photon subtraction processes: detection inefficiencies only lower the probability of success of the scheme without affecting the fidelity of the process itself~\cite{grangier,grangier1}. Likewise, the dark count rate of photo-detectors can generally be neglected in photon subtraction experiments~\cite{grangier,kim2}.

\subsection{Robustness of non-classicality}
\label{robustness}

\begin{figure}[t]
\includegraphics[width=0.5\textwidth]{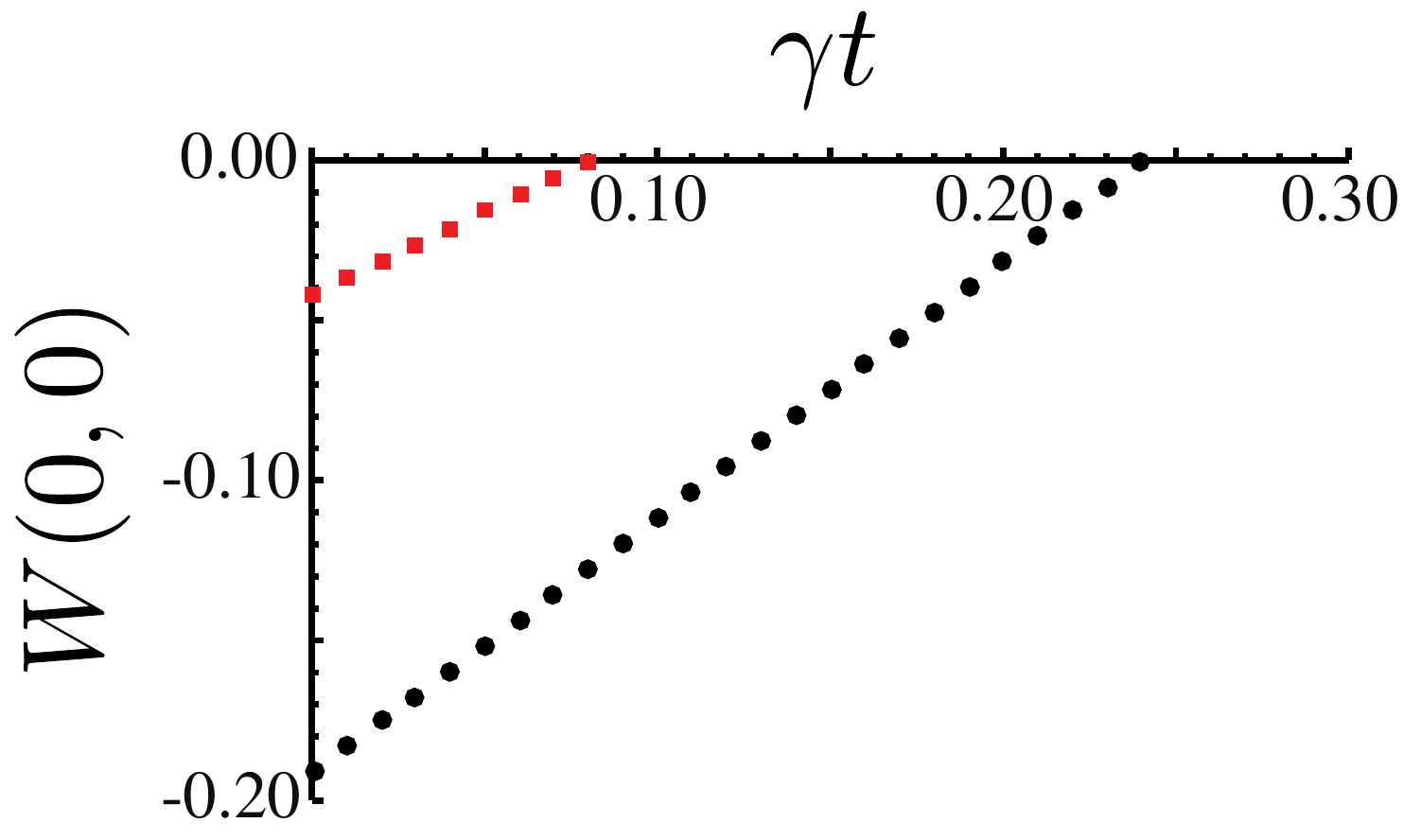}
\caption{Negativity of $W(0,0)$ against the dimensionless time $\gamma t$ for $\Delta/\omega_m=0.1$ and $m=\times10^{-12}$~Kg. We have taken a cavity of length $1$~mm, cavity frequency $\omega_c/2\pi\simeq4{\times}10^{14}$ Hz and finesse $10^4$, pumped with $20$ mW. We have considered $T=0.04$~K (red squared points) and $T=4$~mK (black circle points).}
\label{decado}
\end{figure}

Let us suppose now that a non-classical state of the mechanical mode has been engineered by means of the protocol put forward in this Section. How long would the enforced mechanical non-classicality last? A quantitative answer to this question comes from considering that, as soon as the a photon is revealed at the photo-detector shown in Fig.~\ref{schema}, the pumping of the cavity should be terminated. This means that, from that time on, the mechanical mode would evolve freely yet being subjected to phononic damping at non-zero temperature. This, in general, would be described by a non Markovian dynamics basically corresponding to quantum Brownian motion~\cite{Vitali2007}. However, we consider the experimentally relevant condition of $\omega_m\gg{\gamma}_m$ that makes the Brownian damping equivalent to a Markovian dissipation process such as the one experimenced by a lossy optical mode~\cite{Vitali2007,MauroPRL}. In this case, the density matrix $\rho_m$ of the mechanical mode evolves under the influences of a dissipative thermal bath according to the master equation (in the interaction picture) 
\begin{equation}
\begin{aligned}
\partial_t\rho_{m}&=\gamma(\nbar+1)(2\hat{m}\rho_{m}\hat{m}^\dag-\hat{m}^\dag\hat{m}\rho_m-\rho_m\hat{m}^\dag\hat{m})\\
&+\gamma\nbar(2\hat{m}^\dag\rho_m\hat{m}-\hat{m}\hat{m}^\dag\rho_m-\rho_m\hat{m}\hat{m}^\dag).
\end{aligned}
\end{equation}
This is written as a Fokker-Planck equation for the Wigner function $W(\delta_r,\delta_i,t)$ of the mechanical mode at time $t$ of the evolution as
\begin{equation}
\partial_tW(\sigma_r,\sigma_i,t)=\frac{\gamma}{2}\left[\partial_{\sigma}\sigma+\partial_{\sigma^*}\sigma^*+\Nbar\partial^2_{\sigma\sigma^*}\right]W(\sigma_r,\sigma_i,t)
\label{fp}
\end{equation}
with $\sigma\in\mathbb{C}$ a phase-space variable ($\sigma^*$ is its complex conjugate) and $\Nbar=2\nbar+1$. Eq.~(\ref{fp}) is solved straightforwardly by convoluting $W(\sigma_r,\sigma_i,0)$, {\it i.e.} the Wigner function of the prepared non-classical state, with the Wigner function of a thermal state as
\begin{equation}
W(\delta_r,\delta_i,t)=\frac{2}{\pi{\tau}\Nbar}\int d^2\sigma~W(\sigma_r,\sigma_i,0)e^{-2\frac{|\sigma-\delta\text{e}^{-\gamma t}|^2}{\Nbar{\tau}}}
\end{equation}
where $\tau=1-\exp[-2\gamma t]$. The maximum negative amplitude of the Wigner function is achieved at $\delta=0$ and one can study its behaviour against the evolution time $t$. Typical results are shown in Fig.~\ref{decado}, where we address the case of two different values of $T$. Clearly, as the temperature rises, the time-window within which non-classicality is preserved shrinks, in line with intuition. However, given the small damping rates currently achievable through accurate micro and nano- fabrication processes of the mechanical modes under scrutiny, the width $t^*$ of such window remains quite large with $\gamma t^*=0.08\rightarrow{t^*}=1.3$~ms in the worst case scenario shown in Fig.~\ref{decado} [cfr. red squared points].

\subsection{Hybridisation for photo-subtraction}

Although we have used the language of linear optics to describe the working principles of the scheme presented here, the required intracavity photo-subtraction process can be implemented using alternative experimental techniques. For instance, when a suitable evolution time is chosen, a hybrid architecture incorporating an atomic medium (consisting of either a single particle or a multi-atom one) that interacts resonantly with the cavity field achieves exactly the effect of subtracting excitations from the field. Indeed, let us consider the Hamiltonian coupling the field to the effective dipole of a general spin-like system, such as a two-level atom or a collection of them (collectively coupled to the cavity field). Such resonant interaction can be generically written as 
\begin{equation}
\hat H_{C,spin}=\hbar{\cal G}(\hat a \hat \Sigma^+_{spin}+h.c.)
\end{equation}
with ${\cal G}$ the Rabi frequency of the interaction and $\hat\Sigma^+_{spin}$ the raising operator of the spin-like system. A straightforward calculation shows that the operation that is effectively implemented on the cavity field when the spin (prepared in its fundamental state) is found, at time $t$, in its excited one is proportional to $\sin({\cal G}\sqrt{\hat a^\dag\hat a}t)/\sqrt{\hat a^\dag \hat a}\hat a$. By choosing $t$ such that $\sin({\cal G}\sqrt{\hat a^\dag\hat a}t)/\sqrt{\hat a^\dag \hat a}\simeq1$ for any number of photons in the cavity (a condition satisfied for very small values of the Rabi frequency), this realises the required photo-subtraction process. Notice that the request of weak coupling simply affects the probability to effectively subtract a photon and not the quality of the resulting state. 

This simple example illustrates the benefits of implementing a reliable tripartite configuration incorporating an atomic subsystem into the optomechanical device addressed here. Indeed, the possibilities offered by such a configuration go well beyond the specific case addressed here, as it will be argued in the remainder of this review.

 On the other hand, many are the questions opened by the demonstrated possibility to engineer the state of a mechanical system by conditioning optical modes through photo subtraction stages. A particularly interesting one would be the {\it steering} of the mechanical mode towards a desired target state through the sequential application of photo subtraction and addition processes. 

\section{Hybrid cavity-BEC optomechanics}
\label{BECpaper}

We now move to the addressing of an explicitly hybridised system of much theoretical and experimental appeal. We consider the placement of a Bose-Einstein condendensate (BEC) into an optomechanical cavity as a paradigm for a device combining the handiness of ultra-cold atomic systems to the potential for mesoscopic non-classicality of optomechanical ones. In this Section, we introduce the model and study the phenomenology of mutual back-action dynamics between macroscopic degrees of freedom embodied by physical systems of different nature. We show a non-trivial intertwined dynamics between collective atomic modes, coupled to the cavity field, and the mechanical one, which experiences the radiation-pressure force. 

We start by first focusing on the atom-induced back-action effects over the mechanical device. In particular, our aim is to cool the vibrating mechanical mode, showing that the interaction with the atomic degrees of freedom  (albeit indirect) modifies the cooling capabilities of an optomechanical system quite considerably with respect to the performances predicted and demonstrated experimentally so far~\cite{AspelmeyerRMP}. 
In order to get a picture of the physical situation at hand, let us address the details of the proposed thought experiment. 

\begin{figure}[t]
\psfig{figure=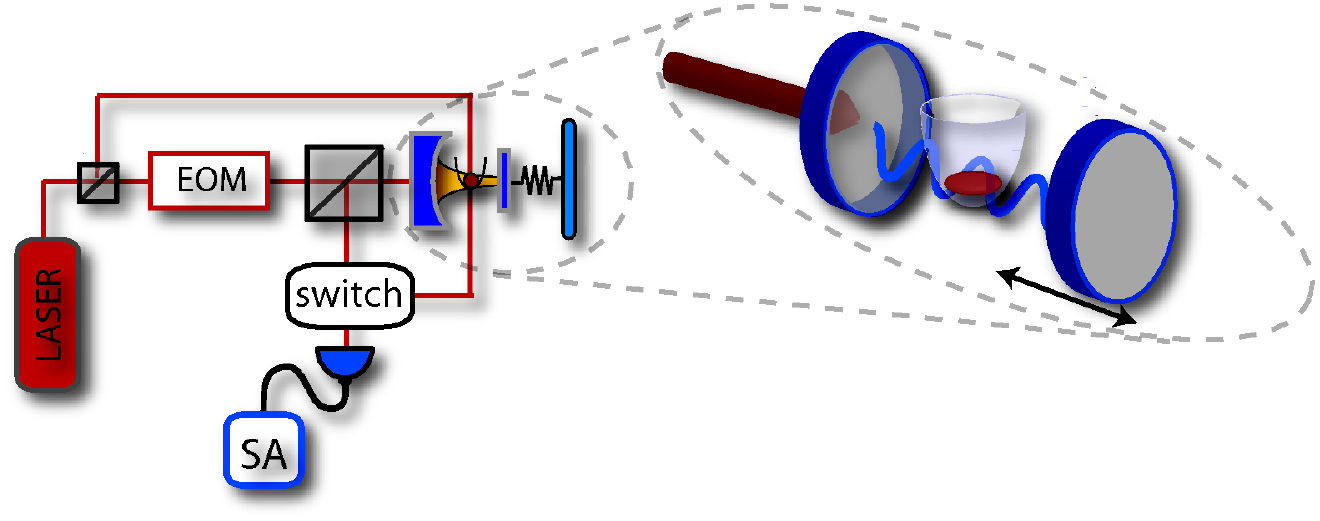,width=\linewidth}
\caption{A laser is split by an unbalanced beam splitter. The transmitted part is phase-modulated and enters the cavity coupled to a BEC. The (weak) reflected part of the pump laser probes the BEC. The signals from the cavity and the BEC go to a detection stage consisting of a switch (selecting the signal to analyze), a photodiode and a spectrum analyzer (SA).}
\label{schemino}
\end{figure}

As an addition to the general optomechanical setting illustrated in Sec.~\ref{Sottraggopaper}, we consider a BEC confined in a large-volume trap within the cavity~\cite{esteve2008,Esslinger2008} [cfr.~Fig.~\ref{schemino}]. Alternatively, the BEC could be sitting in a 1D optical-lattice generated by a trapping mode sustained by a bimodal cavity~\cite{Stamper2008}. The atom-cavity interaction is insensitive to the details of the trapping and our study holds in both cases. In the weakly interacting regime~\cite{StringariPitaevskii}, the atomic field operator can be split into a classical part (the condensate wave function) and a quantum one (the fluctuations) expressed in terms of Bogoliubov modes. Recent experiments coupling a BEC to an optical resonator~\cite{Esslinger2008} suggest that the Bogoliubov modes interacting significantly with the cavity field are those with momentum $\pm2k_c$ ($k_c$ is the cavity-mode momentum) while the condensate can be considered to be initially at zero temperature. As a result, the cavity field excites superpositions of atomic momentum modes giving rise to a periodic density grating sensed by the cavity. 


We write the Hamiltonian of the system made out of the cavity field, the movable mirror and the BEC as
\begin{equation}
\hat{\cal H}=\sum_{j=A,M,C}\hat{\cal H}_j+\hat{\cal H}_{AC}+\hat{\cal H}_{MC}
\end{equation}
where $\hat{\cal H}_{M},\hat{\cal H}_{C}$ and $\hat{\cal H}_{MC}$ have been introduced in Eq.~\eqref{modellobase} and  
\begin{equation}
 \hat{\cal H}_A=\hbar \tilde\omega\hat{c}^\dagger\hat{c}
\end{equation}
describes the free energy of the atomic system, described as a collective Bogoliubov mode with bosonic operators $\hat{c}$ ($\hat{c}^\dag$) and frequency   
$\tilde\omega$ and $\hat{c}$ ($\hat{c}^\dag$). The atoms-cavity interaction reads
\begin{equation}
\label{eq:HAC}
\hat{\cal H}_{AC}=\frac{{\hbar g^2 N_0}}{{2\Delta_a}}\hat{a}^\dagger\hat{a} + \hbar\sqrt{2}\zeta\hat{Q}\hat{a}^\dagger\hat{a}
\end{equation}
and thus contains two contributions: the first one, proportional to the number of condensed atoms $N_0$, comes from the condensate only while the second is related to the position-like operator $\hat{Q}{=}(\hat{c}+\hat{c}^\dag)/\sqrt{2}$ of the Bogoliubov mode (its canonically conjugated operator will be hereafter indicated as $\hat{P}=i(\hat{c}^\dag{-}\hat{c})/\sqrt{2}$). In Eq.~\eqref{eq:HAC}, $g$ is the vacuum Rabi frequency for the dipole-like transition connecting the atomic ground and excited states, $\Delta_a$ is the detuning of the atomic transition from the cavity frequency and the coupling rate ${\zeta{\propto}\sqrt{N_0}{g^2}/\Delta_a}$. A rigorous calculation, which we sketch here, shows that $\zeta$ also depends on the Bogoliubov mode-function and can be conveniently tuned. We work in conditions of far-off resonant coupling between the atoms in the condensate and the field, {\it i.e.} $\Delta_a\gg\{{g},\Gamma_a\}$ (with $\Gamma_a$ the atomic spontaneous-decay rate), so that we can adiabatically eliminate the excited state in the dipolar transition. The resulting cavity-condensate interaction Hamiltonian is  then~\cite{Larson}
\begin{equation}
\hat{\cal H}_{AC} = \hbar U_0\hat{a}^\dagger\hat{a} \int dx \cos^2 (k_c x)\hat{\psi}^\dagger(x)\hat{\psi}(x)
\end{equation}
with $U_0 = g^2/\Delta_a$ 
and $\hat{\psi}(x)$ the atomic field operator, fulfilling bosonic commutation relations. We have restricted the dynamics of the atoms only to the direction $x$ parallel to the cavity axis by assuming tight confinement of the atomics cloud in the transverse directions. We use a Bogoliubov expansion of 
$\hat{\psi}(x)$~\cite{StringariPitaevskii}
\begin{equation}
\hat{\psi}(x) = \sqrt{N_0}\hat{\psi}_0(x) +\sum_{k>0,\sigma=\pm}\left[ u_{k\sigma}(x)\hat{c}_{k\sigma} - v_{k\sigma}^*(x)\hat{c}_{k\sigma}^\dagger  \right]
\end{equation}
where $\hat{\psi}_0(x)$ is the condensate wavefunction that satisfies the Gross-Pitaevskii equation. By assuming a homogeneous system (we have assumed that the condensate wavefunction is not affected by the coupling to the cavity field) we can write $\psi_0(x)=1/\sqrt{V}$. We expand the quantum part of the field operator in real Bogoliubov modes with definite parity
\begin{equation}
\begin{aligned}
u_{k+}(x) &=\alpha_k \sqrt{\frac 2V} \cos(k x),\,\,
u_{k-}(x) = \alpha_k \sqrt{\frac 2V} \sin(k x)\\
v_{k+}(x) &=\beta_k \sqrt{\frac 2V} \cos(k x),\,\,
v_{k-}(x) = \beta_k \sqrt{\frac 2V} \sin(k x)
\end{aligned}
\end{equation}
and 
\begin{equation}
\mu_k=\sqrt{\frac12 \left(\frac{\epsilon_k+n_0 g_B}{E_k} +\text{sign}(\mu)\right )}~~~(\mu=\alpha,\beta)
\end{equation}
where $n_0=N_0/V$ is the condensate density, $g_B$ is the effective unidimensional atom-atom interaction and $\text{sign}(\alpha)=-\text{sign}(\beta)=1$. In these expressions, $\epsilon_k=\hbar^2 k^2/ 2m$ and $E_k$ is the Bogoliubov dispersion relation ${E_k=\sqrt{2\epsilon_k n_0 g_B +\epsilon_k^2}}$. Therefore, the atom-cavity interaction term reads
\begin{equation}
\begin{aligned}
\frac{\hat{\cal H}_{AC}}{\hbar  U_0}
&=\hat{a}^\dagger\hat{a}\left\{
\frac {N_0}{2}\!+\!\frac{\sqrt{2N_0}}{V}\sum_k 
[\alpha_k\hat{c}_{k+}\!\int dx \cos^2(k_c x) \cos(kx)\right.\\
&+ \alpha_k\hat{c}_{k-}\int\!dx \cos^2(k_c x) \sin(kx)\\
&-\beta_k\hat{c}_{k+}^\dagger\int\!dx \cos^2(k_c x) \cos(kx)\\
&\left.-\beta_k\hat{c}_{k-}^\dagger\int\!dx \cos^2(k_c x) \sin(kx)
+h. c.]\right\},
\end{aligned}
\end{equation}
where we have discarded the negligible quadratic terms in the Bogoliubov expansion. Using the orthogonality of the sinusoidal functions, only the even mode $\hat{c}\equiv\hat{c}_{2k_c+}$ with momentum $2k_c$ survives, so that we finally get
\begin{equation}
\hat{\cal H}_{AC} = \hbar U_0\hat{a}^\dagger\hat{a}\left\{
\frac{N_0}{2} +\frac{\sqrt{2N_0}}{4} (\alpha_{2k_c}-\beta_{2k_c})(\hat{c}+\hat{c}^\dagger)
\right\}.
\end{equation}
The coupling constant $\zeta$ between  the cavity and the Bogoliubov mode depends implicitly on ({$g_B$} and is given by 
\begin{equation}
\zeta = U_0\frac{\sqrt{2N_0}}{4}(\alpha_{2k_c}-\beta_{2k_c}).
\end{equation}

Whereas the first term in Eq.~\eqref{eq:HAC} embodies a cavity-frequency pull, the second is formally analogous to $\hat{\cal H}_{MC}$ and shows that, under the above working conditions, the BEC dynamics mimics that of a mechanical mode undergoing radiation-pressure effects. A similar result, for a BEC coupled to a static cavity, has been found in Ref.~\cite{Esslinger2008}. Our approach can be extended to include higher-order momentum modes in the expansion above. 

The dynamical equations of the coupled three-mode system can then be cast into a compact form much along the lines of the approach sketched for a linearised purely optomechanical system in Sec.~\ref{Sottraggopaper}. A Langevin-like equation for the vector of fluctuations of the system's quadrature operators $\hat{\bm \phi}^T_{MCA}=(\delta\hat{x}~\delta\hat{y}~\delta\hat{q}~\delta\hat{p}~\delta\hat{Q}~\delta\hat{P})$ can be easily cast into 
\begin{equation}
\label{eq:dotphi}
\partial_t{\hat{\bm \phi} }_{MCA}={{\cal K}}_{MCA}\hat{\bm \phi}_{MCA}+\hat{\cal {\bm N}}_{MCA},
\end{equation}
where we have introduced the noise vector $\hat{\cal\bm N}^T=(\!\sqrt{\kappa}(\delta\hat{a}^\dag_{in}{+}\hat{a}_{in})~i\sqrt{\kappa}\delta(\hat{a}^\dag_{in}{-}\hat{a}_{in})~0~\hat{\xi}~0~0)$ and the dynamical coupling matrix 
\begin{equation}
{\cal K}_{MCA}=\begin{pmatrix}
-\kappa & \Delta&0&0&0&0
\\
-\Delta & -\kappa&2\chi \alpha_s&0&-2\sqrt 2\zeta \alpha_s&0
\\
0&0&0&\frac 1m&0&0
\\
\hbar \chi \alpha_s&0&-m\omega_m^2&-\gamma&0&0
\\
0&0&0&0&0&\tilde\omega
\\
-\sqrt 2 \zeta \alpha_s &0&0&0&-\tilde\omega&0
\end{pmatrix}.
\end{equation}

 The evolution of the system depends on a few crucial parameters, including the total detuning ${\Delta= \omega_C{-}\omega_L{-}\chi q_s{+}\sqrt 2\zeta Q_s{+}g^2N_0/2\Delta_a}$ between the cavity and the pump laser. This consists of the steady pull-off term in Eq.~\eqref{eq:HAC} as well as both the opto-mechanical contributions proportional to the displaced equilibrium positions  of the mechanical and Bogoliubov modes. These are respectively given by the stationary values ${q_s={\hbar \chi\alpha_s^2}/{m\omega_m^2}}$ and ${Q_s{=-}{\sqrt 2 \zeta\alpha_s^2}/{\tilde\omega}}$, which are in turn determined by the mean intra-cavity field amplitude $\alpha_s={\eta}/\sqrt{{\Delta^2+\kappa^2}}$. The interlaced nature of such stationary parameters (notice the dependence of $\alpha_s$ on the detuning) is at the origin of bistability and chaotic effects~\cite{Esslinger2008,Stamper2008,Meystre2009}. 
 
 Differently from Sec.~\ref{Sottraggopaper}, here it is actually crucial to go through the details of the noise-related part of the dynamics. As done before, we have introduced  $\delta\hat{a}_{in}$ and $\delta\hat{a}^\dag_{in}$ as zero-average [${\langle a_{in}(t)\rangle=\langle a^\dag_{in}(t)\rangle=0}$], delta-correlated [$\langle a_{in}(t) a_{in}^\dagger(t') \rangle=\delta(t{-}t')$] operators describing white noise entering the cavity from the leaky mirror. Dissipation of the mechanical mirror energy is, on the other hand, associated with the decay rate $\gamma$ and the corresponding zero-mean Langevin-force operator $\hat{\xi}(t)$ having non-Markovian correlations ($\beta_B=\hbar/2k_BT$)~\cite{Giovannetti2001} 
 \begin{equation}
\langle\hat{\xi}(t)\hat{\xi}(t') \rangle=({\hbar\gamma m}/{2\pi})\int\!\omega e^{-i \omega(t-t')}[ \coth({\beta_B\omega}){+}1] d\omega.
\end{equation}
 Although the non-Markovianity of the mechanical Brownian motion could be retained in our approach, for large mechanical quality factors ($\gamma{\rightarrow}{0}$), a condition that is met in current experiments on micro-mechanical systems~\cite{gro2009A}, one can take 
 \begin{equation}
 {\langle\hat{\xi}(t)\hat{\xi}(t')\rangle{\simeq}[\hbar\gamma{m}/\beta_B{+}i\partial_t]\delta(t{-}t')}
 \end{equation}
as in Ref.~\cite{Giovannetti2001}. As our analysis relies on symmetrized two-time correlators, the antisymmetric part in the above expression, proportional to $\partial_t\delta(t{-}t')$, is ineffective, thus making our description fully Markovian. Here we show that the model above results in an interesting back-action effect where the state of the mechanical mode is strongly intertwined with the BEC. The physical properties of the mirror are altered by the cavity-BEC coupling. Evidences of such interaction, strong enough to inhibit the cooling capabilities of the radiation-pressure mechanism under scrutiny, are found in the noise properties of the mechanical mode.

We start considering the modification in the mirror dynamics due to the coupling to the cavity and indirectly to the BEC. The Langevin equations are solved in the frequency domain, where we should ensure stability of the solutions. This implies 
negativity of the real part of the eigenvalues of ${\cal K}$. 
Numerically, we have found that stability is given for ${\Delta{>}0}$ and weak coupling of the mirror and the BEC to the cavity, {i.e.} for $\{{\chi\sqrt{\hbar/m\omega_m}, \zeta\}{\ll}\kappa}$, which are conditions fulfilled throughout this section. 
We find the mirror displacement
\begin{equation}
\delta\hat{q}(\omega)=[{\cal A}_M(\omega)\delta\hat{y}_{in}(\omega) + {\cal B}_M(\omega)\delta\hat{x}_{in}(\omega)+{\cal C}_M(\omega)\hat{\xi}(\omega)],
\end{equation}
with 
\begin{equation}
\begin{aligned}
{\cal A}_M(\omega)&=\frac{{\cal B}(\omega)\Delta}{\kappa-i\omega}=-\frac{\hbar\chi\alpha_s{\sqrt{2\kappa}}\Delta}{d_M(\omega)},\\
{\cal C}_M(\omega)&=-\frac{(\omega^2{-}\tilde\omega^2)[(\kappa{-}i\omega)^2{+}\Delta^2]{+}4\tilde\omega\Delta\alpha_s^2\zeta^2}{d_M(\omega)}
\end{aligned}
\end{equation}
and $d_M(\omega)$ that is related to the effective susceptibility function of the mechanical mode.

Let us now get into the detailed procedure for the derivation of the mechanical and atomic density noise spectra (DNSs) and the full expressions for the  associated susceptibility functions. 
For a generic operator $\hat{O}(\omega)$ in the frequency domain, the DNS is defined as 
\begin{equation}
{S\!_{\cal O}(\omega)=\frac{1}{4\pi}\int d\Omega e^{-i(\omega+\Omega)t} \langle\hat{\cal O}(\omega)\hat{\cal O}(\Omega){+}\hat{\cal O}(\Omega)\hat{\cal O}(\omega)\rangle}.
\end{equation}
\begin{figure*}[t]
\center{\hskip0.3cm{\bf (a)}\hskip3.5cm{\bf (b)}\hskip3.7cm{\bf (c)}\hskip3.7cm{\bf (d)}}
\psfig{figure=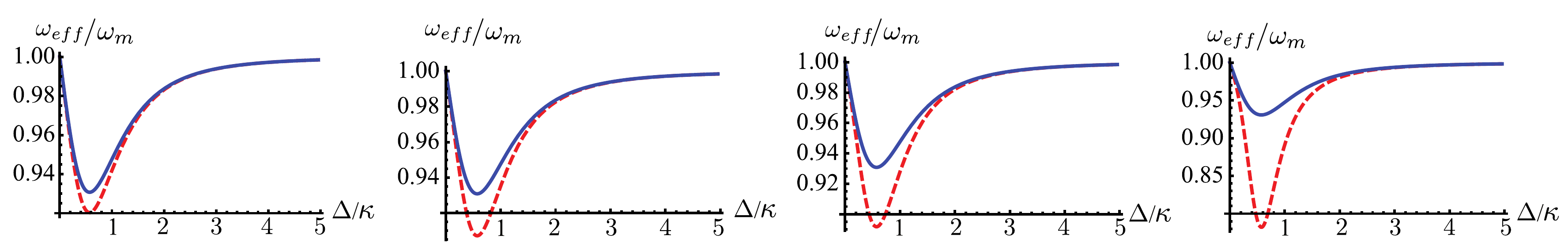,width=\linewidth}
\center{\hskip0.3cm{\bf (e)}\hskip3.5cm{\bf (f)}\hskip3.7cm{\bf (g)}\hskip3.7cm{\bf (h)}}
\psfig{figure=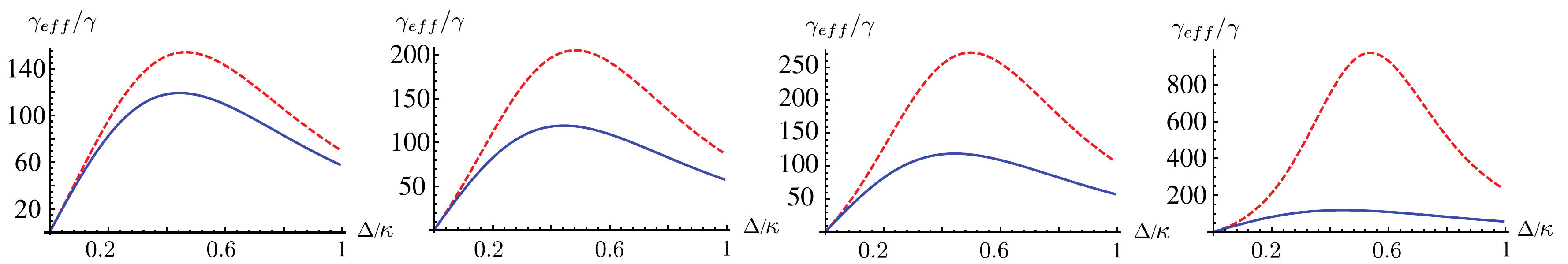,width=\linewidth}
\caption{{\bf (a)}-{\bf (d)}: The effective frequency $\omega_{eff}/\omega_m$ is plotted against the detuning $\Delta/\kappa\in[0,5]$ for the parameters used in the letter. Panel {\bf (a)} corresponds to $\omega\simeq0$ and well represents the trend of this quantity up to $\omega/\omega_m\sim{0.6}$. Panels {\bf (b)}-{\bf (d)} are for $\omega/\omega_m=0.7,0.8$ and $0.9$, respectively. The (dashed) red curve is for a coupled BEC with $\tilde\omega\simeq\omega_m$ and $\zeta=100$Hz. The (solid) blue line is for an empty optomechanical cavity. {\bf (e)}-{\bf (h)}: The effective damping rate $\gamma_{eff}/\omega_m$ is plotted against the detuning $\Delta/\kappa\in[0,1]$ for the parameters used in the letter. Panel {\bf (e)} corresponds to $\omega\simeq0$ and well represents the trend up $\omega/\omega_m\sim{0.6}$. Panels {\bf (f)}-{\bf (h)} show are for $\omega/\omega_m=0.7,0.8$ and $0.9$, respectively. The (dashed) red curve is for a coupled BEC with $\tilde\omega\simeq\omega_m$ and $\zeta=100$Hz. The (solid) blue line is for an empty optomechanical cavity.}
\label{frequenza}
\end{figure*}

By transforming the Langevin equations and solving them in the frequency domain, one can use the correlations~\cite{MauroNJP}
\begin{equation}
\begin{aligned}
&\langle\hat{\xi}_{}(\omega)\hat{\xi}_{}(\Omega)\rangle=2\pi\hbar\gamma{m}\omega[1+\coth(\beta_K\omega)]\delta(\omega+\Omega),\\
&\langle\delta\hat{x}_{in}(\omega)\delta\hat{x}_{in}(\Omega)\rangle=\langle\delta\hat{y}_{in}(\omega)\delta\hat{y}_{in}(\Omega)\rangle=2\pi\delta(\omega+\Omega),\\
&\langle\delta\hat{x}_{in}(\omega)\delta\hat{y}_{in}(\Omega)\rangle=-\langle\delta\hat{y}_{in}(\omega)\delta\hat{x}_{in}(\Omega)\rangle=2\pi{i}\delta(\omega+\Omega),
\end{aligned}
\end{equation}
to write the DNS of the mechanical mode as 
\begin{equation}
S\!_q(\omega,\Delta)=\left|\frac{\varphi(\omega,\Delta)}{d_M(\omega)}\right|^2[S\!_{rp}(\omega,\Delta)+S\!_{th}(\omega)]
\end{equation}
where $S\!_{rp}(\omega,\Delta)$ is the radiation-pressure contribution, modified by the atomic coupling to the cavity field, $S\!_{th}(\omega,\Delta)$ is the thermal part of the DNS, proportional to $(\beta_B\omega)^{-1}$ while the ratio $\varphi(\omega)/d_M(\omega)$ is related, as we show here, to the effective susceptibility of the mechanical mode. Explicitly
\begin{equation}
\begin{aligned}
&S\!_{rp}(\omega,\Delta)=\frac{2\kappa\hbar^2\alpha^2_s\chi^2(\Delta^2+\kappa^2+\omega^2)}{\Delta^4+2\Delta^2(\kappa^2-\omega^2)+(\kappa^2+\omega^2)^2},\\
&S\!_{th}(\omega)=\hbar\gamma{m}\omega\coth(\beta_B\omega)\simeq\hbar\gamma{m}/\beta_B,\\
&d_M(\omega)=2\hbar\Delta\alpha^2_s\chi^2(\omega^2-\tilde\omega^2)+m(\omega^2\!-\!\omega^2_m\!+\!i\gamma\omega)\varphi(\omega,\Delta)\\
\end{aligned}
\end{equation}
with ${\varphi(\omega,\Delta)=4\Delta\alpha^2_s\zeta^2\tilde\omega+[\Delta^2+(\kappa-i\omega)^2](\omega^2-\tilde\omega^2)}$. We call $\eta(\omega,\Delta)\!=\!\varphi(\omega,\Delta)/d_M(\omega)$ the effective susceptibility function of the mechanical mode, which can thus be written as
\begin{equation}
\label{effettivo}
\eta(\omega,\Delta)=-\left[{{m(\omega^2-\omega^2_m+i\gamma\omega)+\frac{2\hbar\Delta\alpha^2_s\chi^2(\omega^2-\tilde\omega^2)}{\varphi(\omega,\Delta)}}}\right]^{-1}.
\end{equation}
Our task here is to write $\eta(\omega,\Delta)$ as the susceptibility of a fictitious harmonic oscillator characterized by frequency $\omega_{eff}$ and a damping rate $\gamma_{eff}$, that is
\begin{equation}
\label{target}
\eta_{eff}(\omega)=-\frac{1}{m(\omega^2-\omega^2_{eff}+i\gamma_{eff}m)}.
\end{equation}
By equating real and imaginary parts of Eqs.~\eqref{effettivo} and \eqref{target}, one easily gets the dependence of the effective frequency and damping rate on $\Delta$ and the opto-mechanical/atomic coupling strengths. These read
\begin{equation}
\label{effective}
\begin{aligned}
\omega_{eff}&=\sqrt{\omega^2_m-\mu_r(\omega,\Delta)},~~\gamma_{eff}&=\gamma+\mu_i(\omega,\Delta)/\omega,
\end{aligned}
\end{equation} 
where we have introduced 
\begin{widetext}
\begin{equation}
\label{effettive}
\begin{aligned}
\mu_r(\omega,\Delta)&=\frac{2\hbar\Delta\chi^2\alpha^2_s(\omega^2-\tilde\omega^2)[(\Delta^2+\kappa^2-\omega^2)(\omega^2-\tilde\omega^2)+4\Delta\tilde\omega\zeta^2\alpha^2_s]}{m\{[\Delta^4+2\Delta^2(\kappa^2-\omega^2)+(\kappa^2+\omega^2)^2](\omega^2-\tilde\omega^2)^2+8\tilde\omega\Delta\zeta^2\alpha^2_s[(\Delta^2+\kappa^2-\omega^2)(\omega^2-\tilde\omega^2)+2\tilde\omega\Delta\zeta^2\alpha^2_s]\}},\\
\mu_i(\omega,\Delta)&=\frac{4\hbar\kappa\Delta\chi^2\alpha^2_s(\omega^2-\tilde\omega^2)^2}{m\{[\Delta^4+2\Delta^2(\kappa^2-\omega^2)+(\kappa^2+\omega^2)^2](\omega^2-\tilde\omega^2)^2+8\tilde\omega\Delta\zeta^2\alpha^2_s[(\Delta^2+\kappa^2-\omega^2)(\omega^2-\tilde\omega^2)+2\tilde\omega\Delta\zeta^2\alpha^2_s]\}}.
\end{aligned}
\end{equation}
\end{widetext}

In Figs.~\ref{frequenza} we show the behaviour of the effective detuning-dependent mechanical frequency and damping rate. A it can be appreciated from Eqs.~\eqref{effective} and \eqref{effettive}, such quantities also depend on the value of the frequency $\omega$ at which we probe the response of the system. Under the operating conditions used in our work, the coupling with the atomic subsystem determines an enhancement of the optical spring effect undergone by the mechanical mode under radiation pressure coupling. In the region of the frequency space associated with $\omega\le\omega_m$, where the red-shift of the mechanical frequency is expected~\cite{MauroNJP}, the effective frequency at $\zeta\neq{0}$ is smaller than the value corresponding to an empty cavity. On the contrary, the effective damping rate is much larger than for an empty cavity. This analysis provides detailed information on the way the system respond to the set of mode couplings entailed by the physical configuration that we are exploring and embodies a useful characterisation of the modification undergone by the key features of the mechanical mode. 
 
We now compute the density noise spectrum of $\delta\hat{q}(\omega)$. 
Using the correlation properties of the input and Brownian noise operators, after a little algebra one gets 
\begin{equation}
\label{spettroM}
S\!_q(\omega)=\sum_{{\cal J}={\cal A},{\cal B}}|{\cal J}_M(\omega)|^2+\hbar\gamma m\left[1\!+\!\coth({\beta_B\omega})\right]|{\cal C}_M(\omega)|^2.
\end{equation}

\begin{figure}[b]
{\bf (a)}\hskip3.5cm{\bf (b)}\\
\psfig{figure=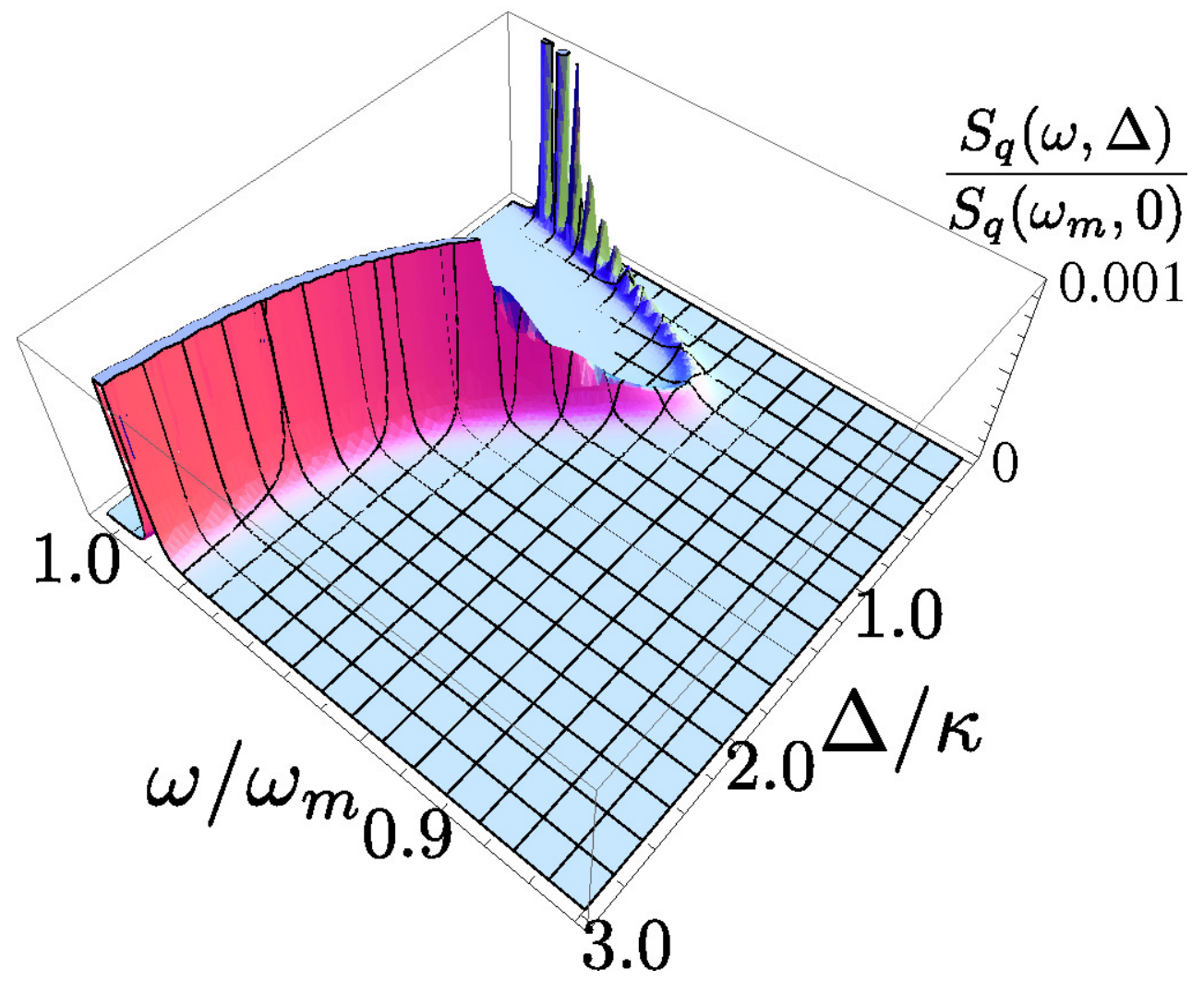,width=0.5\linewidth}\psfig{figure=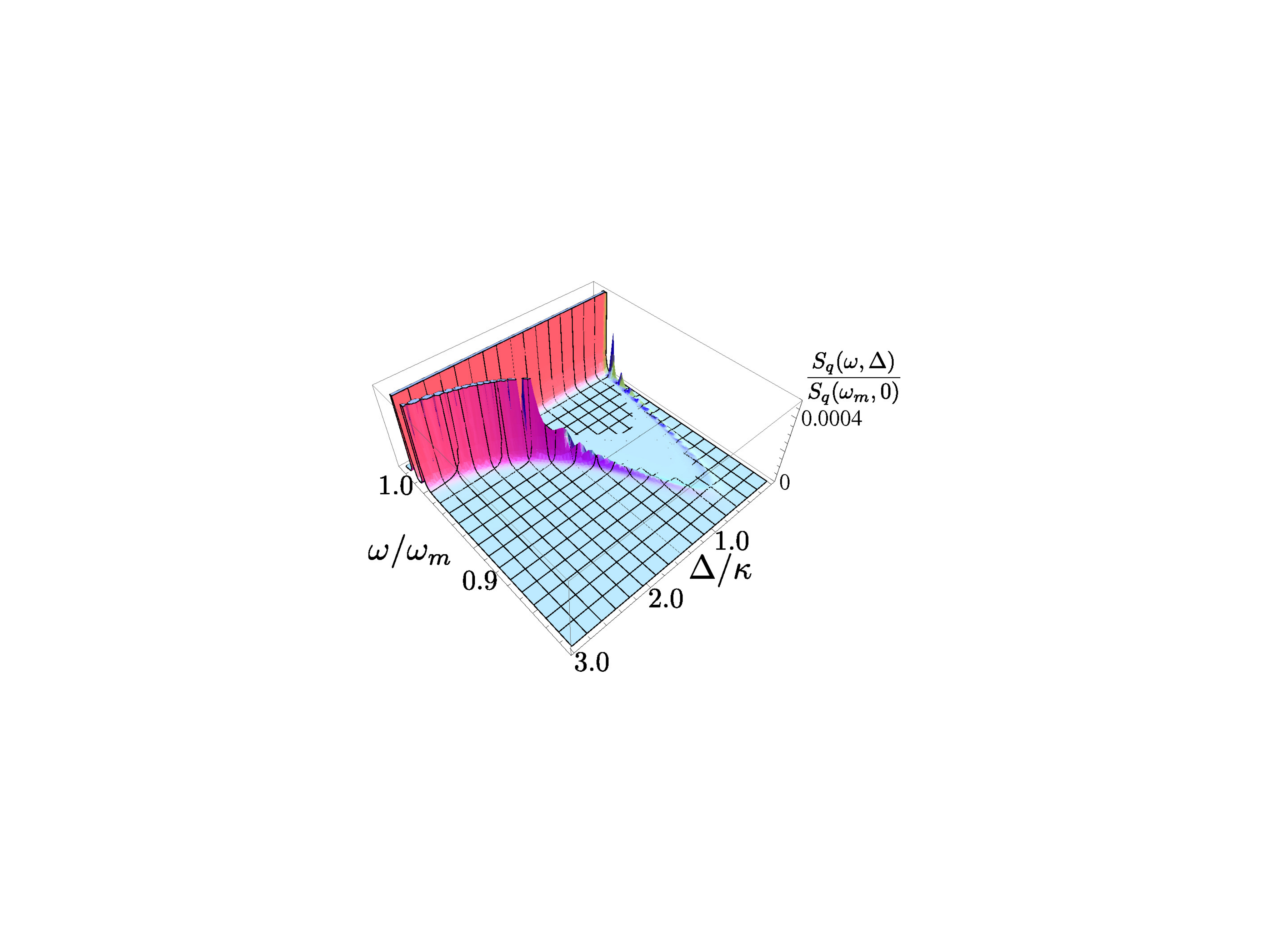,width=0.5\linewidth}
\caption{{\bf (a)} DNS $S_q(\omega,\Delta)$ for an empty cavity against $\Delta$ and $\omega$ for ${\cal L}=25$mm, $m=15$ng, $\omega_m/2\pi=275$KHz, $\gamma=\omega_m/Q$ with $Q=10^5$ and $T=300$K and $\kappa\simeq5$MHz. The pumping light has wavelength $1064$nm and input power ${\cal R}=4$mW. The DNS is rescaled to its value at $\omega=\omega_m$ and $\Delta=0$. {\bf (b)} We include the effects of the atomic coupling by taking $\tilde{\omega}=\omega_m$ and $\zeta\simeq0.7\chi{\sqrt{\hbar/(m\omega_m)}}$. 
}
\label{spettri}
\end{figure}

Some interesting features emerge from the study of $S\!_q(\omega)$. In Fig.~\ref{spettri} we compare  the case of an empty opto-mechanical cavity [panel {\bf (a)}] and one where a weak coupling with the atomic Bogoliubov mode of frequency ${\tilde{\omega}=\omega_{\text{m}}}$ is  included [panel {\bf (b)}]. For an empty cavity, the mechanical-mode spectrum is obviously identical to what has been found in Ref.~\cite{MauroNJP} (the use of that case as a milestone in our quantitative study motivates the choice of the parameters used throughout this work). Both the optical spring effect in a detuned optical cavity and a cooling/heating mechanism are evident: height, width and peak-frequency of $S\!_q(\omega)$ change with the detuning $\Delta$. At ${\Delta\simeq{\kappa/2}}$ optimal cooling is achieved with a considerable shrink in the height of the spectrum. However, as soon as the Bogoliubov modes enter the dynamics, major modifications appear. The optical spring effect is magnified (the red-shift of the peak frequency of $S\!_q(\omega)$ is larger than at $\zeta=0$) and a secondary structure appears in the spectrum, unaffected by any change of $\Delta$. Such a structure is a second Lorentzian peak centered in ${\omega{\simeq}\omega_{\text{m}}}$ and is a signature of the back-action induced by the atoms on the the mirror, an effect that comes from a three-mode coupling and, as discussed later, is {\it determined} by $\tilde{\omega}$ and $\zeta$. In fact, by studying the dependence of $S\!_q(\omega)$ on the frequency of the Bogoliubov mode, we see that the secondary peak identified above is centered at $\tilde{\omega}$. For ${\zeta\ll{\chi}\sqrt{\hbar/(m\omega_m)}}$, {\it i.e.} for weak back-action from the atomic mode onto the mechanical one, the signature of the former in the spectrum of the latter is small. A quantitative assessment reveals, in fact, that it only consists of a tiny structure subjected to negligible detuning-induced changes. The picture changes for $\tilde\omega$ close to the mechanical frequency. In this case, as seen in Fig.~\ref{spettri} {\bf (b)}, the influence of the atomic medium is considerable and present at any value of $\Delta$. While the mechanical mode experiences enhanced optical spring effect (as easily seen by looking at the effective susceptibility of the mechanical mode), the secondary structure persists even at ${\Delta{\sim}{\kappa}/2}$, the working point that for our choice of parameters optimizes the mechanical cooling at empty cavity. However, as demonstrated later on, here the strong optical spring effect is not accompanied by an effective mechanical cooling. 

\begin{figure*}[t]
{\bf (a)}\hskip5cm{\bf (b)}\hskip5cm{\bf (c)}\\
\psfig{figure=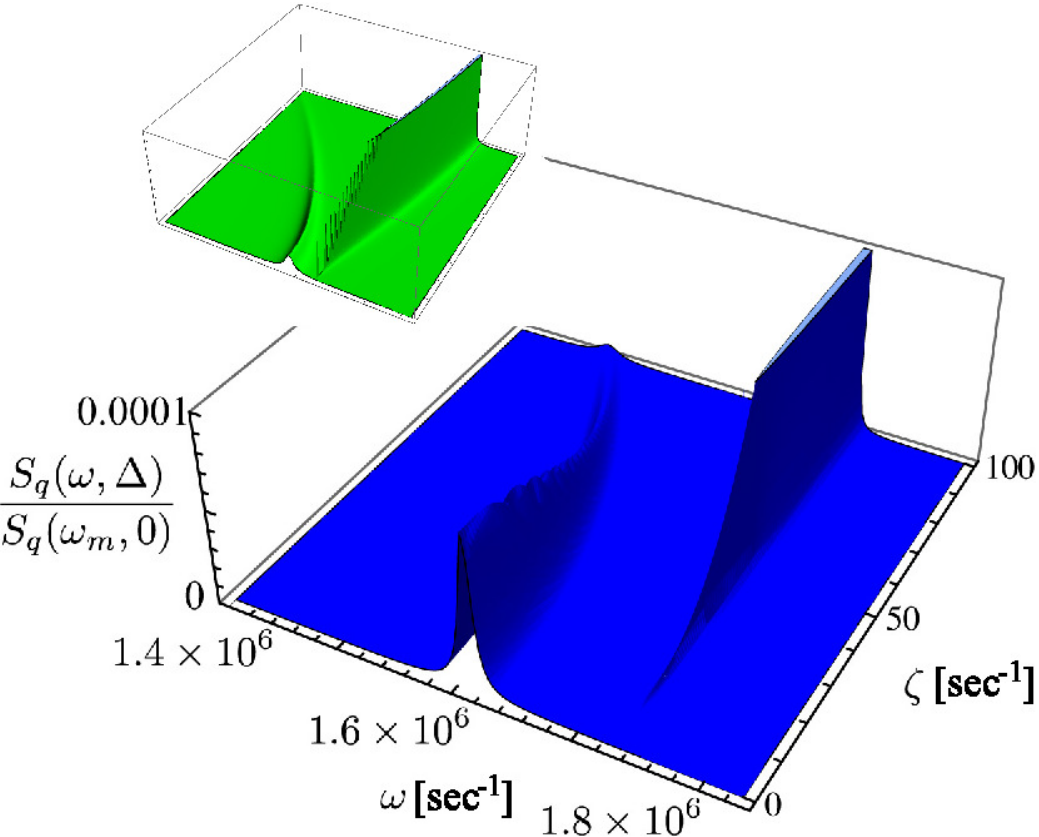,width=0.331\linewidth}~~~~\psfig{figure=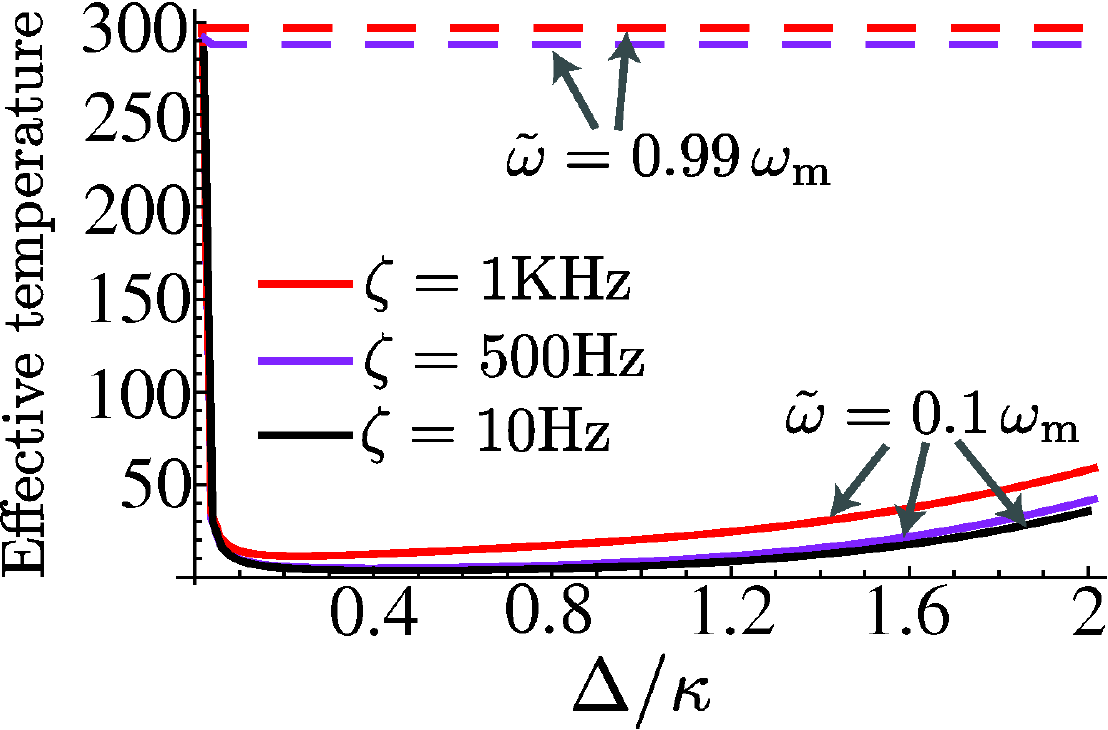,width=.30\linewidth}~~\psfig{figure=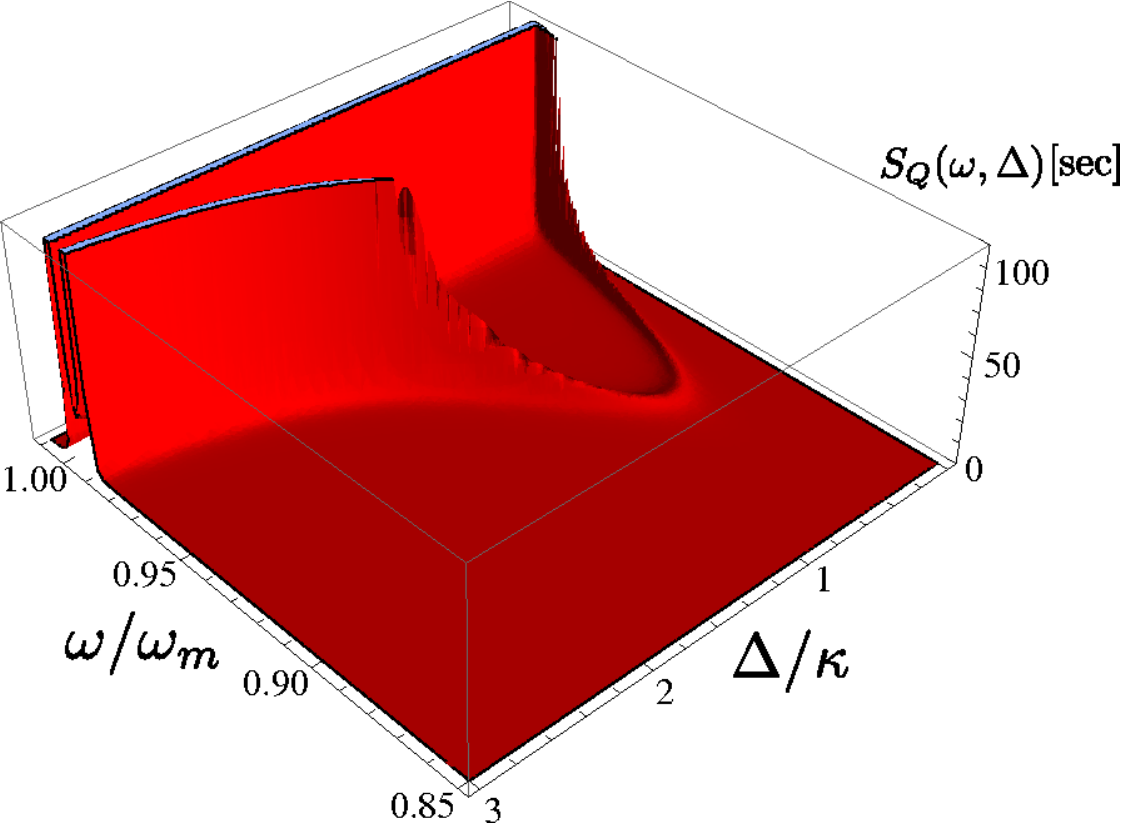,width=.33\linewidth}
\caption{{\bf (a)} DNS of the mechanical mode against $\omega$ and $\zeta$ for ${\tilde{\omega}{\simeq}{\omega_m}}$ and ${\Delta=\kappa/2}$. The structure centered at $\omega_m$ and with amplitude growing with $\zeta$ is due to atomic back-action. 
Inset: same plot for $\tilde{\omega}=0.8\omega_m$. Similar but less important features are found away from the resonance betwen mechanical and atomic mode.  {\bf (b)} Temperature of the mechanical mode against $\Delta/\kappa$. Solid (Dashed) lines are for $\tilde{\omega}=0.1\omega_{\text{m}}$ ($\tilde{\omega}{\simeq}\omega_{\text{m}}$). 
{\bf (c)} $S\!_Q(\omega,\Delta)$ for $\tilde\omega{\simeq}{\omega}_m$ and $\zeta=50$Hz. }
\label{backaction}
\end{figure*}

A better understanding is provided by studying $S\!_q(\omega,\Delta)$ against the atomic opto-mechanical rate $\zeta$ [cfr. Fig.~\ref{backaction} {\bf (a)}]. At the optimal empty-cavity detuning and for $\tilde{\omega}\simeq{\omega}_m$, both the effects highlighted above are clearly seen: the {\it contribution} of the secondary structure centered at $\tilde{\omega}$ grows with $\zeta$ due to the increasing atomic back-action while a large red-shift and shrinking of the mechanical-mode contribution to the DNS shows the enhanced spring effect. An intuitive explanation for all this comes from taking a normal-mode description, where the diagonalization of $\hat{\cal H}$ passes through the introduction of new modes that are linear combinations of the mechanical and Bogoliubov one. The weight of the latter increases with $\zeta$, thus determining a strong influence of the atomic part of the system over the noise properties of the mechanical mode.  

The consequences of the atomic back-action are not restricted to the effects highlighted above. Strikingly, the coupling between the atomic medium and the cavity field acts as a {\it switch} for the cooling experienced by the mechanical mode in an empty cavity  [cfr.~Fig.~\ref{backaction} {\bf (b)}]. That is, the coupling to the collective oscillations of the atomic density is crucial in determining the number of thermal excitations in the state of the mechanical mode, regulating the mean energy of the cavity end-mirror. A way to clearly see it is to consider the effective temperature $T_{\text{eff}}=\langle U\rangle/k_B$, where 
\begin{equation}
{\langle U\rangle=\frac12m\omega^2_m\langle\delta\hat{q}^2\rangle+\frac{\langle\delta\hat{p}^2\rangle}{2m}}
\end{equation}
is the mean energy of the mechanical mode. $\langle U\rangle$ is experimentally easily determined by measuring just the area underneath $S\!_q(\omega,\Delta)$, as acquired by a spectrum analyzer. In fact, we have ${\langle\delta\hat{r}^2\rangle=\int{d}\omega{S}\!_r(\omega,\Delta)~(r=q,p)}$ with ${S\!_p(\omega,\Delta)=m^2\omega^2_mS\!_q(\omega,\Delta)}$. 
Such temperature-regulating mechanism is explained in terms of a simple thermodynamic argument. The exchange of excitations behind passive mechanical cooling~\cite{natures2006A,MauroNJP} occurs at the optical sideband centered at $\omega_m$. When the frequency of the Bogoliubov mode does not match this sideband, mirror and cavity field interact with only minimum disturbance from the BEC. Thus, mechanical cooling occurs as in an empty cavity: even for relatively large values of $\zeta$ the cooling capabilities of the detuned opto-mechanical process are, for all practical purposes, unaffected [see Fig.~\ref{backaction} {\bf (b)}]. However, by tuning $\tilde\omega$ on resonance with the relevant optical sideband, we introduce a {\it well-source mechanism} for the recycling of phonons extracted from the mechanical mode and transferred to the cavity field. The BEC can now {absorb} some excitations taken from the mirror by the field, thus acting as a {\it phononic  well} and {release} them into the field at a frequency matched with $\omega_m$. The mirror can take the excitations back, as in the presence of a {\it phononic source}: thermodynamical equilibrium is established at a temperature set by $\zeta$. For strong atomic back-action, the mirror does not experience any cooling [Fig.~\ref{backaction} {\bf (b)}].

\begin{figure*}[t]
{\bf (a)}\hskip5.cm{\bf (b)}\hskip5.cm{\bf (c)}\\
\psfig{figure=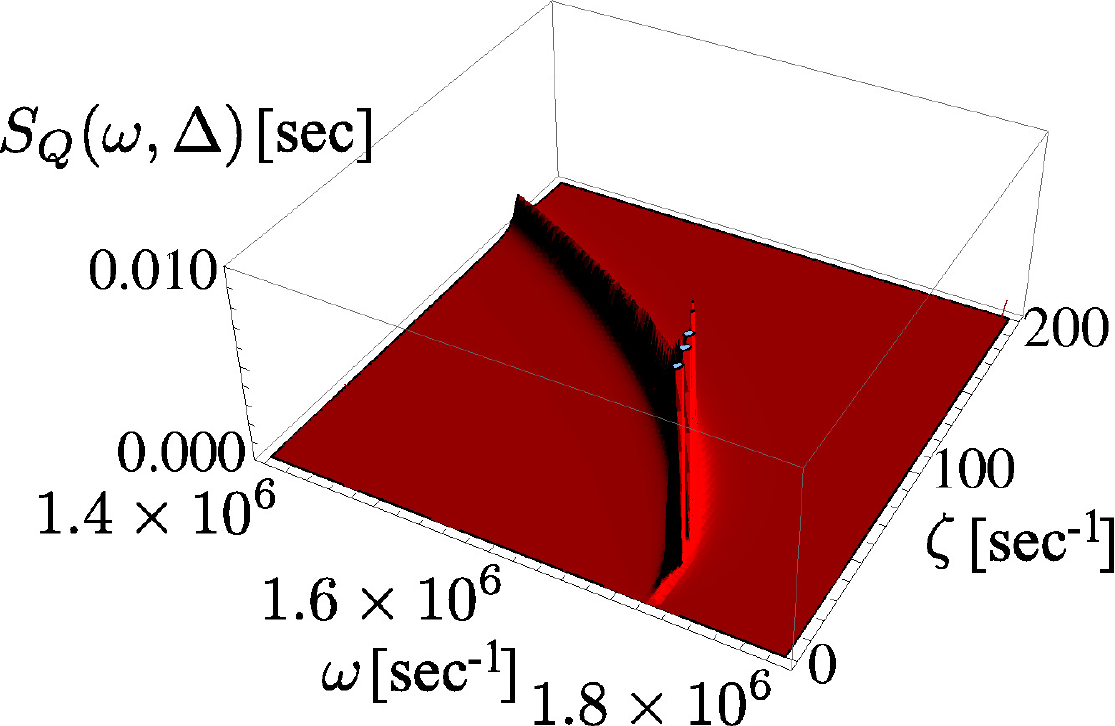,width=.32\linewidth}
\psfig{figure=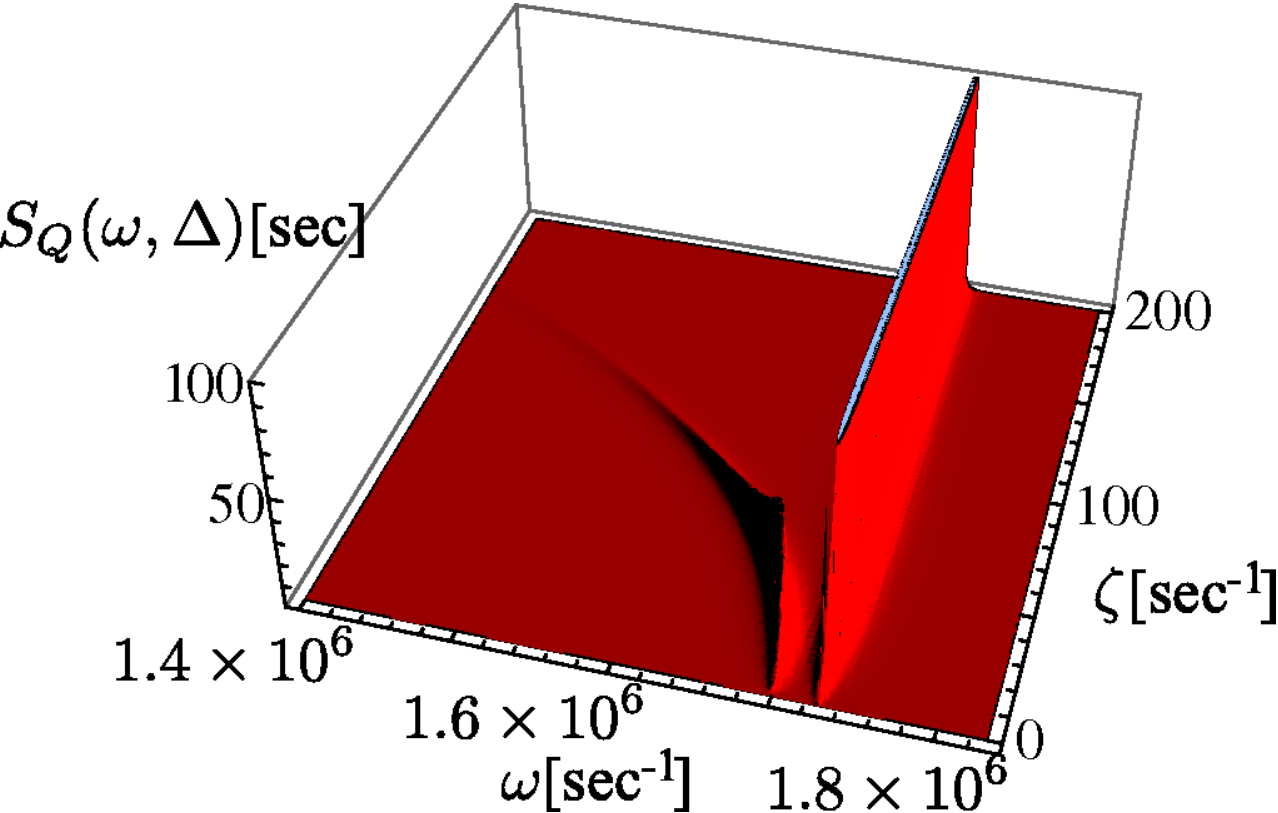,width=.32\linewidth}
\psfig{figure=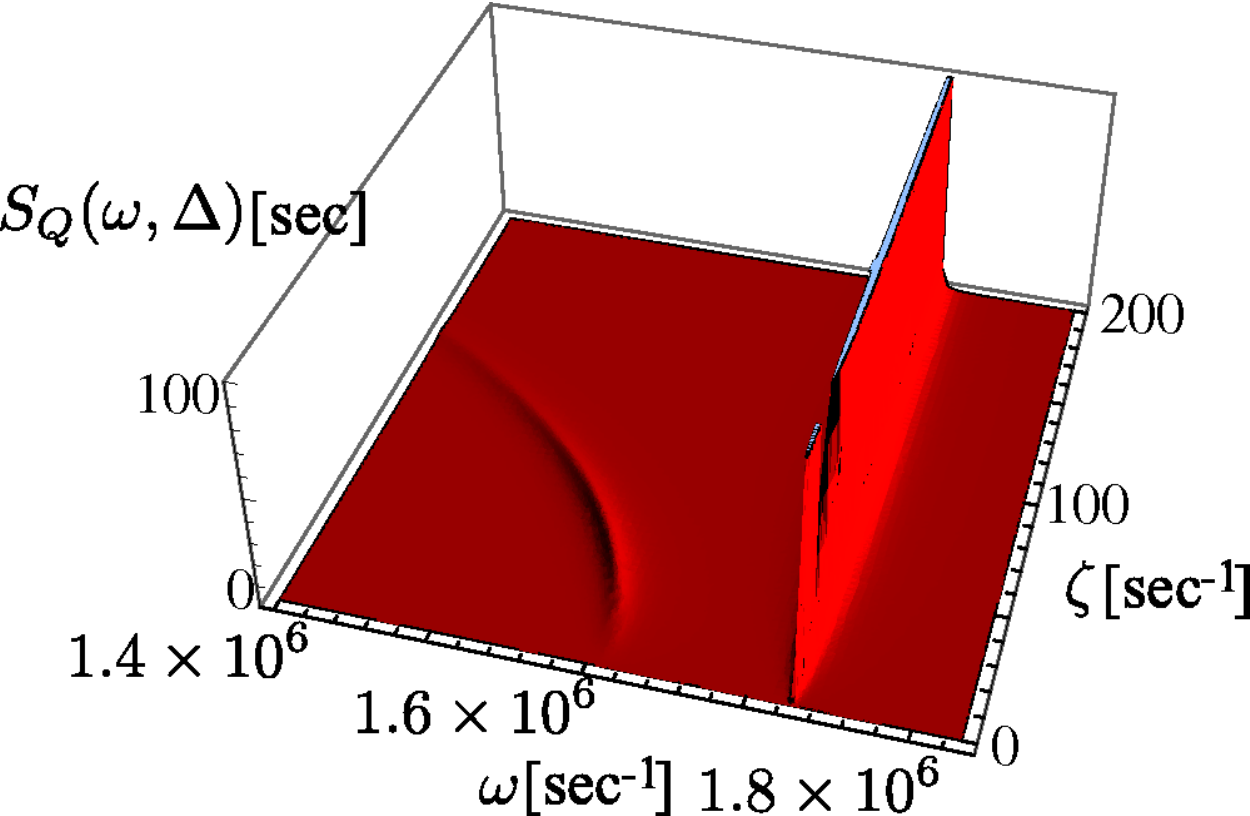,width=.32\linewidth}
\caption{Panels {\bf (a)} to {\bf (c)}: $S\!_Q(\omega,\kappa/2)$ against $\omega$ and $\zeta$ for $\chi=k\omega_C/2{\cal L}~(k=0,1,2)$. The atomic and mechanical part of the spectrum are clearly splitted.}
\label{multiplatomi}
\end{figure*}

Analogously, one finds the atomic DNS associated with the position-like operator of the Bogoliubov mode, which reads 
\begin{equation}
\delta\hat{Q}(\omega)={\cal A}_A(\omega)\delta\hat{y}_{in}(\omega)+{\cal B}_A(\omega)\delta\hat{x}_{in}(\omega)+{\cal C}_A(\omega)\hat{\xi}(\omega)
\end{equation}
 with 
\begin{equation} 
\begin{aligned}
&{\cal A}_A(\omega)=\frac{\Delta{\cal B}(\omega)}{\tilde{\omega}(\kappa{-}i\omega)}=\frac{2i\alpha_s\zeta\Delta\sqrt{\kappa}\omega}{d_A}(i\gamma\omega{+}\omega^2{-}\omega^2_m),\\
&{\cal C}_A(\omega)=-\frac{2\sqrt2i\alpha^2_s\Delta\zeta\chi\omega\tilde{\omega}}{md_A}
\end{aligned}
\end{equation}
 and $d_A$ being rather lengthy. The spectrum ${S}\!_{Q}(\omega,\Delta)$, which is easily determined using the appropriate input-noise correlation functions, is sketched in Fig.~\ref{multiplatomi}. Clearly, in light of the formal equivalence of Eq.~\eqref{eq:HAC} with a radiation pressure mechanism, by setting up the proper working point, the BEC should undergo a cooling dynamics similar to the one experienced by the mirror. The starting temperature of the Bogoliubov mode depends on the values taken by $\tilde{\omega}$ and $\zeta$. At $\zeta=0$, regardless of the atomic-mode frequency, its effective temperature is very low, as it should be. For a set value of $\zeta$, the temperature arises as $\tilde{\omega}\rightarrow\omega_m$. 
The conditions of our investigations are such that weak coupling between the BEC and the cavity field are kept, in a way so as to make the Bogoliubov expansion valid and rigorous. 
The mutually-induced back-action at the center of our discussion is clearly visible in Figs.~\ref{backaction} {\bf (c)} and \ref{multiplatomi}, where features similar to those present in the mechanical DNS appear. For ${\chi=0}$, the atomics DNS at $\tilde{\omega}=\omega_m$ starts from zero (at ${\zeta=0}$) and experiences red shifts and shrinking as the effective opto-mechanical coupling rate grows. Having switched off the coupling between the mechanical mode and the field, the spectrum is single-peaked. This is not the case for $\chi\neq{0}$ where a secondary structure appears, similar to the one in the mechanical DNS. The splitting between mechanical and atomic contributions to $S\!_Q(\omega,\kappa/2)$ grows with $\chi$, a sign of the mechanically-enhanced effect felt by the atomic mode.





\section{Hybrid optomechanical entanglement and its revelation}
\label{Gabrielepaper}

The faithful direct inference of quantum properties of parts of the device that we have addressed in the previous Section is not a simple task to accomplish. This difficulty indeed extends to any mescoscopic/macroscopic system consisting of many interacting parts. Such difficulty pairs up with the current lack of stringent and certifiable theoretical (and experimentally verifiable) criteria for non classicality to make the inference of quantumness at large scales a daunting problem. 

A possible way to get around the problem is the design of techniques to indirectly probing the system of interest by coupling it to fully controllable detection devices~\cite{Probes,MauroPRL}. It has spurred interest in designing techniques for the extraction of information from noisy or only partial data-sets gathered through observations of the detection system. The main problem in such an approach is embodied by the set of extra assumptions that one has to make on the mechanisms to test and the retrodictive nature of the claims that can be made. In fact, one can a posteriori interpret a data-set and infer the properties of a system that is difficult to address by post-processing the outcomes of a few measurements. However, this requires assuming knowledge of the working principles of the effect that we want to probe.

In this sense, it would be highly desirable to conceive indirect detection strategies providing a faithful picture of the property to test by, for instance, ``copying'' it onto the detection device, which would then be subjected to a direct estimate process. By focussing on the same hybrid optomechanical device described in the previous Section, comprising an atomic, mechanical and optical mode, we propose a diagnostic tool for entanglement that operates along the lines of such desiderata.  We show that the entanglement established between the mechanical and optical modes by the means of radiation-pressure coupling can indeed be ``written" into the light-atom subsystem and directly read from it by means of standard, experimentally friendly homodyning. Our analysis goes further to reveal that such a possibility arises from the non-trivial entanglement sharing properties of the system at hand, which is indeed genuinely tripartite entangled. Very interestingly, we find that while any bipartite entanglement is bound to disappear at very low temperature, the multipartite content persists longer against the operating conditions, as a result of entanglement monogamy relations. Our analytical characterization considers all the relevant sources of detrimental effects in the system, is explicitly designed to be readily implemented in the lab and provides a first step into the assessment of multipartite entanglement in a macro-scale quantum system of enormous experimental interest. 

It is also worth stressing that, although our model shares, at first sight, some similarities with the one considered in Ref.~\cite{genes}, it is distinctively different. First, the atomic media utilised in the two cases are different, with Ref.~\cite{genes} considering the bosonised version of the collective-spin operator of an atomic ensemble. Second, the resulting coupling Hamiltonians are built from quite distinct physical mechanisms. Third, the working conditions to be used in the two models in order to achieve interesting structures of shared quantum correlations are very different. We will comment on this latter point later.

In this Section we focus on the steady-state features of the system and analyse the behaviour of stationary entanglement. The main tool of our assessment is the steady-state covariance matrix ${\cal V}_{MCA}$, which satisfies a Lyapunov equation analogous to Eq.~\eqref{Lya} with the replacements ${\cal V}_{MC}\to{\cal V}_{MCA}$, ${\cal K}_{MC}\to{\cal K}_{MCA}$, and ${\cal D}_{MC}\to{\cal D}_{MCA}$. 
with
\begin{equation}
{\cal D}_{MCA}=\text{diag}[\kappa,\kappa,0,\gamma(2\bar n{+}1),0,0]
\end{equation}
and, as before, $\nbar$ the mean phonon number of the mechanical state. 
We quantify the entanglement between any two modes $\alpha$ and $\beta$ ($\alpha,\beta=A,M,C)$ using the logarithmic negativity of the reduced states ${\text E}_{\alpha\beta}=\max[0,{-}\ln 2\nu_{\alpha\beta}]$~\cite{logneg} with $\nu_{\alpha\beta}$ the smallest symplectic eigenvalue of the the matrix ${\cal V}'_{\alpha\beta}=P{\cal V}_{\alpha\beta}P$ and ${\cal V}_{\alpha\beta}$ the reduced covariance matrix of modes $\alpha$ and $\beta$ with elements $({\cal V}_{\alpha\beta})_{ij}$. 

\begin{figure*}[t]
{\bf (a)}\hskip6cm{\bf (b)}\hskip5cm{\bf (c)}
\\
\includegraphics[scale=0.4]{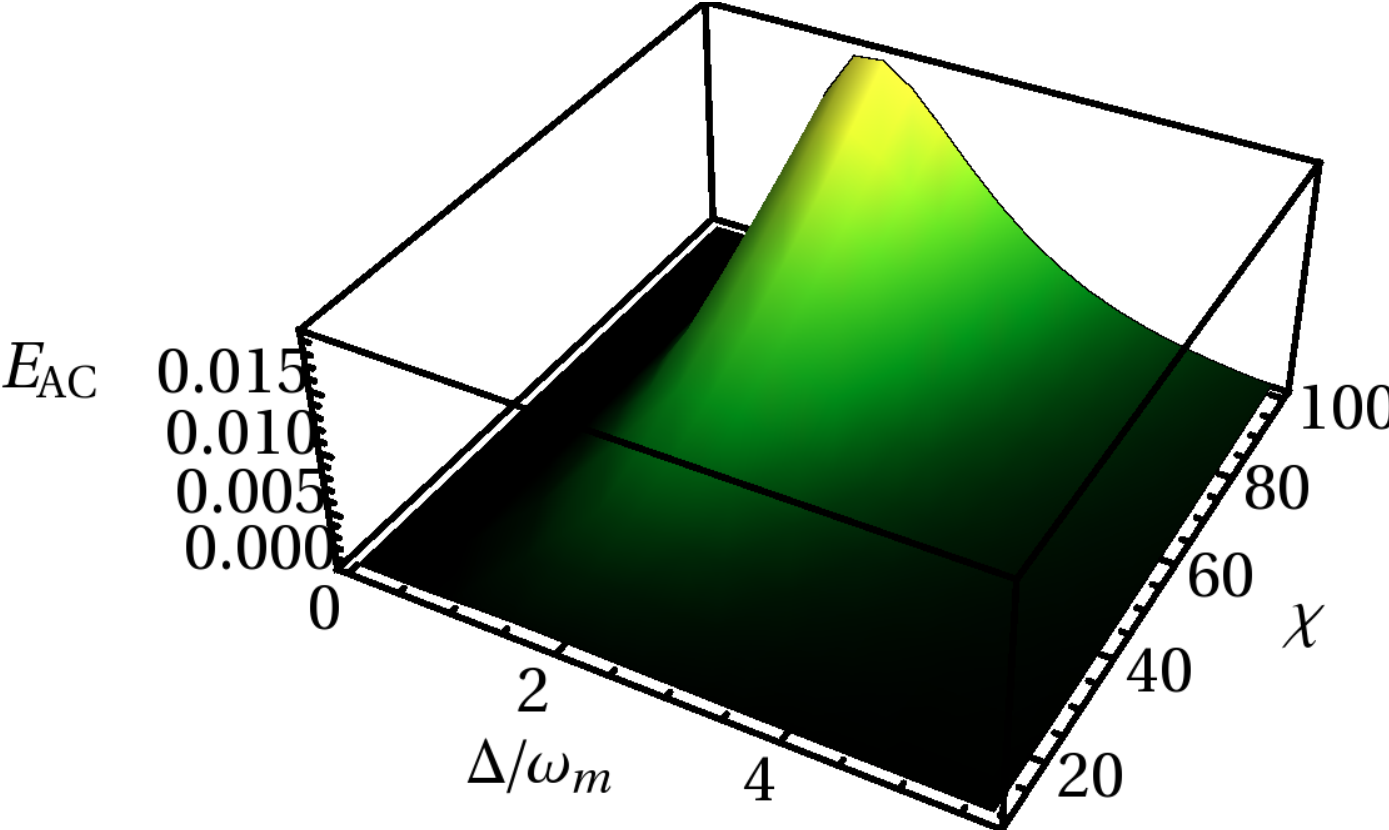}\includegraphics[scale=0.48]{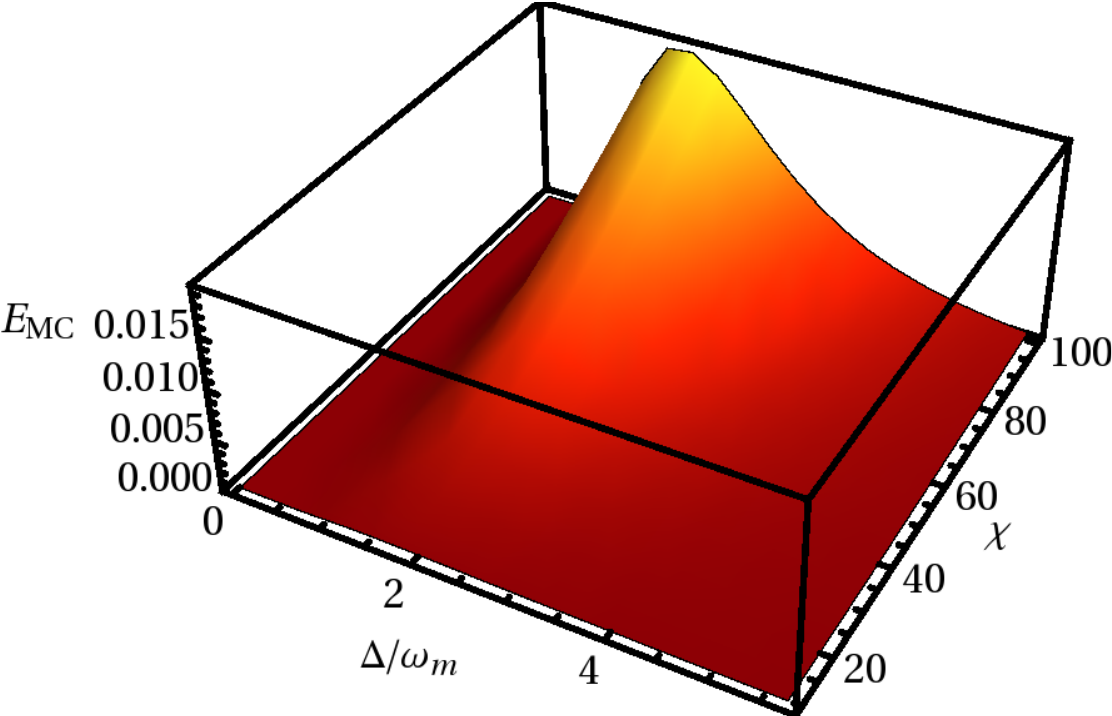}\includegraphics[scale=0.2]{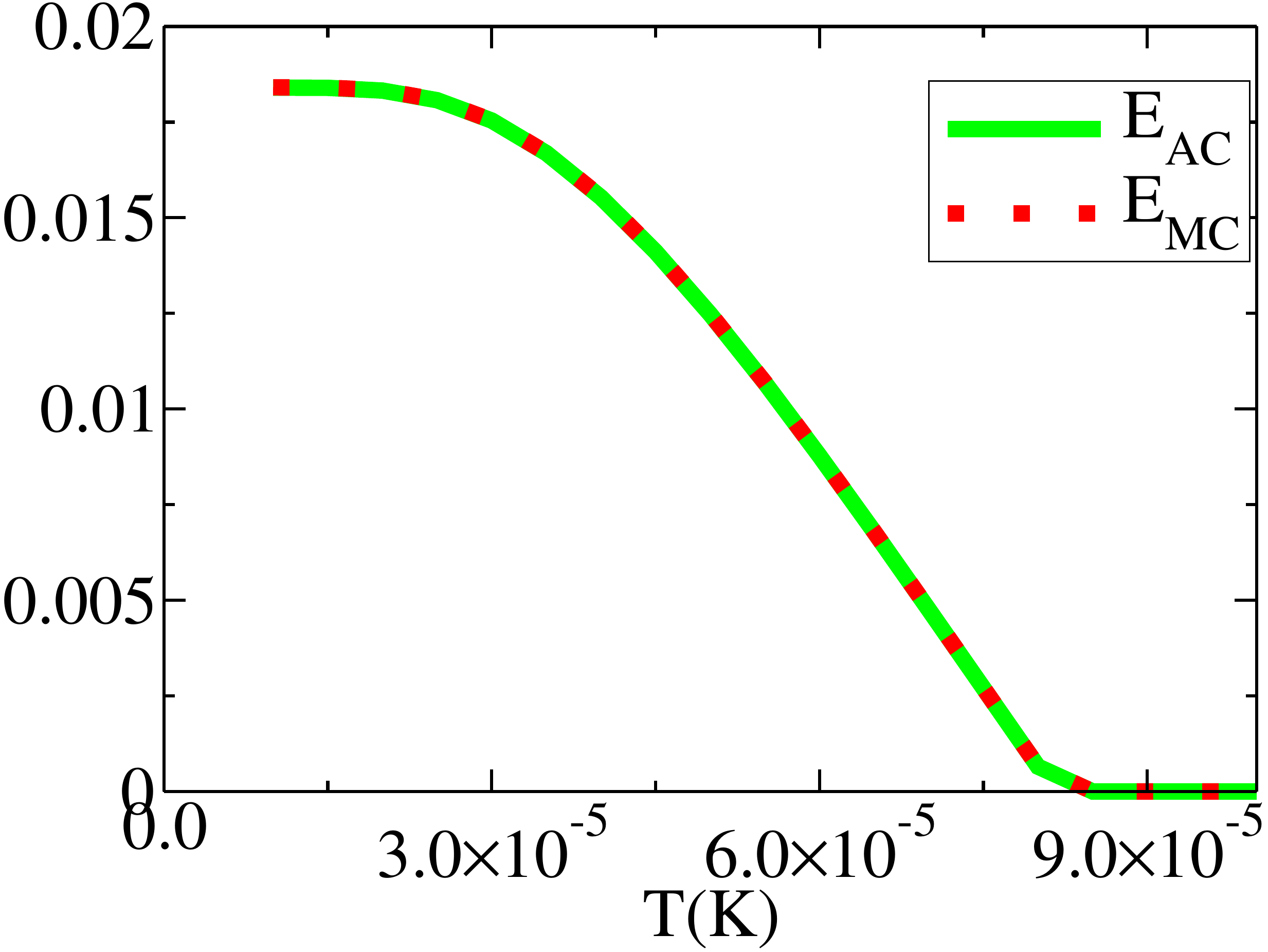}
\caption{
 {\bf (a)} Logarithmic negativity $E_{AC}$ against $\Delta$ (in units of $\omega_m/2\pi={3}{\times}10^6$s$^{-1}$) and the cavity-mirror coupling rate $\chi$. {\bf (b)} Same as in panel {\bf (a)} but for  $E_{MC}$. We have taken $T=10\mu$K, mechanical quality factor $3{\times}10^4$, $m=50$ng, ${\cal R}=50$mW, cavity finesse $F=10^4$ and $\zeta=\chi$.  The cavity decay rate is given by $\kappa=\pi c/2 {\cal L} F$ where $c$ is the speed of light and the cavity length is ${\cal L}=1$mm {\bf (c)} Comparison between $E_{AC}$ (solid) and $E_{MC}$ (dotted) against $T$ for $\Delta=2\omega_m$ and $\chi=100$s$^{-1}$. Other parameters are as in {\bf (a)}.
}
\label{fig:E5}
\end{figure*}

\subsection{Bipartite entanglement}

Here we analyze the stationary entanglement in the three possible bipartitions of the system. 
Let us start with the symmetric situation where $\Omega=\omega_m$ and where the cavity-mirror coupling equals the cavity-atoms coupling ($\zeta=\chi)$. We always restrict ourselves to a stable regime for the linear Langevin system. In this case and for very low initial temperature of the mechanical mode (we take $T=10 \mu$K),  we find that entanglement is generated in the stationary state between the Bogoliubov mode and the cavity field as well within the field-mechanical mode subsystem. Due to the symmetric coupling $E_{AC}$ is very similar to $E_{MC}$, as shown in Fig.~\ref{fig:E5} ({\bf a})-({\bf b}). This provides an interesting {\it diagnostic tool}: the inaccessibility of the mirror mode makes the inference of the optomechanical entanglement a hard task needing cleverly designed, although experimentally challenging, indirect methods~\cite{Vitali2007,MauroPRL}. However, in light of the recent demonstration of controllability of intra-cavity atomic systems~\cite{Esslinger2008,Esslinger2007}, we can think of inferring the pure optomechanical entanglement simply by measuring the more accessible correlations between the optical field and the Bogoliubov mode. In the range of parameters considered here, the mirror-atoms entanglement $E_{AM}$ is always zero. This is in contrast with the results in Ref.~\cite{genes}, where the emergence of $E_{AM}$ is due to the use of an effective negative detuning that regulates the free evolution of the collective atomic quadrature. In our case, such evolution is ruled by the frequency $\Omega$, which is always positive. Therefore, the two models can access quite different regions of the parameter-space. In turn, this implies that, while our system does not allow for hybrid atom-mirror entanglement, Genes {\it et al.} in Ref.~\cite{genes} did not achieve the symmetric $E_{MC}=E_{AC}$ situation revealed above.

\begin{figure*}[t]
{\bf (a)}\hskip5cm{\bf (b)}\hskip5cm{\bf (c)}
\includegraphics[scale=0.21]{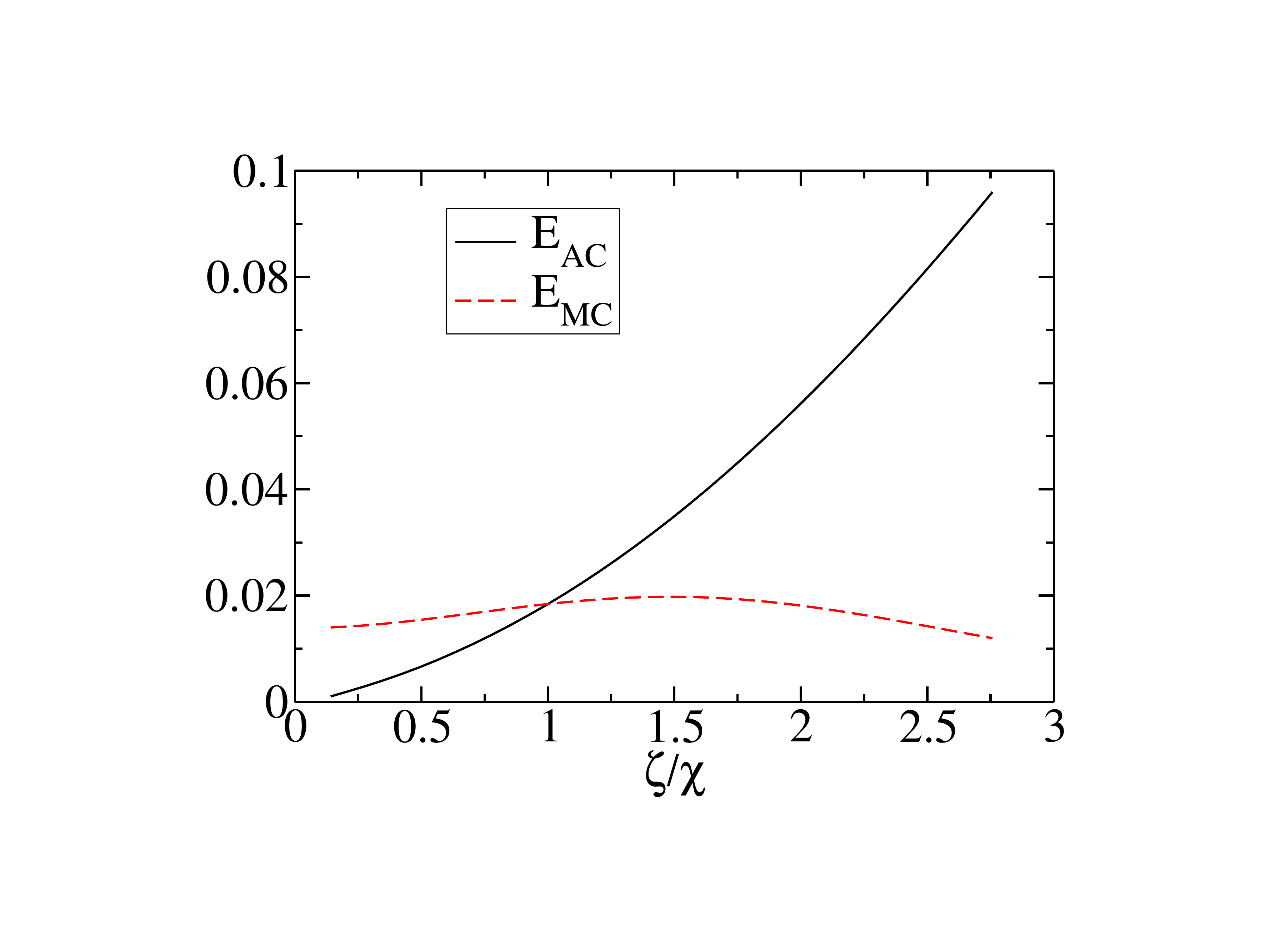}~~\includegraphics[scale=0.21]{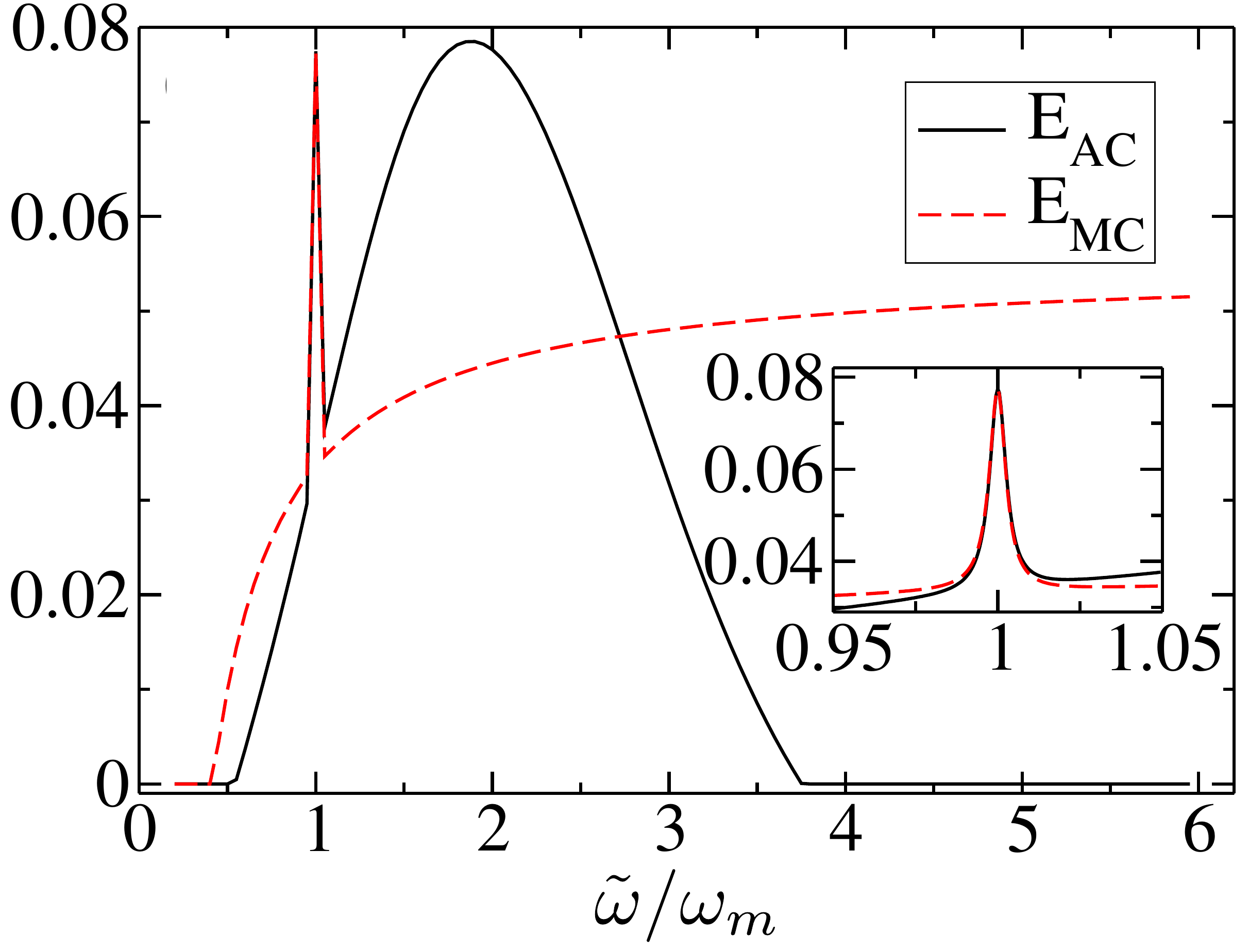}~~\includegraphics[scale=0.21]{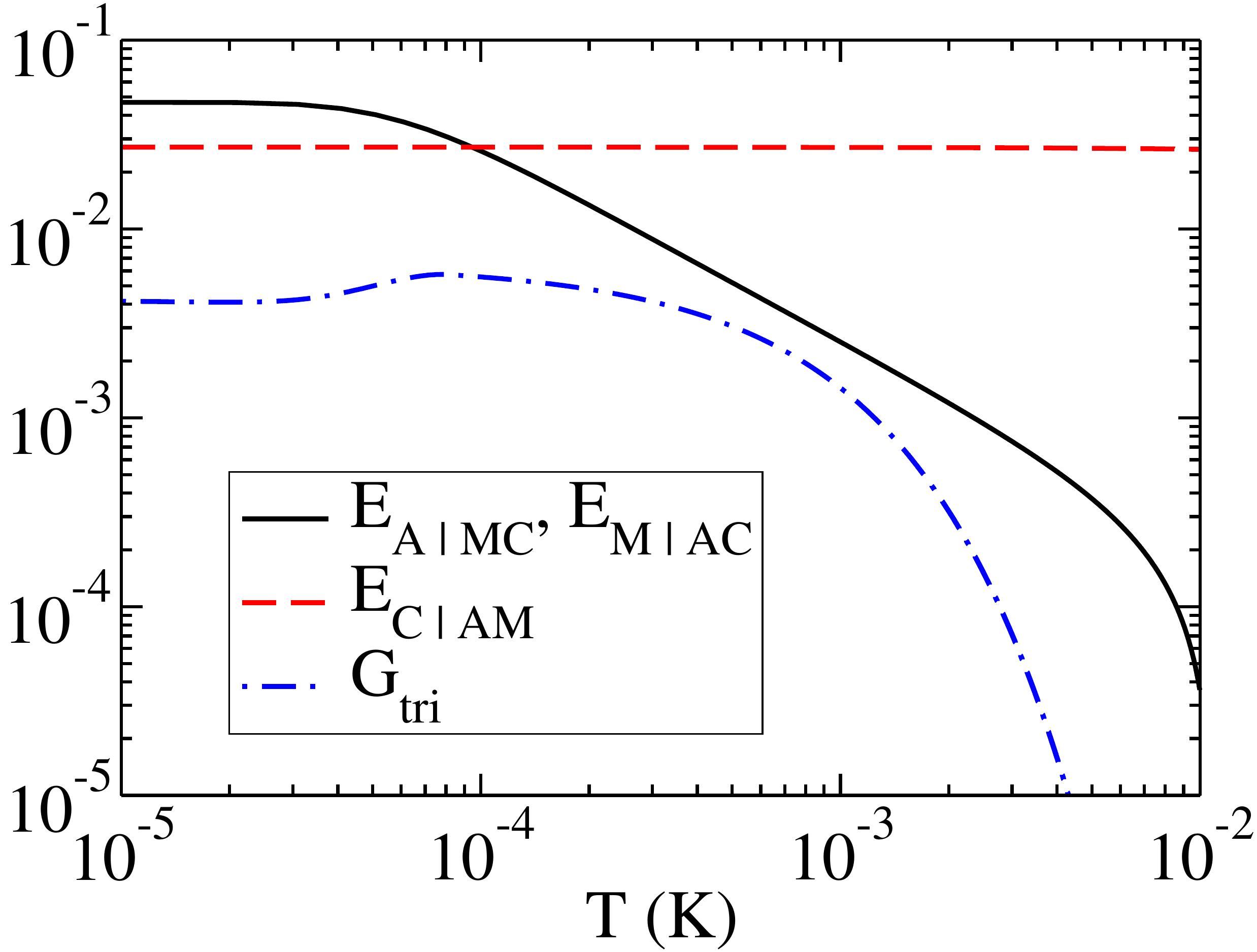}
\caption{{\bf (a)} Logarithmic negativity $E_{AC}$ (solid) and $E_{MC}$ (dashed) against $\zeta/\chi$ for $\Delta=2\omega_m$.
Other parameters are as in Fig.~\ref{fig:E5}. {\bf (b)} We show $E_{AC}$ (line) and $E_{MC}$ (dashed) against $\tilde\omega/\omega_m$. The inset shows the entanglement functions around $\tilde\omega=\omega_m$ for
$T=1\mu$K and $F=4{\times}10^4$. {\bf (c)} We show $E_{i|jk}$  ($i,j,k=A,M,C$) and the genuine tripartite entanglement $G_{tri}$ against $T$ for $\tilde\omega=\omega_m$ and $\chi=\zeta$. We have $E_{A|MC}=E_{M|AC}$ (solid), $E_{C|AM}$ (dashed) and $G_{tri}$ (dot-dashed). Other parameters as in panels {\bf (a)} and {\bf (b)}.}
\label{fig:entasym1}
\end{figure*}

We now study how $E_{AC}$ and $E_{MC}$ decay with $T$. The results plotted in Fig.~\ref{fig:E5} {\bf (c)} show that the two entanglements are indistinguishable and therefore the BEC can still be used as an entanglement probe. As expected the entanglement decays when increasing the environment temperature and disappears for $T>0.1$mK. Higher critical temperatures for the disappearance of entanglement are found for larger cavity quality factors and more intense pumps.
We consider two different regimes, where we relax the symmetric conditions between the mirror and the Bogoliubov mode. In the first situation, we take $\omega_m=\tilde\omega$ and vary the coupling rates. In the second, we take symmetric couplings $\zeta=\chi$ and change the frequencies. For  $\omega_m=\tilde\omega$, $E_{AC}$ and $E_{MC}$ are strongly affected by the change in $\zeta$ and $\chi$. As shown in Fig.~\ref{fig:entasym1} {\bf (a)}, $E_{AC}$ grows continuously for larger $\zeta$ while $E_{MC}$ decreases slightly.
In this regime, it is clear that small inaccuracies of the ratio $\zeta/\chi$ do not affect the mirror-cavity entanglement, therefore confirming the role of the BEC as a minimal disturbance probe.
A more involved situation occurs when we keep $\omega_m$ fixed and change $\tilde\omega$. We take $T=1\mu$K, cavity finesse $F=4{\times}10^4$ and find a sharp peak at $\tilde\omega=\omega_m$ where $E_{MC}=E_{AC}$, see [Fig.~\ref{fig:entasym1} {\bf (b)}]. While $E_{MC}$ increases slowly with $\Omega/\omega_m$ (the Bogoliubov mode goes out of resonance from the cavity and decouples from the rest of the system), $E_{AC}$ reaches its  maximum at $\tilde\omega/\omega_m\approx2$.
These results show that by changing the Bogoliubov frequency, for example varying the longitudinal trapping frequency, the BEC acts as a switch for the mirror-cavity entanglement, inhibiting or enhancing it.

\subsection{Genuine multipartite entanglement}
\label{sec:multi}

We now study the existence of genuine multipartite entanglement in our system by looking at the logarithmic negativity in each one-vs-two-mode bipartition. According to Ref.~\cite{giedke}, if in a tripartite state all such bipartitions are inseparable, genuine multipartite entanglement is shared. For the systems at hand, we indeed find inseparability of the bipartitions $A|MC$, $M|AC$ and $C|AM$ [see Fig.~\ref{fig:entasym1} {\bf (c)}], which strongly confirms the presence of tripartite entanglement up to very high temperatures. Evidently, multipartite entanglement persists up to  $T\sim{0.01}$K, which is much larger than the critical one for the disappearance of bipartite entanglement ($\sim8\times10^{-5}$K). Such a result spurs the quantitative study of the genuine tripartite-entanglement content of the overall state~\cite{contangle,contangle2}. For pure multipartite states, this is based on {monogamy inequalities} valid for the {squared logarithmic negativity}, which turns out to be a proper entanglement monotone. For mixed states, the convex-roof extension of such a measure is required, albeit restricted to the class of Gaussian states. More explicitly, we aim at determining~\cite{contangle2}
\begin{equation}
\label{cota}
{\text G}_{\text{tri}}=\min_{\Pi({ijk})}[{{\text G}_{i|jk}{-}{\text G}_{i|j}{-}{\text G}_{i|k}}]
\end{equation}
with $\text{G}_{i|j}$ the convex roof of the squared logarithmic negativity for the $i$-vs-$j$ bipartition of a system and $\Pi(ijk)$ the permutation of indices $i,j,k=A,M,C$. More explicitly, given a (generally mixed) Gaussian state with covariance matrix ${\bm\sigma}$, we have
\begin{equation}
\text{G}({\bm\sigma})=\inf\overline{(E^{p}_{i|j})^2}=\inf_{{\bm\sigma}_p\le{\bm\sigma}}(E^p_{i|j})^2
\end{equation}
with $(E^{p}_{i|j})^2$ the squared logarithmic negativity of a pure Gaussian state with covariance matrix ${\bm\sigma}_p$ whose eigenvalues are all smaller or equal than those of ${\bm \sigma}$, which is the covariance matrix of the state to assess, and $\overline{z}$ the average value of quantity $z$. In general, the evaluation of $\text{G}_{\text{tri}}$ is very demanding. However, for $\omega_m=\Omega$ and $\chi=\zeta$, the covariance matrix $\mathcal V$ is symmetric under the permutation of $A$ and $M$, which greatly simplifies the calculations: The residual tripartite entanglement can be thus determined against the effects of $T$, showing a non monotonic behaviour. To understand this, we take $i=C$, $j=M$ and $k=A$ (any other combination could be considered). Similar to $E_{AC}$ and $E_{MC}$, $\text{G}_{C|A,M}$ decays very quickly as $T$ increases, thus biasing the competition between ${\text G}_{C|MA}$ and ${\text G}_{C|A}+{\text G}_{C|M}$. Analogously to $E_{C|AM}$, ${\text G}_{C|A}+{\text G}_{C|M}$ decreases very slowly with $T$, thus determining an overall increase of the residual entanglement. However, as $T$ is raised further, the system tends toward two-mode biseparability and any tripartite entanglement is washed out [see Fig.~\ref{fig:entasym1} {\bf (c)}].


\subsection{Probing experimentally the stationary entanglement}
\label{sec:det}
\begin{figure}[b]
\includegraphics[scale=1.3]{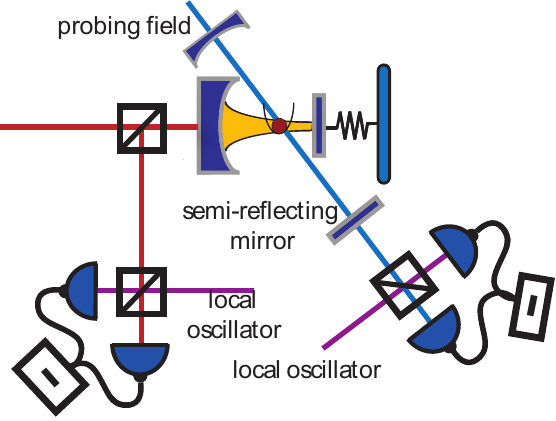}
\caption{
Sketch of the detection scheme. The probe field has the same periodicity of the cavity field. It impinges on a semi-reflecting mirror, creating a standing wave. Part of the probe light is transmitted and then measured with the output light of the cavity. 
}
\label{fig:sketch}
\end{figure}

We can now discuss the direct observation of the entanglement between the Bogoliubov mode and the cavity field. The idea is to shine the BEC with a probe laser in a standing wave configuration
~\cite{RitschNatPhys}  such that the laser beam forms a small angle with respect to the cavity axis, as sketched in Fig.~\ref{fig:sketch}.
For an off-resonant probe field with polarization perpendicular to that of the primary cavity field, 
the additional Hamiltonian term reads
\begin{equation}
\hat{\cal H}_P=\hbar\left[\Delta_P{+} U_P \int d^3 {\bf r} \cos^2(k_c x{+}k_z z)
\hat \Psi^\dagger\hat \Psi\right]\hat a_P^\dagger \hat a_P^{},
\end{equation}
where $\Delta_P$ is the detuning of the probe field from the pumping laser,
$\hat a^{}_P$ is the corresponding annihilation operator, $U_P=g_P^2/\Delta_{AP}$ is the light-atom coupling constant, $k_z$ is the wave-vector of the probing field along a direction orthogonal to the cavity axis and $\hat\Psi$ is the condensate field operator. For a BEC  strongly confined in the plane orthogonal to the cavity axis, we can neglect excitations of transverse modes and factorize the field operator as 
\begin{equation}
\label{eq:psifactor}
\Psi=\psi(x) f(y) f(z),
\end{equation}
where $f(y)$ is a function peaked around the cavity axis and characterized by a width $\sigma$. We then expand $\psi(x)$ as~\cite{StringariPitaevskii}
\begin{equation}
\psi(x)=\sqrt{N_0} \psi_0(x){+}\sum_{k>0,\sigma=\pm}[u_{k\sigma}(x) c_{k\sigma}{-}v_{k\sigma}^*(x) c_{k\sigma}^\dagger].
\end{equation}
We assume that the width of the BEC in the transverse direction is much smaller than the periodicity of the probe field along the $z$ direction. This implies that transverse excitations are suppressed.
Neglecting terms ${\cal O}(k_z^2\sigma^2)$ we get
\begin{equation}
\hat{\cal H}_{AP}\simeq\hbar\left(\Delta_P+\frac{N_0 U_P}{2}{+}\zeta_P \hat Q \right)\hat a_P^\dagger \hat a^{}_P-i\hbar \eta_P(\hat a^{}_P-\hat a^\dagger_P),
\end{equation}
where $\zeta_P$ is the same as $\zeta$ with $U_0{\rightarrow}U_P$.
The Langevin equation for the probe field reads
\begin{equation}
{\partial}_t \hat{a}^{}_P=\eta_P{-}i\left[\Delta_P{+}\frac{N_0 U_P}{2}+\zeta_P \hat Q\right] \hat a^{}_P
{+}\sqrt{2\kappa} \hat a_{Pin}{-}\kappa \hat a_P.
\end{equation}
Assuming again an intense pump field, the steady intensity of the probe field is $|\alpha_P|^2={\eta_P^2}/(\tilde\Delta_P^2+\kappa_P^2)$,
where 
\begin{equation}
\tilde\Delta_P=\Delta_P{+}\frac{U_P N_0}{2}{+}\zeta_P Q_s.
\end{equation}
We also assume that $\zeta_P\ll\Omega,\zeta$ and $\alpha_P\ll\alpha_S$.
The equation of motion for the fluctuations of the probe field is
\begin{equation}
\partial_t\delta \hat a_P=-i \tilde \Delta_P \delta \hat a_P{-}i\zeta_P\alpha_p\delta \hat Q{+}\sqrt{2\kappa}\delta \hat a_{Pin}{-}\kappa_P\delta \hat a_P.
\end{equation}
In order to map the Bogoliuobov mode onto the probe field, following the same technique as in Ref.~\cite{Vitali2007}, we choose $\tilde\Delta_P=\Omega\gg\kappa_P, \zeta_P\alpha_P$. The equations of motion for the slowly varying variables $\hat{\tilde o}(t)=e^{i\Omega t} \hat o(t)$ thus read
\begin{equation}
\partial_t\delta\hat{\tilde a}=-i\frac{\zeta_P \alpha_P}{\sqrt 2}\hat{ \tilde c}{+}\sqrt{2\kappa}\delta\hat{\tilde a}_{Pin}{-}\kappa_P\delta \hat{\tilde a}_P.
\end{equation}
If the decay of the probe cavity is faster than the dynamics of the Bogoliubov modes, the probe field follows adiabatically the dynamics of the latter. Using the cavity input-output relations~\cite{Walls} we get
\begin{equation}
\delta\hat{\tilde a}_{Pout}={-}i\frac{\zeta_P \alpha_P}{\sqrt \kappa_P}\hat{\tilde c}{+}\delta\hat{\tilde a}_{Pin}
\end{equation}
which shows how the Bogoliubov mode is mapped onto the output probe field. To measure the entanglement between the field of the primary cavity and the Bogoliubov mode one can homodyne the cavity field and rotate the quadrature of the probe. Analogously, one could map $\hat{q}$ onto a further field so that the light-matter correlations are changed into amenable light-light ones~\cite{Vitali2007}.


\section{Embedding optimal control in hybrid quantum optomechanics}
\label{Benpaper}

Having proven the possibility offered by hybridisation for the observation of entanglement in a BEC-assisted cavity optomechanical setting, we now explore a route for overcoming the limitations enforced by a steady-state analysis. In particular, we are interested in discussing whether or not it is possible to observe entanglement between the BEC and the mechanical system during the short-time dynamics of the overall device. 

In order to assess this problem, here we review a proposal for the active driving control of the hybrid system here at hand realized by time-modulating the intensity of the driving field. Inspired by recent works on the control of optomechanical devices~\cite{mari,schmidt}, we show that by using a monochromatic modulation, entanglement between two mesoscopic systems, the mirror and the BEC, can be created and controlled~\cite{farace}. Furthermore, by borrowing ideas from the theory of optimal control~\cite{doria} we show that, with respect to the unmodulated case, we achieve a sixfold improvement in the degree of generated entanglement. We interpret such performance in terms of the occurrence of a special resonance at which the building up of entanglement  is {\it favored}. This strongly suggests the viability of the optimal control-empowered manipulation of open mesoscopic systems for the achievement of strong quantum effects, even in the hybrid context, of which the system that we study is a significant representative.

\begin{figure*}[t]
{\bf (a)}\hskip7cm{\bf (b)}
\includegraphics[width=0.45\linewidth]{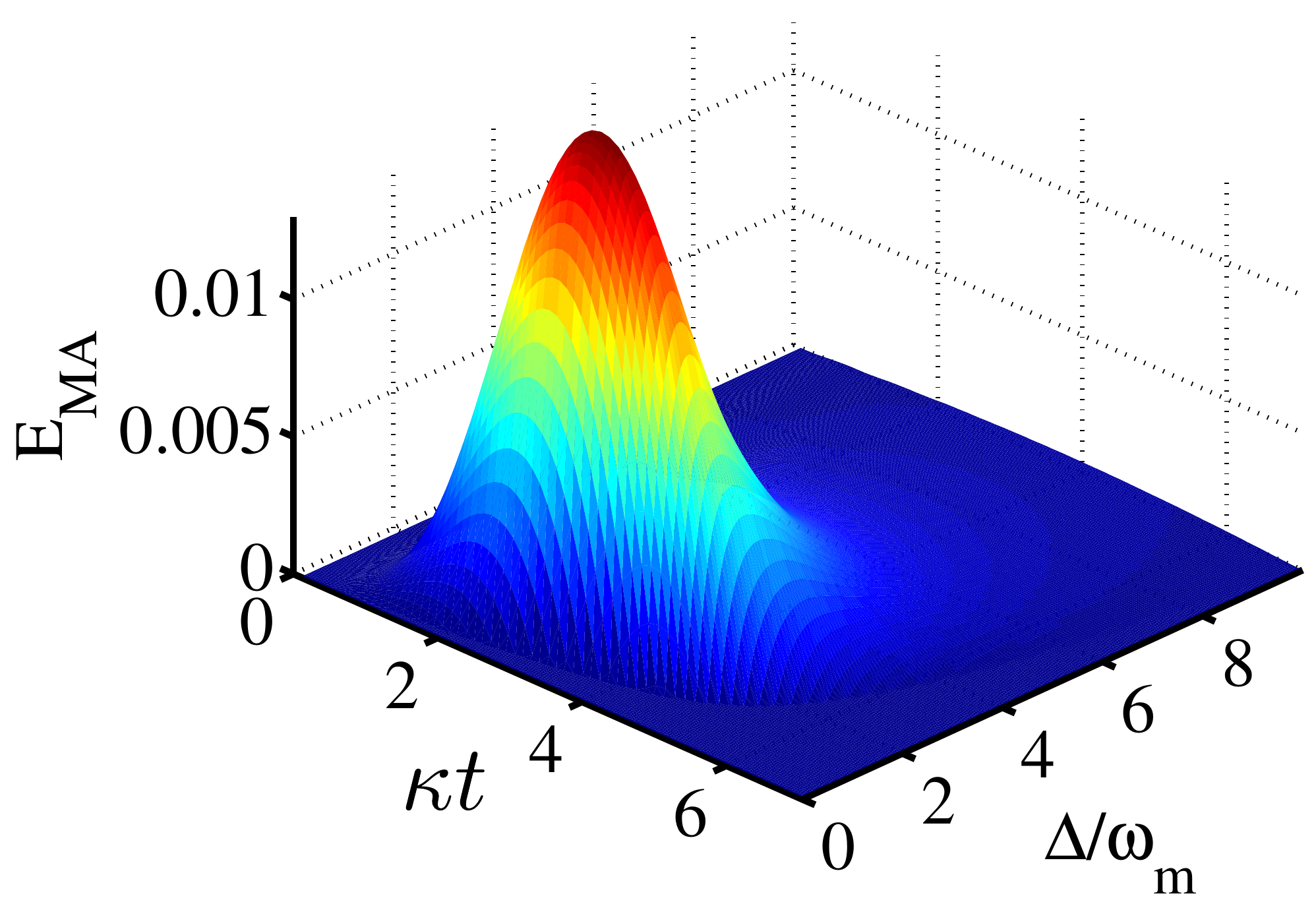}~~\includegraphics[width=1.1\columnwidth]{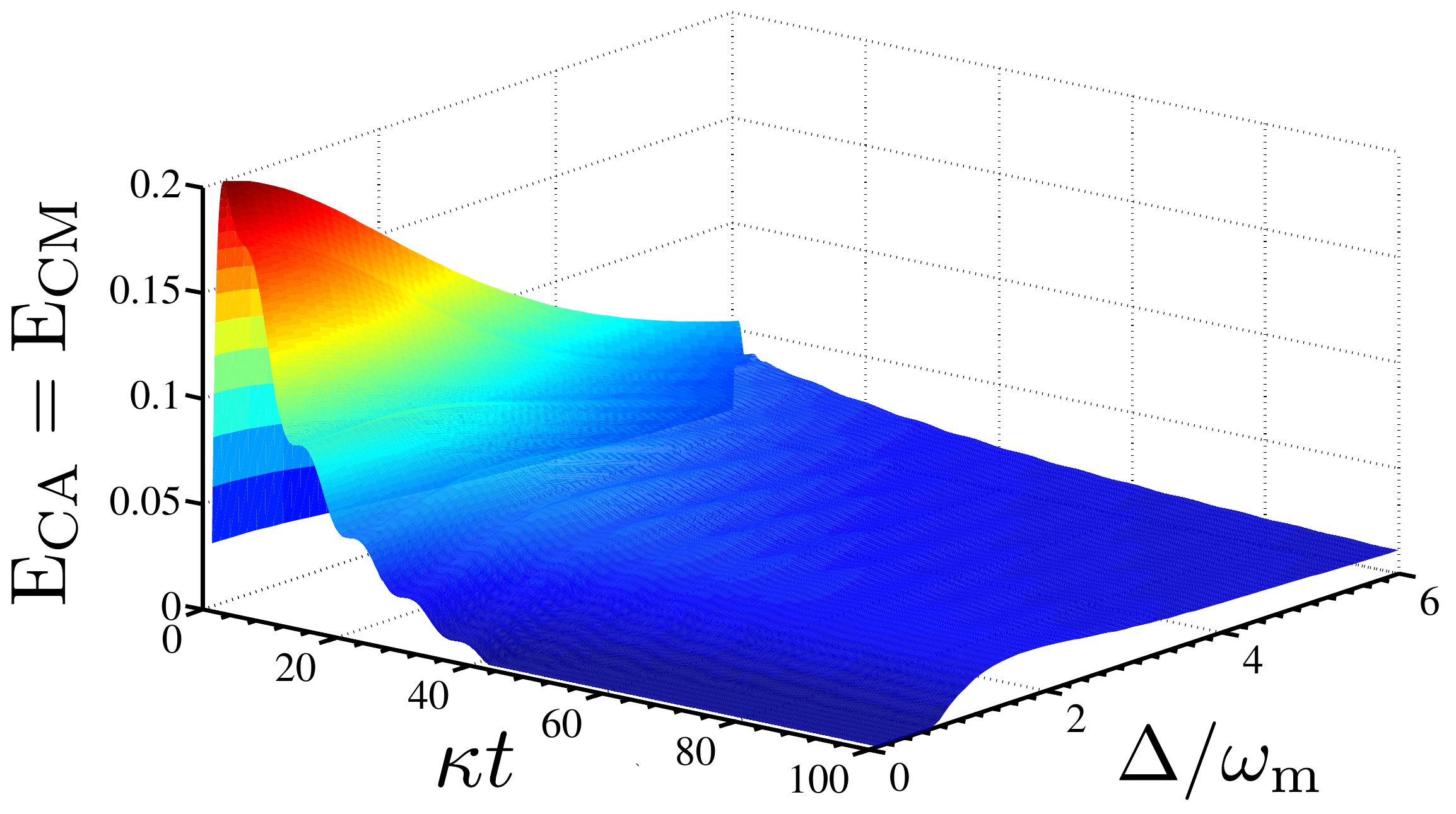}
\caption{{\bf (a)} Entanglement $E_{MA}(t)$ between mirror and atoms against $\kappa t$  and  $\Delta/\omega_m$. {\bf (b)}  Same as panel {\bf (a)} for  $E_{CM}(t)=E_{CA}(t)$.
Parameters: $\omega_m/2\pi=3\times10^6$s$^{-1}$; $T=10\mu$K; $Q=3\times10^4$; $m=50$ng, ${\cal R}=50$mW, cavity finesse $F=10^4$, and $\zeta=\chi$; cavity length is $1$mm from which $\kappa=\pi c/2{\cal L}F$ can be determined ($c$ is the speed of light).}
\label{fig:deltaplots}
\end{figure*}

\subsection{Time-resolved entanglement in the hybrid system}
In Ref.~\cite{DeChiara2011} and in the previous Section, the stationary entanglement within the hybrid optomechanical system has been considered. Here we focus on the dynamical regime where the evolution of the entanglement is resolved in time.  We consider the fully symmetric regime encompassed by equal frequencies for the Bogoliubov and mechanical modes (i.e. $\omega_b=\omega_m$) and identical coupling strengths in the bipartite cavity-mirror and cavity-atom subsystems (that is, we take $\zeta=\chi$). We are particularly interested in the emergence of atom-mirror entanglement at short interaction times. The analysis conducted in Sec.~\eqref{Gabrielepaper} has shown it to be absent at the steady state~\cite{DeChiara2011}. 
%
%
However, this might not well be the case for the time-resolved dynamics of the system, which can be assessed by using the dynamical version of Eq.~\eqref{Lya}, much along the lines of Eq.~\eqref{Lyapunovdue} 
%
\begin{equation}
\label{eq:Vdot}
\dot{{\cal V}}_{MCA} = {\cal K}_{MCA}{\cal V}_{MCA}+{\cal V}_{MCA}{\cal K}^T_{MCA} + {\cal D}_{MCA}.
\end{equation}
%
Eq.~(\ref{eq:Vdot}) is solved assuming the initial conditions ${\cal V}_{MCA}(0) = \mathrm{diag}[{1},1,2\bar{n}{+}{1},2\bar{n}{+}{1},1,1]/2$, which describes the vacuum state of both the cavity field and the BEC mode and the thermal state 
of the mechanical system. Physicality of the covariance matrix has been thoroughly checked by considering the fulfilment of the Heisenberg-Robinson uncertainty principle and checking that the minimum symplectic eigenvalue $\nu_{MCA}=\min\mathrm{eig}(i\omega{\cal V}_{MCA})$ is such that $|\nu_{MCA}|\ge\frac{1}{2}$ (we have used the $6 \times 6$ symplectic matrix $\omega=\oplus^3_{j=1}i\sigma_y$ with $\sigma_y$ the y-Pauli matrix). 
Using again the logarithmic negativity we find that the atom-mirror entanglement $E_{MA}(t)$, whose time evolution is shown in Fig.~\ref{fig:deltaplots} {\bf (a)} for different values of the effective detuning $\Delta$, gradually develops and reaches its peak value as the cavity-atom and cavity-mirror entanglement drop to a quasi-stationary value.  As no direct atom-mirror interaction exists in this system, mediation through the cavity mode essentially results  in a delay: quantum correlations between the atoms (the mirror) and the cavity mode must build up before any atom-mirror correlation can appear. This is clearly shown in Fig.~\ref{fig:deltaplots} {\bf (b)}, where $E_{CM}(t)=E_{CA}(t)$ reach their maximum well before $E_{MA}(t)$ starts to grow.
The atom-mirror entanglement is non-zero only within a very short time window, signalling the fragility of the entanglement resulting from only a second-order interaction between the BEC and the cavity end-mirror. These results go far beyond the limitations of the steady-state analysis conducted in Sec.~\ref{Gabrielepaper} and Ref.~\cite{DeChiara2011}, and prove the existence of a regime where all the reductions obtained by tracing out one of the modes from the overall system are inseparable, a situation that is typical only of a time-resolved picture.

\subsection{Optimal control of the early-time entanglement}
We now consider the effects of time-modulating the external pump power ${\cal R}$, which is now considered a function of time. In turn, this implies that we now take $\eta\rightarrow\eta(t)$ and study the time behaviour of the entanglement $E_{MA}$ set between the atoms and the mirror. We will show that a properly optimised $\eta(t)$ can increase the maximum value of $E_{MA}(t)$ for values of $t$ within the same time interval $\tau$ where atom-mirror entanglement has been shown to emerge in the unmodulated case.
We assume to vary $\eta(t)$ slowly in time, so that the classical mean values $\phi_s$ adiabatically follow the change in $\eta(t)$. This approximation is valid as long as the number of intra-cavity photons is large enough to retain the validity of the linearization procedure and the time-variation of $\eta(t)$ is slow compared to the time taken by the mean values to reach their stationary values. For all cases considered here we have verified the validity of such assumptions. The dynamics of the covariance matrix is thus still governed by Eq.~\eqref{eq:Vdot} with the replacement ${\cal K}\rightarrow{\cal K}(t)$. In the following, we use the value of $\Delta=2.7\omega_m$ which maximises the short time entanglement $E_{MA}$.

\begin{figure*}[t]
{\bf (a)}\hskip5cm{\bf (b)}\hskip5cm{\bf (c)}
\includegraphics[width=0.34\linewidth]{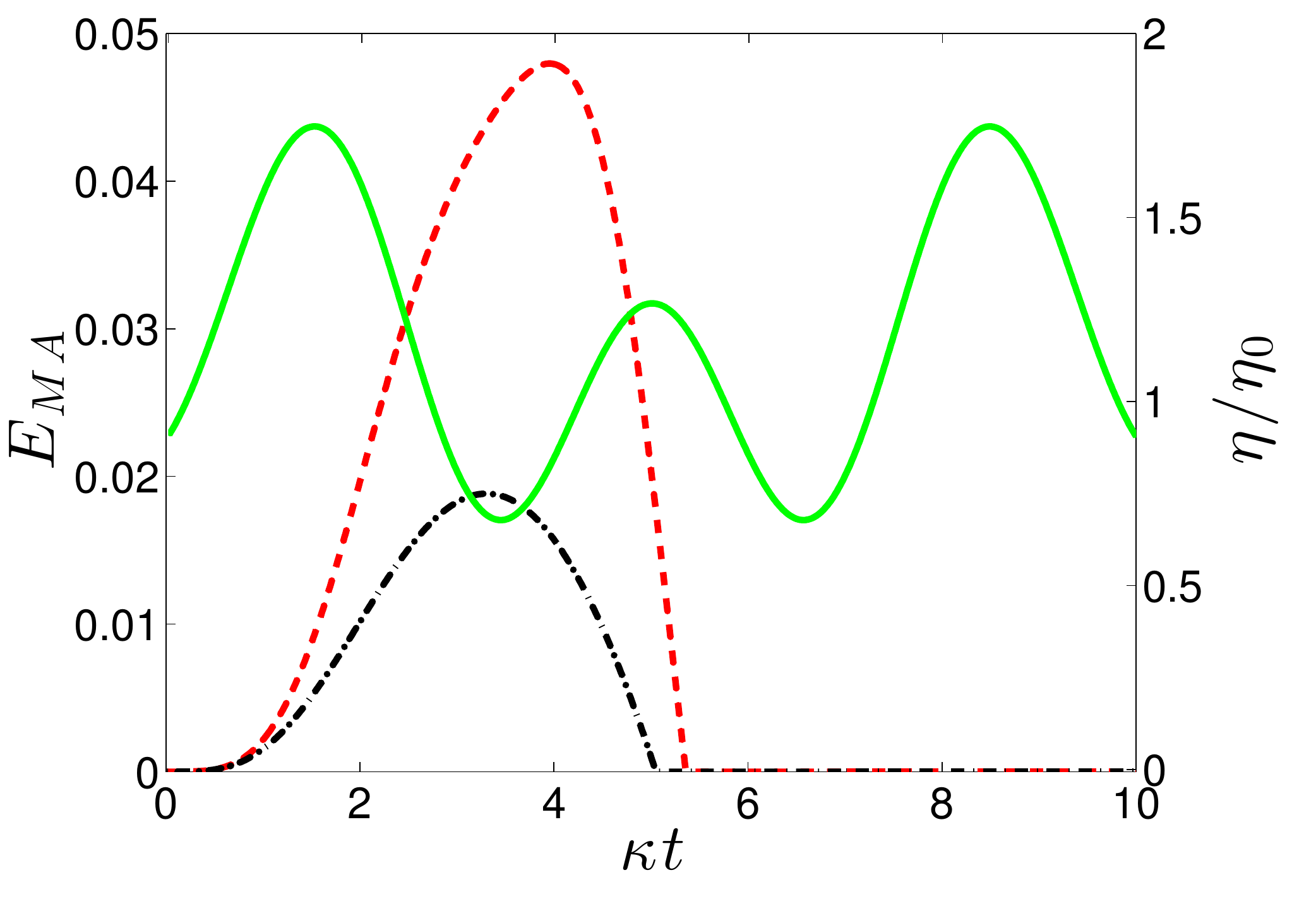}\includegraphics[width=0.33\linewidth]{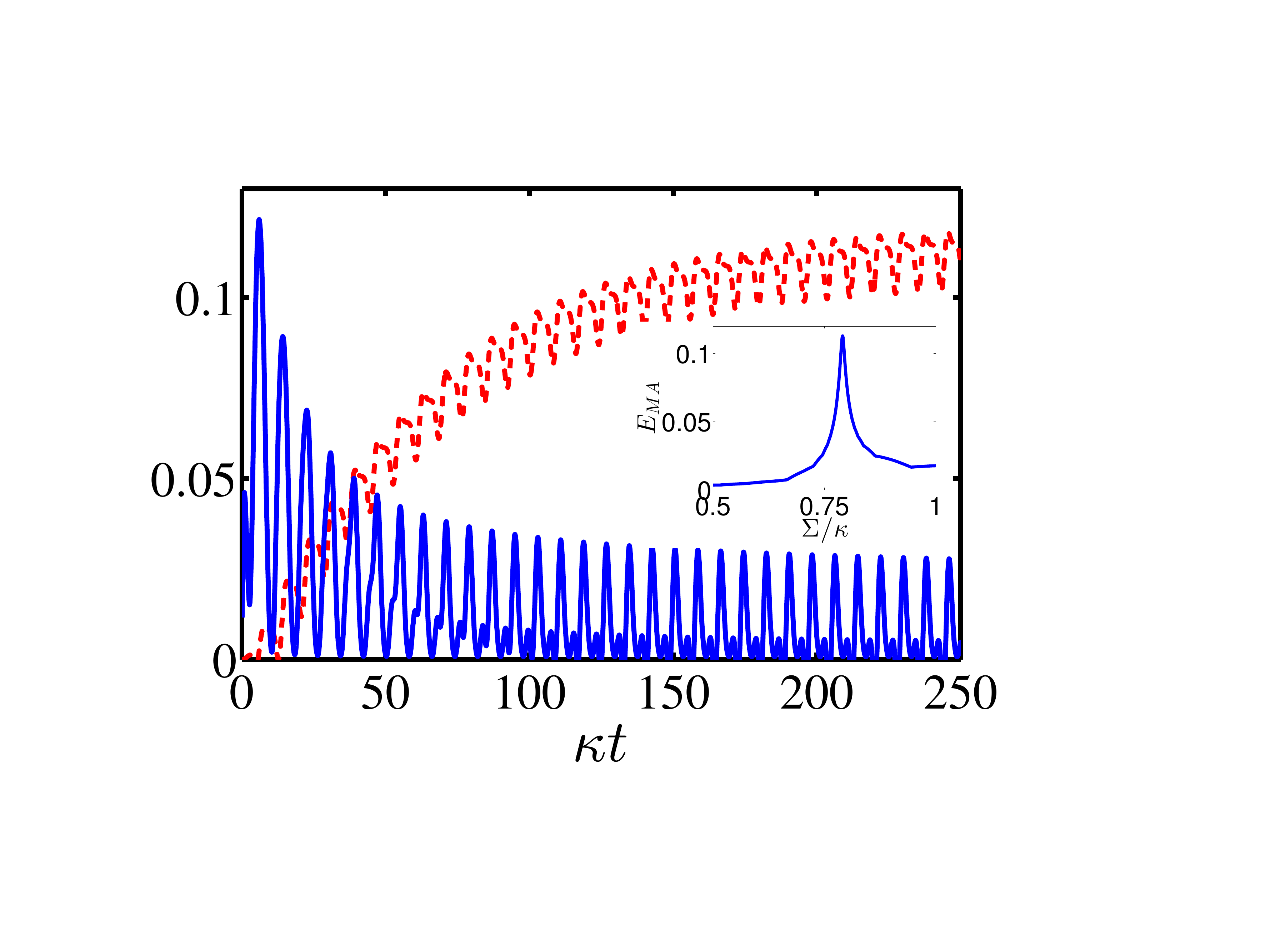}\includegraphics[width=0.35\linewidth]{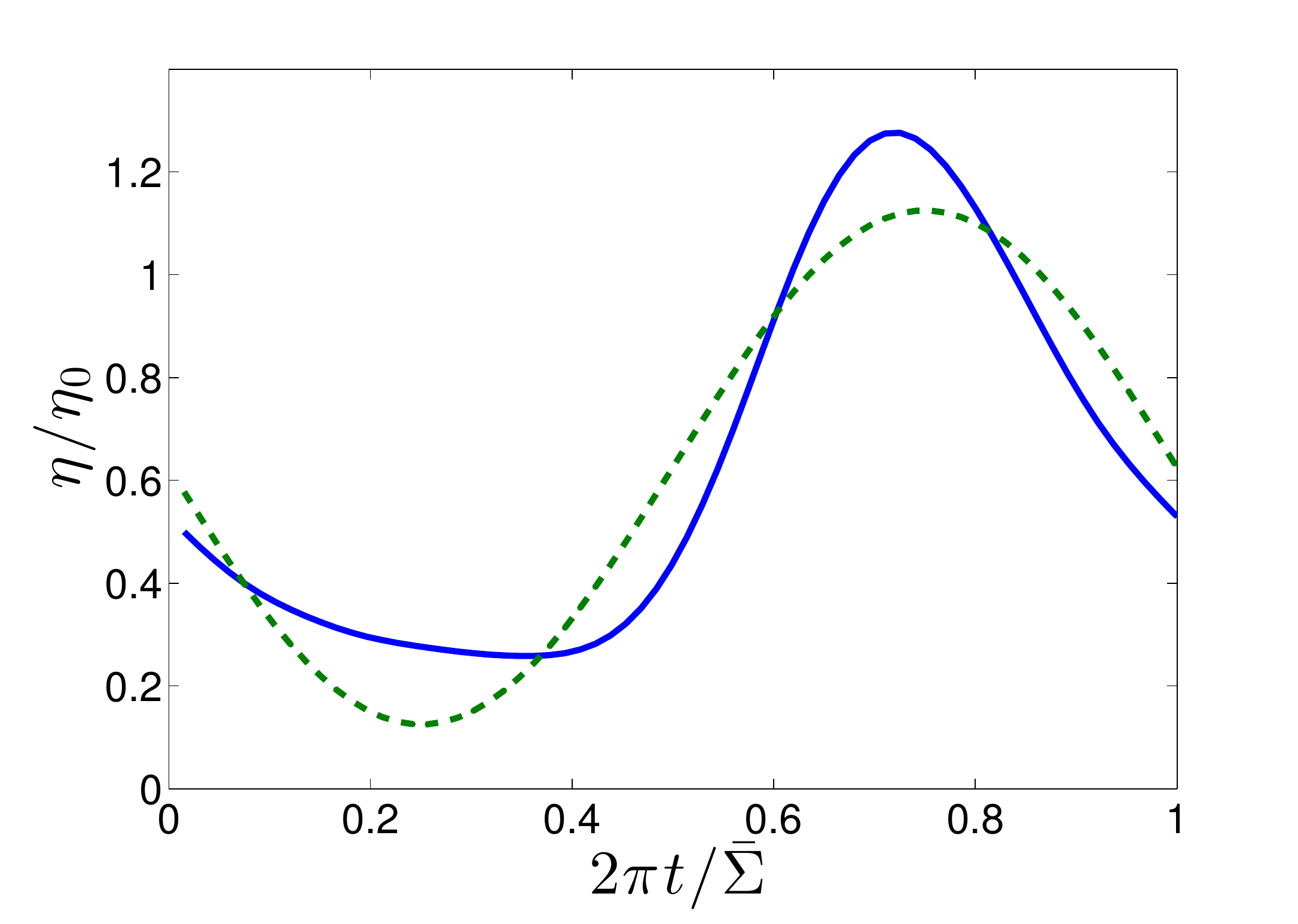}
\caption{\label{fig:periodic}
{\bf (a)} Entanglement dynamics $E_{MA}(t)$ (dashed line) with the optimal laser intensity modulation $\eta(t)$ (solid line). The entanglement $E_{MA}(t)$ for constant $\eta(t)=\eta_0$ is also shown (dashed-dotted line). {\bf (b)} Dynamics of the cavity-mirror and cavity-atoms entanglement $E_{CM,CA}$ (solid line) and atoms-mirror entanglement $E_{MA}$ (dashed line) for $\Sigma=\bar\Sigma\sim 0.79 \kappa$. Inset: Maximum $E_{MA}$ for long times with a periodic modulation as a function of the frequency $\Sigma$. {\bf (c)} Optimal periodic modulation $\eta(t)$ (solid line) for one period of time $2\pi/\bar\Sigma$ compared to the monochromatic modulation (dashed line).}
\end{figure*}

 Inspired by the techniques for dynamical optimization proposed in~\cite{doria}, we call $\eta_0$ the unmodulated value of $\eta$ and take
\begin{equation}
\eta(t) = \eta_0 + \sum_{j=1}^{j_{max}}\left[ A_j \cos(\omega_j t)+B_j\sin(\omega_j t) \right],
\end{equation}
where $\omega_j = 2\pi j/\tau+\delta_j$ are the harmonics and $\delta_j$ is a small random shift. The coefficients $A_j$ and $B_j$ are chosen in a way that the total energy brought in by the time-modulated field is the same as the one associated with the unmodulated case.
We then set the time-window so that $\tau=3.4\kappa^{-1}$, when we observe the maximum value of $E_{MA}$ in the unmodulated instance. The other parameters are as in Fig.~\ref{fig:deltaplots}. We then look for the parameters $A_j$ and $B_j$ that  maximise the value of $E_{MA}(\tau)$ for a given set of random shifts $\delta_j$. We use standard optimisation routines to find a (local) maximum of $E_{MA}(\tau)$. We repeat the search of the optimal coefficients  for different values of $\delta_j$ and take the overall maximum. The corresponding results are presented in Fig.~\ref{fig:periodic} {\bf (a)}, where we show the optimal modulation $\eta(t)$ and the optimal $E_{MA}(t)$. These findings are also compared to the case without modulation. The maximum value attained in the interval $[0;\tau]$ is $E_{MA}(t)\simeq0.05$ which is 2.5 times larger than the case without modulation, thus demonstrating the effectiveness of our approach. 
%
\noindent
\subsection{Periodic modulation: long time entanglement}
In the two situations analyzed so far [constant laser intensity $\eta$ and optimally modulated $\eta(t)$], the atom-mirror entanglement $E_{MA}(t)$ is destined to disappear at long times. A complementary approach based on a periodic modulation of the laser intensity $\eta(t)$ was used in the pure optomechanical setting \cite{mari} to increase the long-time light-mirror entanglement. Here we use a similar approach by assuming the monochromatic modulation of the laser-cavity coupling
%
$\eta(t) =\eta''_0+{\eta'_0}\left[1 - \sin(\Sigma t)\right]$,
%
where $\Sigma$ is the frequency of the harmonic modulation, $\eta'_0=4\eta''_0={\eta_0}/2$, $\eta_0$ being the same constant coupling parameter taken before. These choices ensure that the approximations used in the dynamical analysis are valid. After the transient dynamics, the covariance matrix and, in turn, $E_{MA}(t)$ become periodic functions of time. In order to achieve the best possible performance at long times, we compute the maximum of $E_{MA}(t)$ after the transient behaviour, scanning the values of $\Sigma$. The result is shown in Fig.~\ref{fig:periodic} {\bf (b)} (inset) revealing a sharp resonance with a maximum value of $E_{MA}\simeq 0.12$ for $\Sigma=\bar\Sigma\sim 0.79 \kappa$ (no further peak appears beyond this interval). This arises as a result of the effective interaction between atoms and mirror mediated by the cavity field. As shown in Fig.~\ref{fig:deltaplots} {\bf (a)} the entanglement dynamics strongly depend on the effective detuning giving rise to such optimal behaviour. Similar results have also been observed in Ref.~\cite{mari}.

The analysis of the evolution of $E_{CM(CA)}$ and $E_{MA}$, shown in the main panel of Fig.~\ref{fig:periodic} {\bf (b)}, reveals that while $E_{CM(CA)}$ develops very quickly due to the direct cavity-atoms and cavity-mirror couplings, $E_{MA}$ grows in a longer time lapse, during which the cavity disentangles from the dynamics. The quasi-asymptotic value achieved by $E_{MA}$ reveals a sixfold increase with respect to the unmodulated case. This behaviour is worth commenting as it strengthens our intuition that any atom-mirror entanglement has to result from a process that effectively couples such subsystems, bypassing any mechanism giving rise to multipartite entanglement within the overall system. 
Finally, we discuss the results achieved in the long time case by adopting an optimal-control technique similar to the one used for enhancing the short time entanglement. We have considered the periodic modulation of the intensity at the frequency $\bar\Sigma$ given by
\begin{equation}
\label{eq:mod}
\eta(t) = {\eta''_0}+{\eta'_0}\left[1 - \sum_{n=1}^{n_{max}} A_n\sin(n\bar\Sigma t)+B_n\cos(n\bar\Sigma t)\right],
\end{equation}
and looked for the coefficients $\{A_n,B_n\}$ optimizing $E_{MA}(t)$ at long times with the power constraint: $\sum_{n=1}^{n_{max}}( A_n^2+B_n^2)\le 1$, ensuring that no instability is introduced in the dynamics of the overall system. In our simulation, $n_{max}=8$ has been taken to limit the complexity of the modulated signal. The resulting optimal coefficients give the periodic modulation shown in Fig.~\ref{fig:periodic} {\bf (c)} and a maximum entanglement $E_{MA}\simeq0.17$ which is about $30\%$ larger than the results obtained for the monochromatic modulation. This demonstrates the powerful nature of our scheme. Our extensive multiple-harmonic  analysis is able to outperform quite significantly the single-frequency driving scheme addressed above and discussed in Ref.~\cite{mari}, proving strikingly the sub-optimality of the monochromatic modulation, both at short and, surprisingly, at long times of the system dynamics. This legitimates  experimental efforts directed towards the use of time-modulated driving signals for the optimal control of the hybrid mesoscopic systems addressed here and similar ones based, for instance, on the use of a vibrating membrane~\cite{thompson} or a levitated nano-sphere~\cite{sphere} instead of the BEC.  

It should be noted that an adiabatic approach is used in Ref.~\cite{mari} to find an effective Hamiltonian.  When performed in our hybrid optomechnical scheme, such technique would remove the dynamics of the Bogoliubov modes of the atomic subsystem.

\subsection{Robustness to inaccuracies}

We now make a last comment on this scheme, addressing the robustness of the protocol with respect to inaccuracy in the control of the value $\chi$. After finding the best periodic modulation assuming $\chi=\xi$, the simulations have been run again using the same modulation with $\chi=1.1 \xi$ and $\chi=0.9 \xi$, i.e. a 10\% inaccuracy in the nominal values of $\chi$ and $\xi$. Notwithstanding the substantial entity of such inaccuracy, one can see that the corresponding maximum entanglement is at most only 3\% less than the original value, thus confirming the stability of the results illustrated above. Notice also that if the imbalance between $\chi$ and $\xi$ is known, for example by a calibration measurement, we can in principle run the optimization including the actual values of $\chi$ and $\xi$ therefore aiming at a larger entanglement value.


\section{Hybrid BEC optomechanics as a probe of mechanical quantum coherences}
\label{Nicolapaper}

The problems addressed so far have shown that the BEC-based hybridization
of optomechanical settings enriches considerably the range of interesting physical effects that 
can be induced in the system. However, they have also raised (and partially addressed) the 
issue of the revelation of the quantum features of mechanical systems of difficult addressability.
In this Section we attack this problem directly by designing an hybrid optomechanical setup for the inference of 
quantum coherences in the state of a mechanical oscillator. Our method is again based on the interaction with a BEC, 
although the regime that we adopt here is rather different from the one used in Secs.~\ref{BECpaper}-\ref{Benpaper}.

\begin{figure}[t]
  \includegraphics[width=\linewidth]{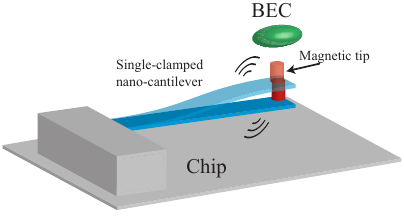}
  \caption{Sketch of the setup for BEC-based probing
    of mechanical coherences.  A BEC is placed in close proximity to a
    nano-mechanical cantilever endowed with a magnetic tip. The
    coupling between the magnetic field generated by the mechanical
    {\it quantum antenna} and the ultra-cold atoms embodies a
    mechanism for the effective probing of coherences in the state of
    the mechanical system.}
\label{fig:setup}
\end{figure}

We consider the setup sketched in Fig.~\ref{fig:setup}, which consists
of an on-chip single-clamped cantilever and a spinor BEC trapped in
close proximity to the chip and the cantilever. The latter is assumed
to be manufactured so as to accommodate at its free-standing end a
single-domain magnetic molecule (or {\it tip}). Technical details on
the fabrication methods of similar devices can be found in
Refs.~\cite{tip,hunger}, which have also been found to have very large
quality factors, which guarantee a good resolution of the rich variety
of modes in the cantilever's spectrum.
%
%
At room temperature, thermal fluctuations are able to (incoherently)
excite all flexural and torsional modes and in the following we assume
that a filtering process is put in place, restricting our observation
to a narrow frequency window, so as to select only a single mechanical
mode.


The second key element of our setup is a BEC of ${}^{87}$Rb atoms held
in an (tight) optical trap and prepared in the hyperfine level
$\ket{F=1}$.  As we assume the trapping to be optical, there is no
distinction between atoms with different quantum numbers $m_F=0,\pm1$
of the projections of the total spin along the quantization axis.
Moreover, for a moderate number of atoms in the condensate and a tight
trap, we can invoke the so-called single-mode approximation
(SMA)~\cite{SMAfail}, which amounts to considering the same spatial
distribution for all spin states.  These approximations will be made
rigorous and formal in the next Subsections.

\subsection{Hamiltonian of the system}

We now briefly review the mapping of a spinor BEC into a rotor~\cite{rotor}.  The most effective way is to start from the second-quantization version of the Hamiltonian of a BEC trapped in an anisotropic potential, which reads~\cite{spinorham}
\begin{equation}
  \label{eq:ham}
  \begin{aligned}
    \hat{H}&=\sum_{\alpha} \int d^3{\bm r}\,\hat{\psi}_{\alpha}^{\dagger}({\bm r})\hat{H}_{\alpha}^{0}\hat{\psi}_{\alpha}({\bm r})\\
    &+\sum_{\alpha,\beta,\mu,\nu}\!\!G_{\alpha,\beta,\mu,\nu}\!\!\int d^3{\bm r}\,\hat{\psi}_{\alpha}^{\dagger}({\bm r})\hat{\psi}_{\beta}^{\dagger} ({\bm r})\hat{\psi}_{\mu}({\bm r})\hat{\psi}_{\nu}({\bm r}),
  \end{aligned}
\end{equation}
where the second line of equation describes the particle-particle
scattering mechanism, ${\bm r}=(x,y,z)$, and 
\begin{equation}
\hat{H}_{\alpha}^{0}{=}-\frac{\hbar^2}{2
m}{\bm \nabla}^2+\frac 12 m_a \left[\omega^2(x^2+y^2)+\omega_z^2 z^2\right].
\end{equation}
 Here, $m_a$ is the mass of
the Rb atoms and $(\omega,\omega,\omega_z)$ is the vector of frequencies of the trapping potential. 
The subscripts $\alpha, \beta,\mu,\nu$ refer to different z-components of the single-atom spin states. Since
the scattering between two particles does neither change the total
spin nor its z-component, we can link the coefficients
$G_{\alpha,\beta,\mu,\nu}$ to the scattering lengths for the channels
with total angular momentum $F_{T}=0,2$. Thus, by making use of the
Clebsch-Gordan coefficients,
the full BEC Hamiltonian can be re-written as
\begin{equation}
  \label{eq:hamspin}
  \begin{aligned}
    &\hat{H}=\sum_{\alpha} \int d{\bm r} \,
    \hat{\psi}_{\alpha}^{\dagger}({\bm r})\hat{H}_{\alpha}^{0}\hat{\psi}_{\alpha}({\bm r})\\
    &+\frac{c_s}{2} \sum_{\alpha,\beta} \int d{\bm r} \, \hat{\psi}_{\alpha}^{\dagger}({\bm r})\hat{\psi}_{\beta}^{\dagger} ({\bm r})\hat{\psi}_{\alpha}({\bm r})\hat{\psi}_{\beta}({\bm r})\\
    &+\frac{c_a}{2}\!\!\sum_{\alpha,\beta,\alpha{'},\beta{'}}\!\!\int
    d{\bm r} \, \hat{\psi}_{\alpha}^{\dagger}({\bm
      r})\hat{\psi}_{\beta}^{\dagger} ({\bm r})({\bm
      F}_{\alpha,\beta}{\cdot}{\bm
      F}_{\alpha'\!,\beta})\hat{\psi}_{\alpha{'}}({\bm
      r})\hat{\psi}_{\beta{'}}({\bm r}),
  \end{aligned}
\end{equation}
where we have set $c_{s}{=}(g_0{+}2 g_2)/3$ and $c_{a}{=}(g_2{-}g_0)/3$ with
$g_{2j}{=}4 \pi \hbar^2 a_{2j}/m$ $(j{=}0,1)$ ($a_{2j}$ is the
scattering length for the $F_T=2j$ channel~\cite{scatteringlengths}). Here ${\bm F}$ is the vector of 
spin-$1$ matrices obeying the commutation relation $[{\bm F}^{i},{\bm
  F}^{j} ]=i\epsilon_{ijk} {\bm F}^{k}$ with $\epsilon_{ijk}$ the Levi-Civita tensor.

If $c_a\approx 0$ (i.e. if $g_0\approx g_2$) and/or the number of atoms is not too
large, the total Hamiltonian is symmetric in the three spin
components.  By assuming a strong enough optical confinement and a BEC
of a few thousand atoms, one can thus think of the condensate field operator as having a constant spatial distribution for all the three
species $m_F=0,\pm1$ and write
$\opx{\psi}{\bm r}{\alpha}{}{=}\funcx{\psi}{\bm r}{}{}\hat{d}_{\alpha}{}$.
This is the so-called SMA~\cite{SMAfail,spindynamics}, which leaves the Hamiltonian in the
form
\begin{equation}
  \label{eq:single}
  \begin{aligned}
    \hat{H}&= \sum_{\alpha} \op{c}{\alpha}{\dagger}\op{c}{\alpha}{}
    +\frac{c'_s}{2}\sum_{\alpha,\beta}
    \op{c}{\alpha}{\dagger}\op{c}{\beta}{\dagger}
    \op{c}{\alpha}{}\op{c}{\beta}{}\\
    &+ \frac{c'_a}{2}\!\!\sum_{\alpha,\beta,\alpha{'},\beta{'}}\!\!
    ({\bm F}_{\alpha,\beta}\cdot{\bm F}_{\alpha',\beta'})\,
    \op{c}{\alpha}{\dagger}\op{c}{\alpha{'}}{\dagger}
    \op{c}{\beta}{}\op{c}{\beta{'}}{},
  \end{aligned}
\end{equation}
where we have defined $c'_{i}{=}c_{i} \int d^3{\bm r}\modulo{{\psi}({\bm r})}^{4}$.  As the distance $z_0$ between the
BEC and the magnetic tip can be in the range of a few $\mu$m (we take
$z_0=1.5 \mu$m in what follow) and the spatial dimensions of the BEC
are typically between tenths and hundredths of $\mu$m (we considered
$a_z=0.25 \mu$m and $a_r=0.09 \mu$m), the relative correction to the
magnetic field across the sample is of the order of
0.2, 
which is small enough to justify the SMA.  Moreover, in the
configuration assumed here, the system will be mounted on an atomic
chip, where the static magnetic field can be tuned by adding magnets
and/or flowing currents passing through side wires. Such a design can
compensate distortions to the trapping potential induced by the
tip.

By introducing
$\hat{N}{=}\sum_{\alpha}\hat{c}^\dag_\alpha\hat{c}_\alpha$
%
%
and the angular momentum operators
$\op{L}{+}{}{=}\sqrt{2}(\op{c}{0}{\dagger}\op{c}{-1}{}+\op{c}{1}{\dagger}\op{c}{0}{})$
and
$\op{L}{z}{}{=}(\op{c}{1}{\dagger}\op{c}{1}{-}\op{c}{-1}{\dagger}\op{c}{-1}{})$,
we can rewrite Eq.~\eqref{eq:single} as
$\hat{H}{=}\op{H}{A}{}+\op{H}{S}{}$, where we have explicitly
identified a symmetric part $\op{H}{S}{}=\mu_{cp}
\hat{N}-c'_{s}\hat{N}(\hat{N}-1)$ (with $\mu_{cp}$ the chemical potential) and an antisymmetric one
$\op{H}{A}{}=c'_{a}(\hat{\bm L}^2-2\hat{N})$ with $\hat{\bm L}=(\hat L_x,\hat L_y,\hat L_z)$ and $\hat L_{\pm}=(\hat L_x\pm\hat L_y)/2$.
It is important to remember that such a mapping is possible due to the
assumption of a common spatial wave function for the three spin
components. As long as the antisymmetric term is small enough, this is
not a strict constraint.  By exploiting Feshbach resonances~\cite{feshbach}, it is
possible to adjust the couplings $g_0$ {and} $g_2$ in such a way that
$g_0\approx g_2$, which allows for the possibility to increase the
number of atoms in the BEC, still remaining within the validity of the SMA.

We now consider the BEC interaction Hamiltonian when an external
magnetic field is present.  Due to its magnetic tip, the cantilever
produces a magnetic field and we assume that only one mechanical mode
is excited, so that the cantilever can be modelled as a single quantum 
harmonic oscillator
whose annihilation (creation) operator we call $\op{b}{}{}$
($\op{b}{}{\dagger}$), in line with the notation used so far. By allowing the tip to have an intrinsic
magnetization, we can split the magnetic field into a (classical) static 
contribution ${\bm B^0}$ and an (operatorial) oscillating one $\delta \op{{\bm
    B}}{}{}$ that arises from the oscillatory behaviour of the
mechanical mode. The physical mechanism of interaction is
Zeeman-like, i.e. each atom experiences a torque which tends to
align its total magnetic moment to the external magnetic field. The
Hamiltonian for a single atom can be written as
\begin{equation}
 \label{eq:zee}
 \op{H}{Z}{(1)}=-{\bm \mu}{\cdot}{\bm B}= \frac{g\mu_{B}}{\hbar}\op{{\bm S}}{}{(1)}{\cdot}{\bm B},
\end{equation}
where $\mu_{B}$ is the Bohr magneton, $\op{{\bm S}}{}{(1)}$ is the
spin operator vector for a single atom and $g$ is the gyromagnetic
ratio. In line with Ref.~\cite{gyro}, we adopt the convention that $g$
and ${\bm \mu}$ have opposite signs. The total interaction
Hamiltonian is then given by the sum over all the atoms.
By taking the direction of ${\bm B}^{0}$ as the quantization axis
(z-axis) and the x-axis along the direction of $\mean{\delta \op{{\bm
      B}}{}{}}{}$, the magnetic Zeeman-like Hamiltonian is
\begin{equation}
 \label{eq:zeeang}
 \op{H}{Z}{}=g\mu_{B}B_{z}^{0}\op{L}{z}{}+g\mu_{B}G_ca_{c}(\op{b}{}{\dagger}+\op{b}{}{})\op{L}{x}{},
\end{equation}
where we have used $\delta \op{{\bm B}}{}{}=(\delta\op{{\bf B}}{x}{},0,0)$ with $\delta\op{{\bf B}}{x}{}= G_c
a_{c}(\op{b}{}{\dagger}+\op{b}{}{})\overline{{\bf x}}$, $G_c=3\mu_0|{\boldsymbol \mu}_c|/(4\pi z_0^4)$ 
the gradient of the magnetic field produced by the tip at a distance
$z_0$,
$\overline{{\bf x}}$ the unit vector along the x-axis, and
$a_{c}=\sqrt{\hbar/(2 m \omega_m)}$.
As in the previous Sections, we have used the symbols $m$ and $\omega_m$ to indicate the mass of the mechanical oscillator and its frequency.
The full Hamiltonian of the BEC-cantilever system is thus $\op{H}{}{}=
\op{H}{BEC}{0}+\op{H}{c}{0}+\op{H}{I}{}$ with
\begin{equation}
 \label{eq:totham}
 \begin{aligned}
  &\op{H}{BEC}{0}=\mu_{cp} \hat{N}{-}c'_{s}\hat{N}(\hat{N}-1){+}
 c'_{a}(\hat{\bm L}^2-2\hat{N}){+}g\mu_{B}B_{z}^{0}\op{L}{z}{},\\
 &\op{H}{c}{0}=\hbar\omega_{c}\op{b}{c}{\dagger}\op{b}{c}{},~~\op{H}{I}{}=g\mu_{B}G_c a_{c}(\op{b}{}{\dagger}+\op{b}{}{})\op{L}{x}{}.
 \end{aligned}
\end{equation}
It has been shown in Refs.~\cite{spinorham,spindynamics} that
$\op{H}{BEC}{0}$ with $B_{z}^{0}=0$ allows for an interesting dynamics
of the populations of the three spin states, which undergo Rabi-like
oscillations, thus witnessing the coherence properties of the BEC.


\subsection{Mapping into a rotor}

The Hamiltonian $\hat H$ with components as in Eqs.~\eqref{eq:totham} should be further manipulated in order to cast it 
in a form that fits with our needs. 
This is accomplished by implementing a formal mapping of the BEC into a quantum rotor, in line with  
Ref.~\cite{rotor}. As we work with a fixed number of particles,
the state of the BEC can be decomposed as
\begin{equation} 
  \label{gen}
\sum_{n'_{0,\pm1}}C_{n'_{0,\pm1}}(\op{c}{1}{\dagger})^{n'_{1}}(\op{c}{0}
  {\dagger})^{n'_{0}}(\op{c}{-1}{\dagger})^{n'_{-1}}\ket{0},
\end{equation}
where the sum is performed over all sets of labels $\{n'_{0,\pm1}\}$
such that $n'_{0}{+}n'_{-1}{+}n'_{1}{=}N$.  Let us now introduce the
Schwinger-like operators
$\op{s}{x}{}{=}(\op{c}{-1}{}{-}\op{c}{1}{})/\sqrt{2}$,
$\op{s}{y}{}{=}(\op{c}{1}{}{+}\op{c}{-1}{})/(i \sqrt{2})$,
$\op{s}{z}{}{=}\op{c}{0}{}$ such that
$[\op{s}{\alpha}{},\op{s}{\beta}{}]{=}0$,
$[\op{s}{\alpha}{},\op{s}{\beta}{\dagger}]{=}\delta_{\alpha,\beta}$~\cite{rotor}.
The generic state of the BEC in Eq.~\eqref{gen} can now be written as
$\ket{{\bm \Omega}_{N}}{=}\frac{1}{\sqrt{N!}}({\bm \Omega}{\cdot}\hat{\bm s}^\dag)^{N}\ket{0}$ with ${\bm
  \Omega}{=}(\cos\phi\sin\theta,\sin\phi\sin\theta,\cos\theta)$ a unit vector whose direction is determined by the set of polar coordinates $(\theta, \phi)$. 
  By varying the direction of ${\bm \Omega}$ on the unit sphere it is possible to recover any
superposition for the state of a single atom among the states with
$m_z=0,\pm 1$. 

Any state with a fixed number of particles in the
bosonic Hilbert space can then be written as $\ket{\Psi}=\int
d\Omega\, \varphi({\bf \Omega})\ket{{\bf \Omega}_N}$ where $\varphi({\bf
  \Omega})$ is the wave function of the rotor we are looking for to
complete the mapping. 
This is accomplished by introducing the following 
components of the angular momentum operator along the $x$ and $z$ directions
\begin{equation}
  \begin{aligned}
    \op{L}{z}{}&=-i
    (\op{s}{x}{\dagger}\op{s}{y}{}-\op{s}{y}{\dagger}\op{s}{x}{})=
    -i \overline{{\bf z}}\cdot({\bf \Omega}\times{\bm \nabla})=\frac{1}{\hbar}\overline{{\bf z}} \cdot \op{\mathcal{\bm L}}{}{}=-i \partial_{\phi},\\
    \op{L}{x}{}&=\frac{1}{2}(\op{s}{z}{\dagger}\op{s}{x}{}-\op{s}{x}{\dagger}\op{s}{z}{})+\frac{i}{2} \;(\op{s}{z}{\dagger}\op{s}{y}{}-\op{s}{y}{\dagger}\op{s}{z}{})\\
    &=-i\overline{{\bf x}}\cdot({\bf \Omega}\times{\bm \nabla})=\frac{1}{\hbar}\overline{{\bf x}} \cdot
    \op{\mathcal{\bm L}}{}{}=i(\sin\phi\;\partial_{\theta}{+}\cot\theta\cos\phi\partial_{\phi}).
  \end{aligned}
\end{equation}
After discarding an inessential constant term, the Hamiltonian that we will need reads
$\op{\mathcal{H}}{}{}=\op{\mathcal{H}}{R}{0}+\op{\mathcal{H}}{c}{0}+\op{\mathcal{H}}{I}{}$
with
\begin{equation}
 \begin{aligned}
 \label{eq:rotorham}
	\op{\mathcal{H}}{R}{0}&= \;c'_{a}\op{\mathcal{L}}{}{2}+\frac{g\mu_{B}}{\hbar}B_{z}^{0}\;\op{\mathcal{L}}{z}{},\\
	\op{\mathcal{H}}{c}{0}&=\frac{\op{p}{}{2}}{2m}+\frac12m\omega_{m}^{2}\op{q}{}{2},~\op{\mathcal{H}}{I}{}=\frac{g\mu_{B}G_c}{\hbar}\op{q}{}{}\op{\mathcal{L}}{x}{}.
 \end{aligned}
\end{equation}
We are now in a position to look at the BEC-cantilever joint dynamics.  In
particular we will focus on the detection of the cantilever properties
by looking at the BEC spin dynamics.

\subsection{Probing mechanical quantum coherences}
\label{meas}


The form of the interaction Hamiltonian $\op{\mathcal{H}}{I}{}$ allows
for the measurement of any observable whose corresponding operator on
the Hilbert space can be expressed as a function of $\op{q}{}{}$ and
$\op{p}{}{}$ with no backaction on the cantilever dynamics.
Moreover, when there is no magnetic field, the ground state of a
``ferromagnetic'' ({\it i.e.}~$c_2<0$) spinor BEC is such that all the
atomic spins are aligned along a direction resulting from a
spontaneous symmetry breaking process~\cite{spinorham}. Under the
effects of the cantilever antenna, two preferred directions are
introduced in the system: the $z$-direction along which we have the
static magnetic field and the $x$-direction defined by the oscillatory
component.  The interplay between these two competing magnetic fields
is responsible for a ``gyroscopic'' motion of the rotor about the
$z$-axis, exactly as in a classical spinning top. By looking at the
way the rotor undergoes such a gyromagnetic motion, we can gather
information on the properties of the state of the cantilever. We notice
that an approach similar to the one considered here has been used to show the resonant coupling of
an atomic sample of ${}^{87}$Rb atoms with a magnetic tip~\cite{magntip}. 
 
In order to understand the mechanism governed by Eqs.~\eqref{eq:rotorham}, let us look at the time
evolution of the operator $\opt{\mathcal{L}}{t}{x}{}$.  We take an
initial state of the form
\begin{equation}
  \label{eq:initial}
  \ket{in}=\sum_{n}C_{n}\ket{E_{n}}\otimes\int_{\Sigma_1}d\Omega\;\varphi(\Omega)\ket{\Omega},
\end{equation}
where $\Sigma_{1}$ is the unit sphere and $\ket{E_{n}}$ are the energy
eigenstates of the mechanical oscillator (such that
$\op{\mathcal{H}}{c}{0}\ket{E_{n}}{=}E_{n}\ket{E_{n}}$). In the
Heisenberg picture, the mean value of the x-component of the angular
momentum is
\begin{widetext}
 \begin{equation}
   \begin{aligned}
     \label{eq:mean}
     \mean{\opt{\mathcal{L}}{t}{x}{}}{}&=\braketM{in}{e^{i  \frac{\op{\mathcal{H}}{}{}}{\hbar}t}\opt{\mathcal{L}}{0}{x}{}e^{-i  \frac{\op{\mathcal{H}}{}{}}{\hbar}t}}{in}
     =\int_{q}dq \sum_{n,p}C_{p}^{*}C_{n}e^{-i \omega_{n,p}
       t}\phi_{p}^{*}(q)\phi_{n}(q) \int_{\Sigma_1}d\Omega
     \;\varphi^{*}(\Omega)\bigg({e^{i
         \frac{\op{\mathcal{H}_I}{}{}+\op{\mathcal{H}}{R}{0}}{\hbar}t}\opt{\mathcal{L}}{0}{x}{}e^{-i
         \frac{\op{\mathcal{H}_I}{}{}+\op{\mathcal{H}}{R}{0}}{\hbar}t}}\bigg)\varphi(\Omega),
   \end{aligned}
 \end{equation}
\end{widetext}
where we have used the position eigenstates $\ket{q}$ of the mechanical oscillator 
with associated wavefunction $\phi_{n}(q){=}\braket{q}{E_{n}}$ and
$\omega_{n,p}{=}\omega_{m}(n-p)$.  By setting
$\Omega_q{=}\sqrt{(g\mu_B/\hbar)^2[(B_z^0)^2+G_c^2 q^2]}$, the
time-evolved x-component of the angular momentum operator is
\begin{equation}
  \label{eq:lx}
\begin{aligned}
  &\opt{\mathcal{L}}{t}{x}{}
  =\frac{g^2 \mu_B^2}{\hbar^2\Omega^{2}_q}\left[(B_{z}^{0})^{2}\cos(\Omega_qt)+G_{c}^{2}q^{2}\right]\opt{\mathcal{L}}{0}{x}{}\\
  &{+}\frac{g \mu_B B_z^0 }{\hbar\Omega_q}\sin(\Omega_qt)\opt{\mathcal{L}}{0}{y}{}{+}\frac{g^2 \mu_B^2 B_z^0 G_c q}{\hbar^2\Omega^{2}_q}\left[1{-}\cos(\Omega_qt)\right]\opt{\mathcal{L}}{0}{z}{}\\
  &=a_1(q,t)\opt{\mathcal{L}}{0}{x}{}+a_2(q,t)\opt{\mathcal{L}}{0}{y}{}+a_3(q,t)\opt{\mathcal{L}}{0}{z}{}.
\end{aligned}
\end{equation}
Comparing Eqs.~\eqref{eq:mean} and \eqref{eq:lx} we find
$\langle{\opt{\mathcal{L}}{t}{x}{}}\rangle{=}\sum_{j=x,y,z}A_j(t)L_j^0
$,
where
\begin{equation}
  \begin{aligned}
    L_j^{0}&=\int_{\Sigma_1}d\Omega \;\varphi^{*}(\Omega)\opt{\mathcal{L}}{0}{j}{}\varphi(\Omega),\\
    A_j(t)&=\sum_{n,p}C_{p}^{*}C_{n}e^{-i \omega_{n,p} t}\int_{q}dq
    \phi_{p}^{*}(q)\phi_{n}(q) a_j(q,t).
\end{aligned}
\end{equation}
If the cantilever is initially prepared in the general mixed state $\rho_c(0)=\sum_{n}C_{n,m}\ketbra{E_n}{E_m}$,
a similar expression for the mean value of $\opt{\mathcal{L}}{t}{x}{}$ is found.

\begin{figure*}[t]
{\bf (a)}\hskip7cm{\bf (b)}
  \includegraphics[width=0.49\linewidth]{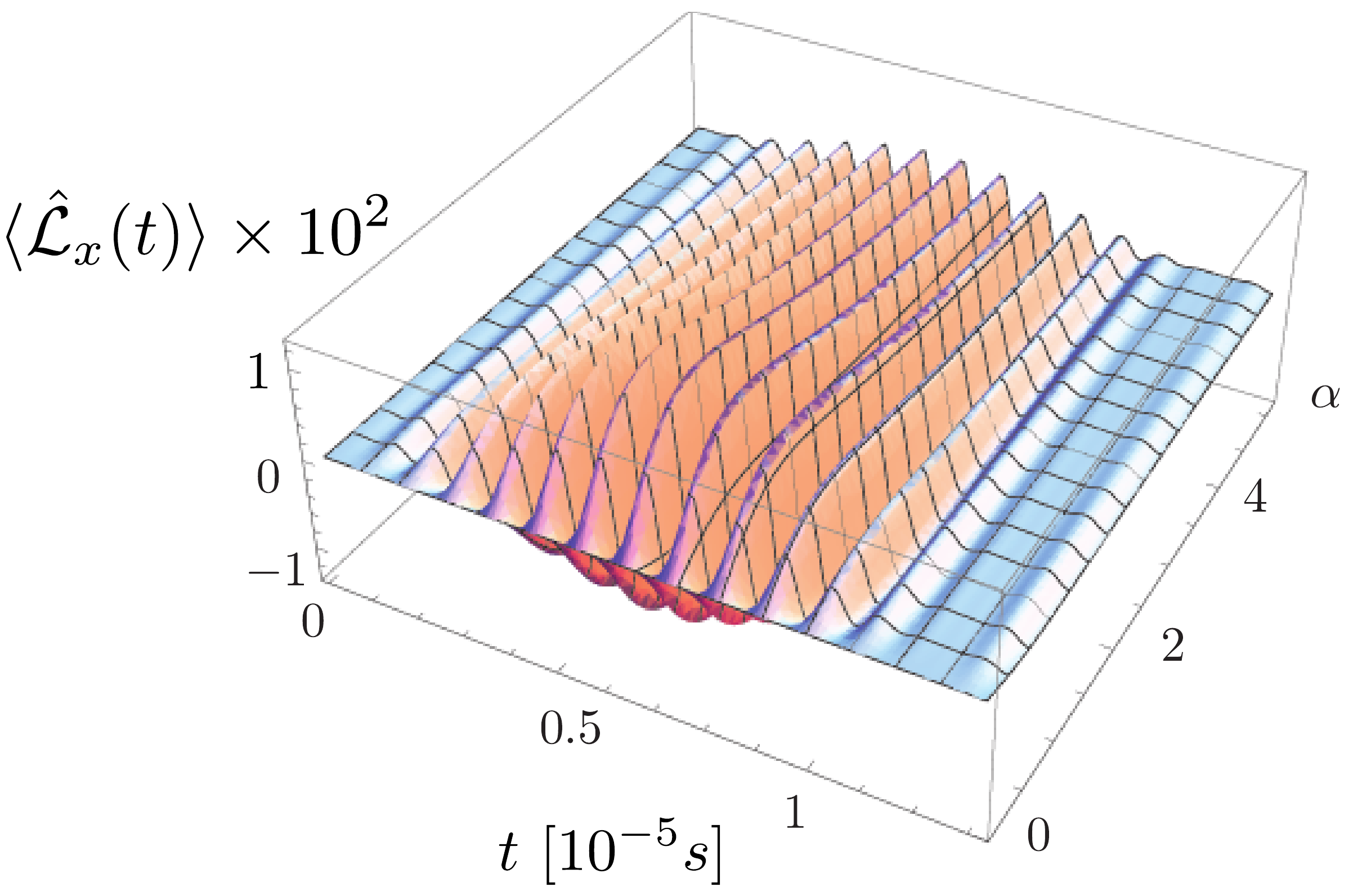}\includegraphics[width=0.51\linewidth]{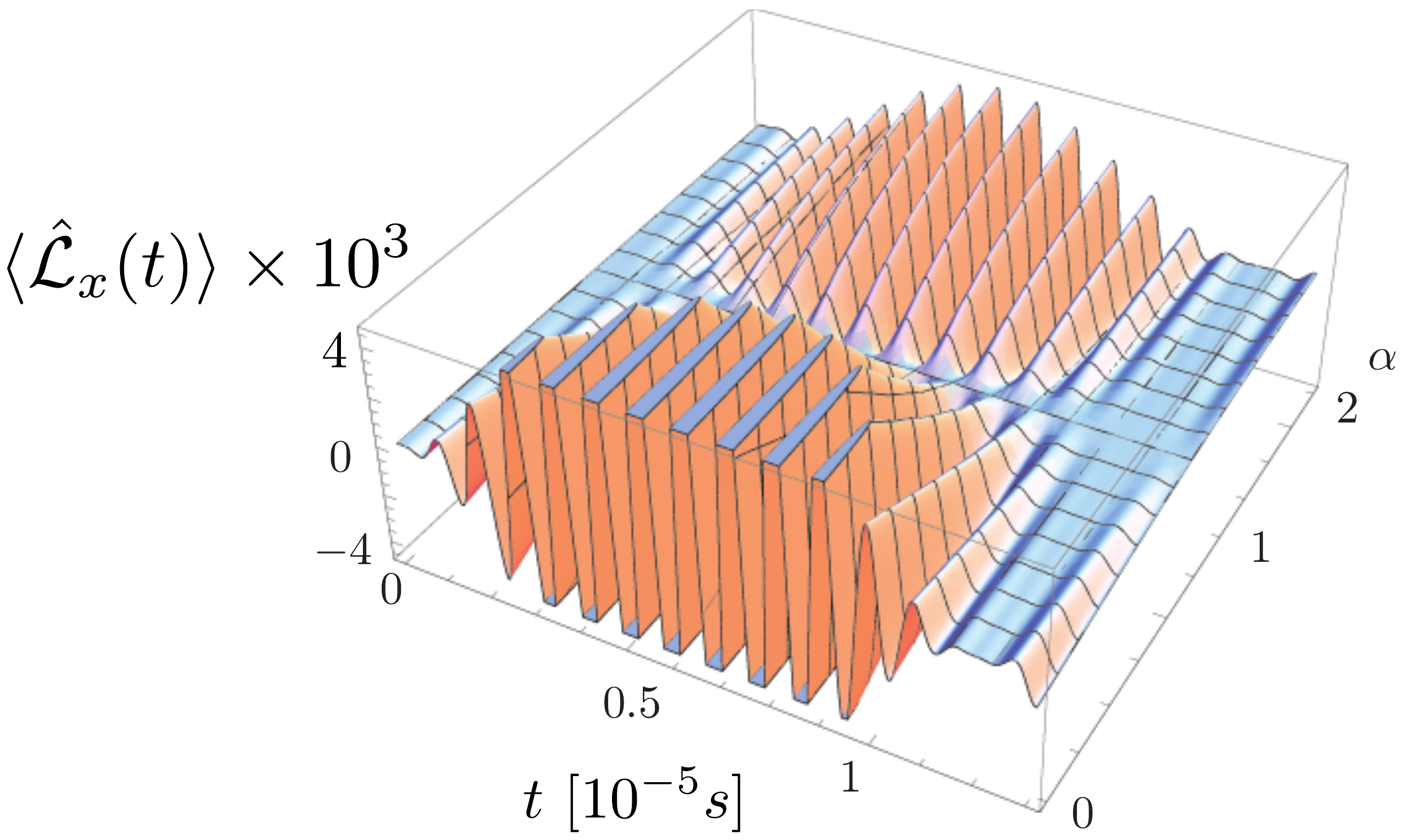}
  \caption{{\bf (a)} Mean value of $\opt{\mathcal{L}}{t}{x}{}$
    for a cantilever in the initial state as given by
    Eq.~(\ref{eq:initial}) with $C_0{=}C_1/\alpha{=}1/\sqrt{1+\alpha^2}$
    and $C_n=0$ otherwise.  The BEC consists of $N{=}10^3$
    ${}^{87}\text{Rb}$ atoms and
    $\langle{\opt{\mathcal{L}}{0}{x,y}{}}\rangle=0, \;
    \langle{\opt{\mathcal{L}}{0}{z}{}}\rangle=100$. 
    {We have used $B_z^0{=}3\times 10^{-6} \mu T$ 
            and $G_c\approx 1.8 \times 10^3 \mu T/\mu$m}. {\bf (b)} Mean value of $\opt{\mathcal{L}}{t}{x}{}$
    for a cantilever in the initial state as given by
    Eq.~(\ref{eq:initial}) with $C_0=C_1=C_2/\alpha=1/\sqrt{2+\alpha^2}$ 
    and $C_n=0$ otherwise. The BEC parameters are the same as in
    Fig.~\ref{fig:coh01}. The inset shows that the change in $|\alpha|$ amounts to a shift of the oscillations [we have taken $=e^{i \pi/6}(0.5, 1, 2)$].}
\label{fig:coh01}
\end{figure*}
As the qualitative conclusions of our analysis do not depend upon the
initial value of the angular momentum component of the spinor, in what
follows we shall concentrate on an illustrative example that allows us
to clearly display our results. We thus consider, without affecting
the generality of our discussions,
$\mean{\opt{\mathcal{L}}{0}{x,y}{}}{}
=0$ {and} $\mean{\opt{\mathcal{L}}{0}{z}{}}{}=100$.  When the
cantilever and the BEC are uncoupled, we should expect
$\mean{\opt{\mathcal{L}}{t}{x}{}}{}$ to oscillate at the Larmor
frequency $\omega_L=g \mu_B B_z^0$ and with an amplitude independent
of $\mean{\opt{\mathcal{L}}{0}{x}{}}{}$.  The BEC-cantilever coupling
introduces a modulation of such oscillations and in the following we
will demonstrate that the analysis of such oscillatory behaviour is
indeed useful to extract information on the state of the cantilever.

We first consider the case of a cantilever initially prepared in a
superposition of a few eigenstates of the free Hamiltonian
$\op{\mathcal{H}}{c}{0}$, as in Eq.~(\ref{eq:initial}).  In
Fig.~\ref{fig:coh01} we show the mean value of
$\opt{\mathcal{L}}{t}{x}{}$ as a function of the coherence between the
states with quantum number $n=0$ and $n=1$, {\it i.e.}~a state having
$C_0=C_1/\alpha=1/\sqrt{1+\alpha^2}$ and $C_n=0$ otherwise.  One can
see a clear modulation of the behaviour of $\langle\hat{\cal
  L}_x(t)\rangle$: a close inspection reveals that the carrier
frequency $\omega_L$ is modulated by the frequency $\omega_{0,1}$. In
reality, the Larmor frequency is renormalized as can be seen by the
expression for $\Omega_q$. However, as we have taken $G_c a_c\ll B_z^0$,
one can safely assume that the carrier frequency is very close to $
\omega_L$. Moreover, the maximum of the function is found at
$C_{0,1}=1/\sqrt{2}$, which maximizes the coherence between the two
states and thus the effect of the modulation.  For symmetry reasons,
the modulation described is not visible if the cantilever is prepared
in a superposition of phonon eigenstates whose quantum numbers are all
of the the same parity (such as a single-mode squeezed state).  In
this case, in fact, the function entering the integral over $q$ in
$A_3$ is antisymmetric, thus making it vanish.  In
Fig.~\ref{fig:coh01} {\bf (b)}, $\langle\opt{\mathcal{L}}{t}{x}{}\rangle$ is
shown for an initial state of the cantilever having
$C_{0,1}=C_2/\alpha=1/\sqrt{2+\alpha^2}$ and $C_n=0$ otherwise.  It is
worth noticing that one can identify two regions of oscillations
separated by the line of nodes at $\alpha=1$ where $C_0=C_1=C_2$. We
can understand this behaviour by studying the amplitudes of oscillation
in three $\alpha$-dependent regions. For $\alpha<1$, the main
modulation frequency is given by $\omega_{0,1}$ and the role of the
third state is to modify the amplitude of the oscillations [see
Fig.~\ref{fig:coh01} {\bf (b)}].  At $\alpha=1$ a destructive interference
takes place and the amplitude drops down.  For $\alpha>1$ the
frequency $\omega_{1,2}$ enters into the evolution of
$\langle\opt{\mathcal{L}}{t}{x}{}\rangle$ (for parity reasons, the
term with frequency $\omega_{0,2}$ has no role) and determines a phase
shift of the oscillation
fringes.
It is interesting to observe that if the initial state of the
cantilever is purely thermal, $\mean{\opt{\mathcal{L}}{t}{x}{}}{}$
does not oscillate: only quantum coherence in the state of the
mechanical system gives rise to oscillatory behaviours and their
presence is well signaled by the pattern followed by the angular
momentum of the spinor-BEC.

\begin{figure*}[t]
{\bf (a)}\hskip3.5cm{\bf (b)}\hskip3.5cm{\bf (c)}\hskip3.5cm{\bf (d)}
\includegraphics[width=0.25\linewidth]{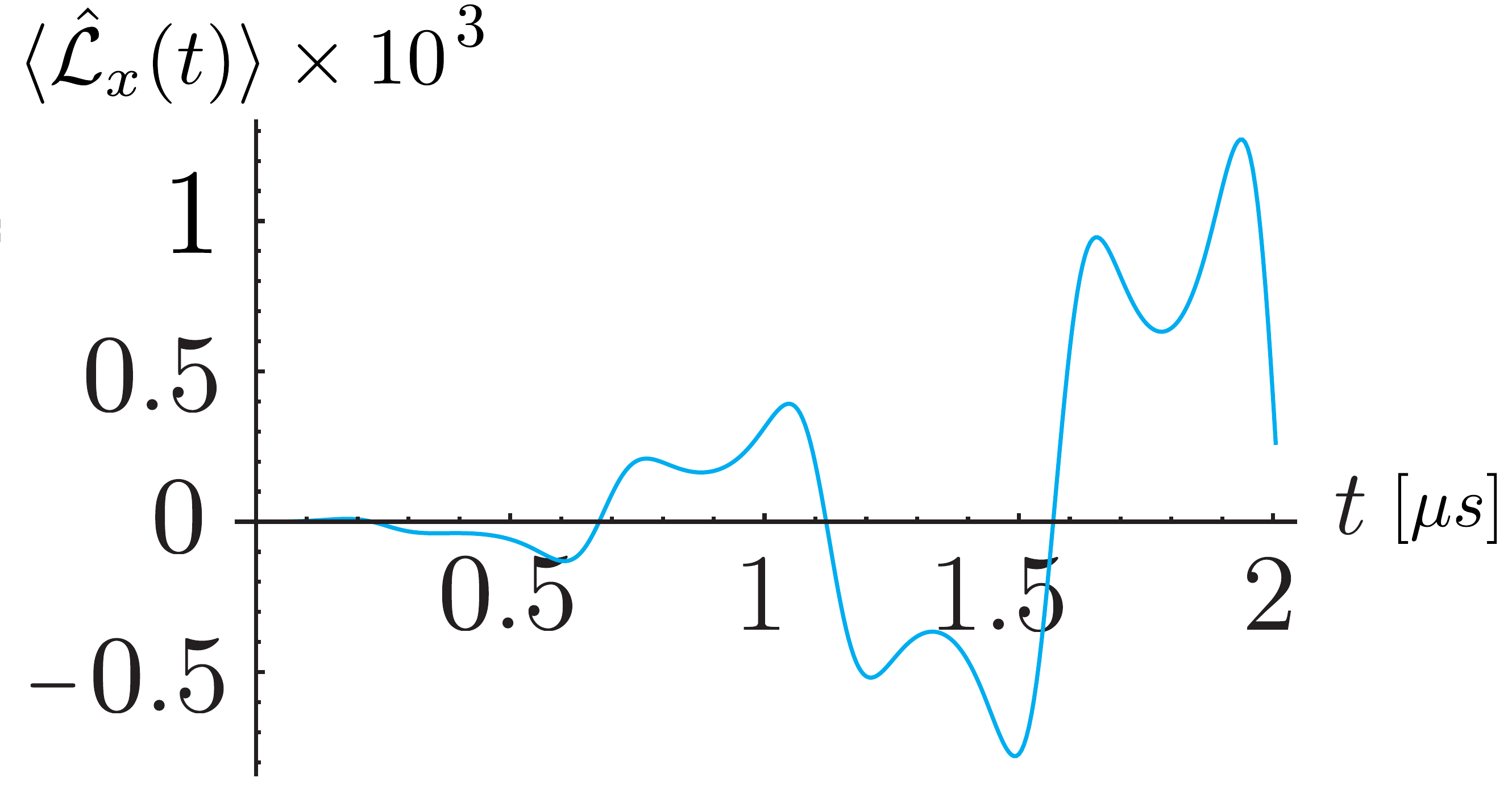}\includegraphics[width=0.25\linewidth]{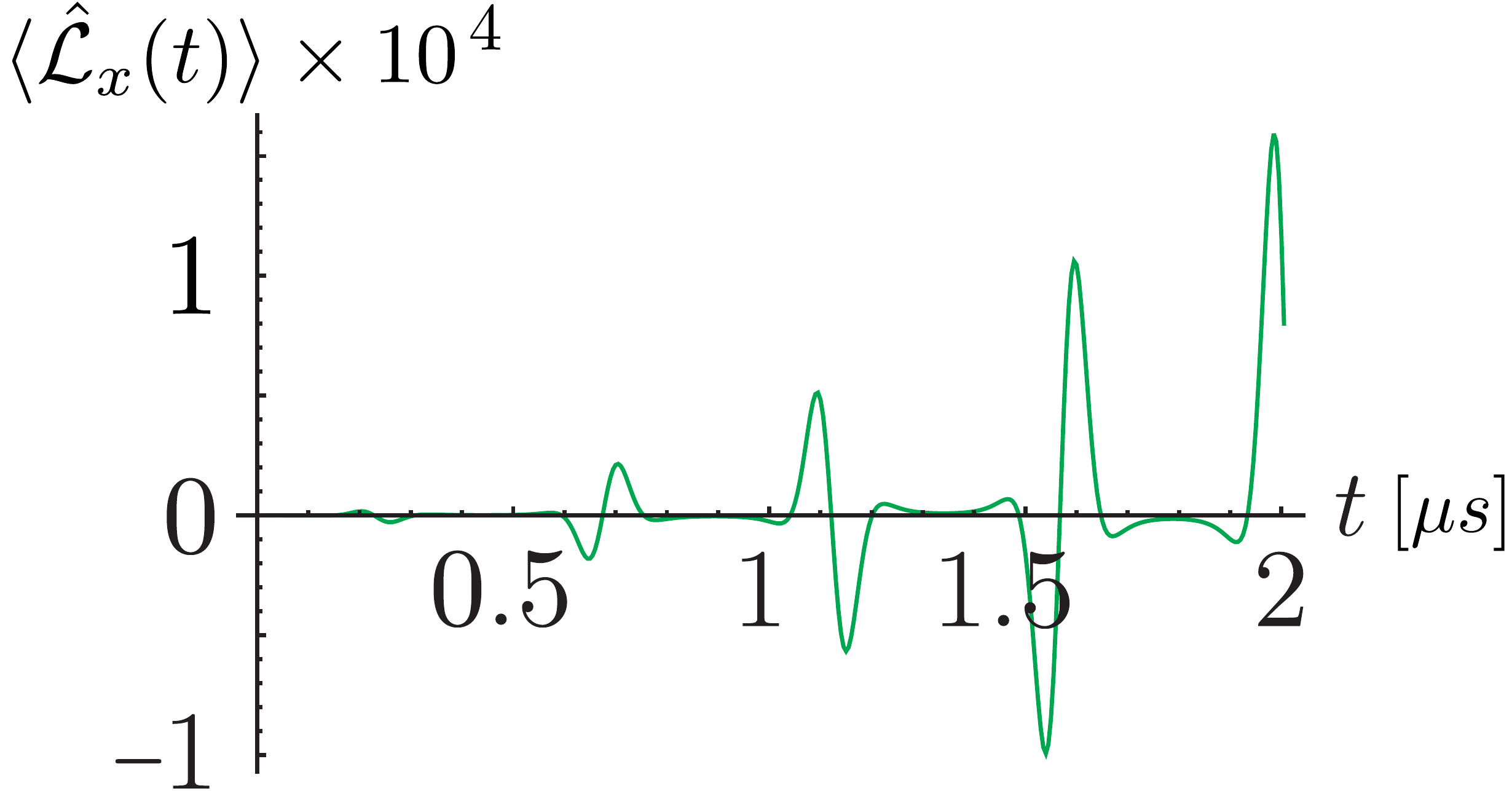}\includegraphics[width=0.25\linewidth]{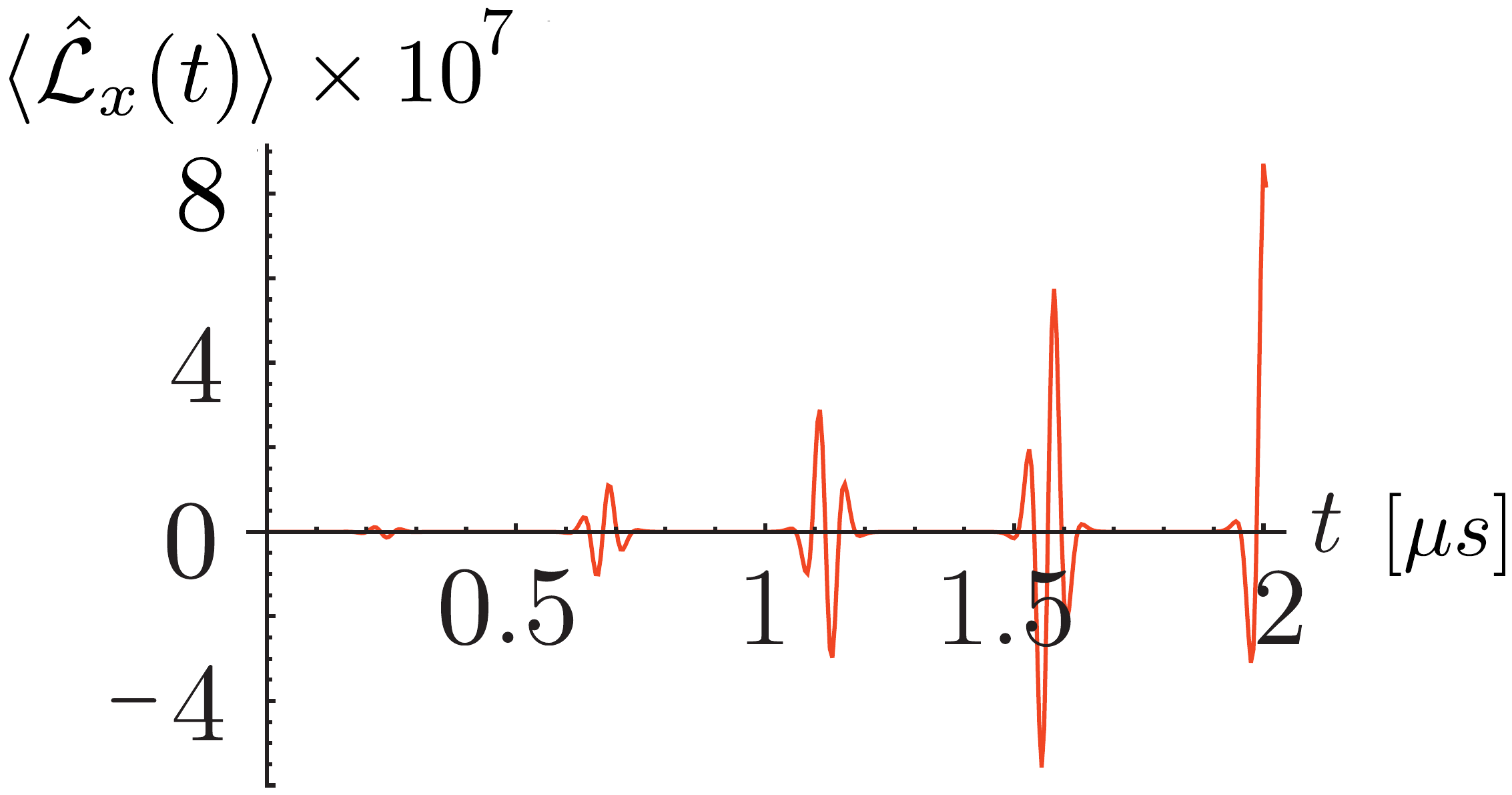}\includegraphics[width=0.25\linewidth]{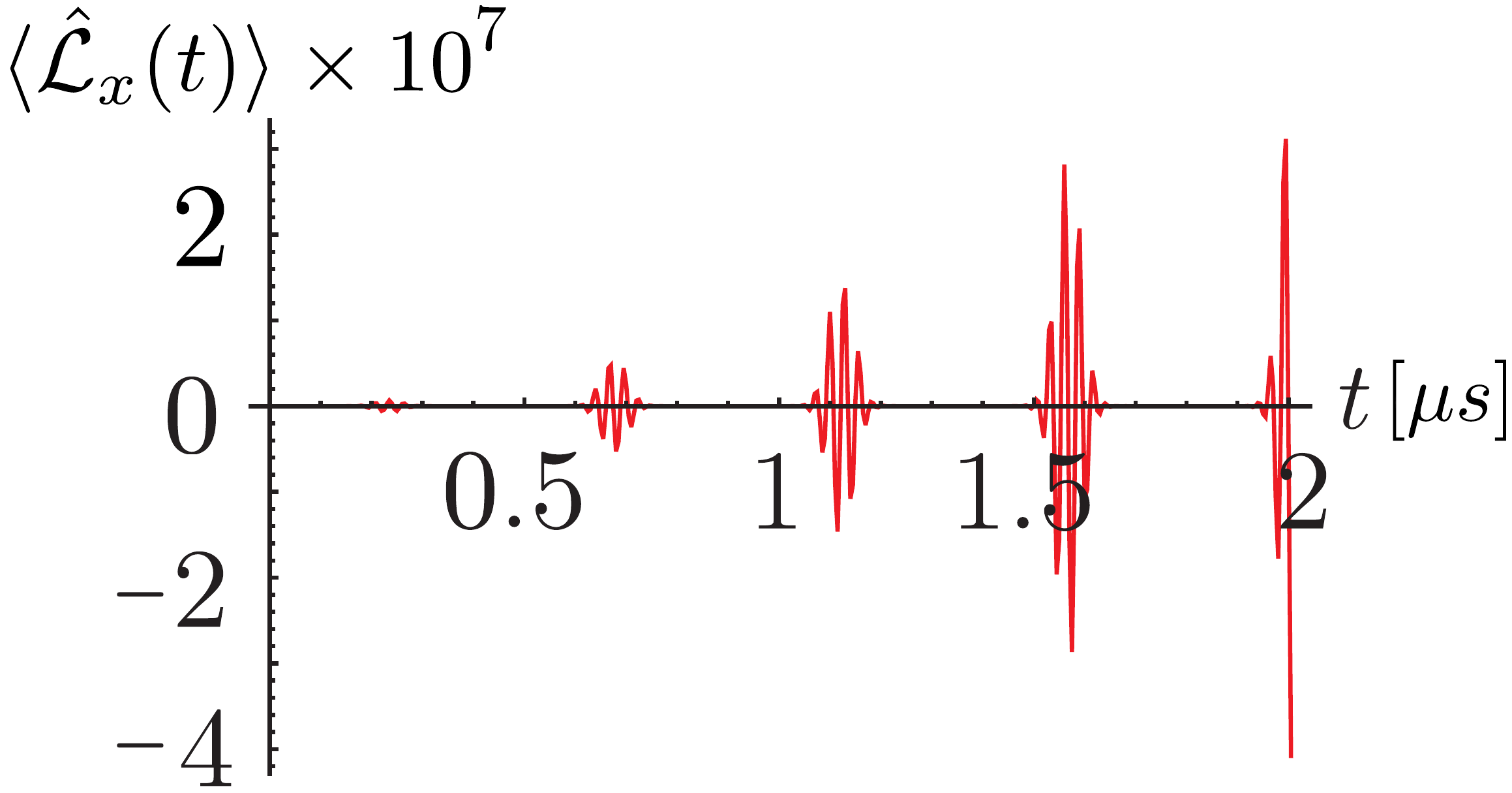}
\caption{Time evolution of $\langle\op{\mathcal{L}}{x}{}(t)\rangle$ for
  a coherent initial state of the cantilever with $|\alpha|^2=1\;
  \;\text{[panel {\bf (a)}]},5\;\text{[panel {\bf (b)}]},15\;\text{panel [{\bf (c)}]},20\;
  \text{[panel {\bf (d)}]}$.  
  }
\label{fig:coherentABCD}
\end{figure*}

Although the examples considered so far have been instrumental in
explaining the connections between the properties of the cantilever
and the dynamics of the spinor's degrees of freedom, they are
unfortunately currently far from being realistic. We will therefore
now consider closer-to-reality example of a pure state that is likely
to be achieved soon. Given the impressive advances in the control and
state-engineering of micro and nano-mechanical systems, we will
consider the cantilever to be prepared in a coherent state with an
average phonon number $n_{ph}$ .
In Fig.~\ref{fig:coherentABCD} we show the time evolution of
$\langle\opt{\mathcal{L}}{t}{x}{}\rangle$ for $|\alpha|^2=1$ [panel \text{{\bf
    (a)}}], $5$ [panel \text{{\bf (b)}}], $15$ [panel \text{{\bf (c)}}],
and $20$ [panel \text{{\bf (d)}}]. One can see that, depending on the
mean number of phonons initially present in the mechanical state, new
frequencies are introduced in the dynamics of the device: the larger
$|\alpha|^2$, the larger the number of frequencies involved due to the
Poissonian nature of the occupation probability distribution of a
coherent state.

\subsection{Detection scheme}
\label{dete}

To read out the information imprinted on the rotor, one can make use
of the Faraday-rotation effect, which allows to measure one component
of the the angular momentum of the BEC with only a negligible back
action on the condensate itself. It is well-known from classical
optics that the linear polarization of an electromagnetic field
propagating across an active medium rotates with respect to the
direction it had when entering the medium itself. This is the essence
of the Faraday-rotation effect, which can be understood by decomposing
the initial polarization in terms of two opposite circularly polarized
components experiencing different refractive
indices~\cite{polartensor}: by going through the medium, the two
components acquire different phases, thus tilting the resulting
polarization.

In the case of an ultra-cold gas, an analogous rotation of the
polarization of a laser field propagating across the BEC is due to the
interaction of light with the atomic spins. If the spins are randomly
oriented the net effect is null, while for spins organized in
clusters, the effect can indeed be measured. It has been shown in
Refs.~\cite{faradayQNM, faradayback} that the back-action on the BEC
induced by this sort of measurements is rather negligible. In recent
experiments non destructive measurements on a single BEC of
${}^{23}$Na atoms have been used to show the dynamical transition
between two different regions of the stability diagram of the
system~\cite{faradayNa}. This method can thus be effectively used to
determine the dynamics of the angular momentum components of the rotor
BEC and thus indirectly witness the presence of coherences in the
state of the cantilever. Moreover, as shown in
  Ref.~\cite{faradayback}, the signal to noise ratio is
  proportional to $\sqrt{\tau_{pd}/\tau_s}$ where $\tau_{pd}$ is the
  characteristic time for the response of the photo-detector and
  $\tau_s$ is the average time between consecutive photon-scattering
  events. In order to be able to detect two distinct events on a time
  scale $\tau$ we thus need $\tau_{pd}<\tau<\tau_s$ to hold. This
  condition states that the number of scattered photons has to be
  small enough during the time $\tau$ over which the dynamics we want
  to resolve occurs. On the other hand the detector ``death time''
  should be smaller than the typical evolution time.  While $\tau_s$
  can be easily tuned by adjusting the experimental working point,
  ultrafast photo-detectors of the latest generation have response
  time $\tau_{pd}$ of a few $ps$. As in our scheme we have
  $\tau\in[10^{-8},10^{-5}]$s, the proposed coherence-probing method
  appears to be within reach.


\section{Single-spin aided optomechanics for quantum non-locality}
\label{Giovannipaper}

Having analysed in detail the case of a collective atomic system embodying the element of hybridisation of an otherwise pure optomechanical device, we now change perspective and explore the potential arising from a single-body matter-like system. In particular, we introduce and study a system comprising an optomechanical cavity containing a single spin-like system, embodied for instance by a three-level atom. We show how the state of the system reveals strong non-classical features such as non-local correlations between the atom and the mirror and negative values of the Wigner function of the mirror, even in presence of dissipative processes and non-zero temperature. We focus on the correlations established between the two systems as well as the non-classical features induced on the state of the mirror. The general aim is precisely to prove how  non-classical behaviours can be induced in massive mesoscopic  systems out of the reach of direct addressability, even by the means of microscopic quantum-inducing ancillae that might be only weakly coupled to the mechanical subsystem.   
The scenario addressed in this Section deals with non-classical features (such as non-local correlations  and negative values of the Wigner function) that are truly mesoscopic  and thus different from more extensively studied nano-scale setups~\cite{armour,rabl,tian,rodrigues}, well-controllable and, although close to experimental capabilities in the fields of optomechanics and light-matter interaction, yet unexplored.


\subsection{The Model}
\label{subsectionOMModel}
\begin{figure*}[t]
{\bf (a)}\hskip4.5cm{\bf (b)}\hskip4.5cm{\bf (c)}
\includegraphics[width=0.6\textwidth]{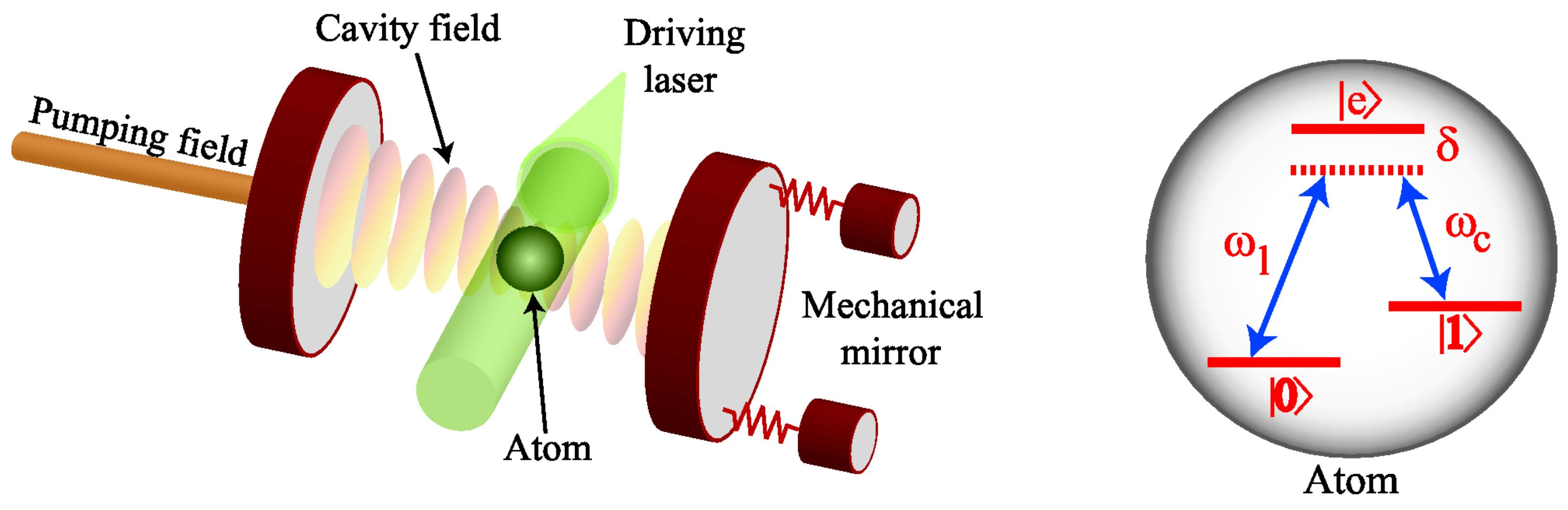}~~~~\includegraphics[width=0.4\textwidth]{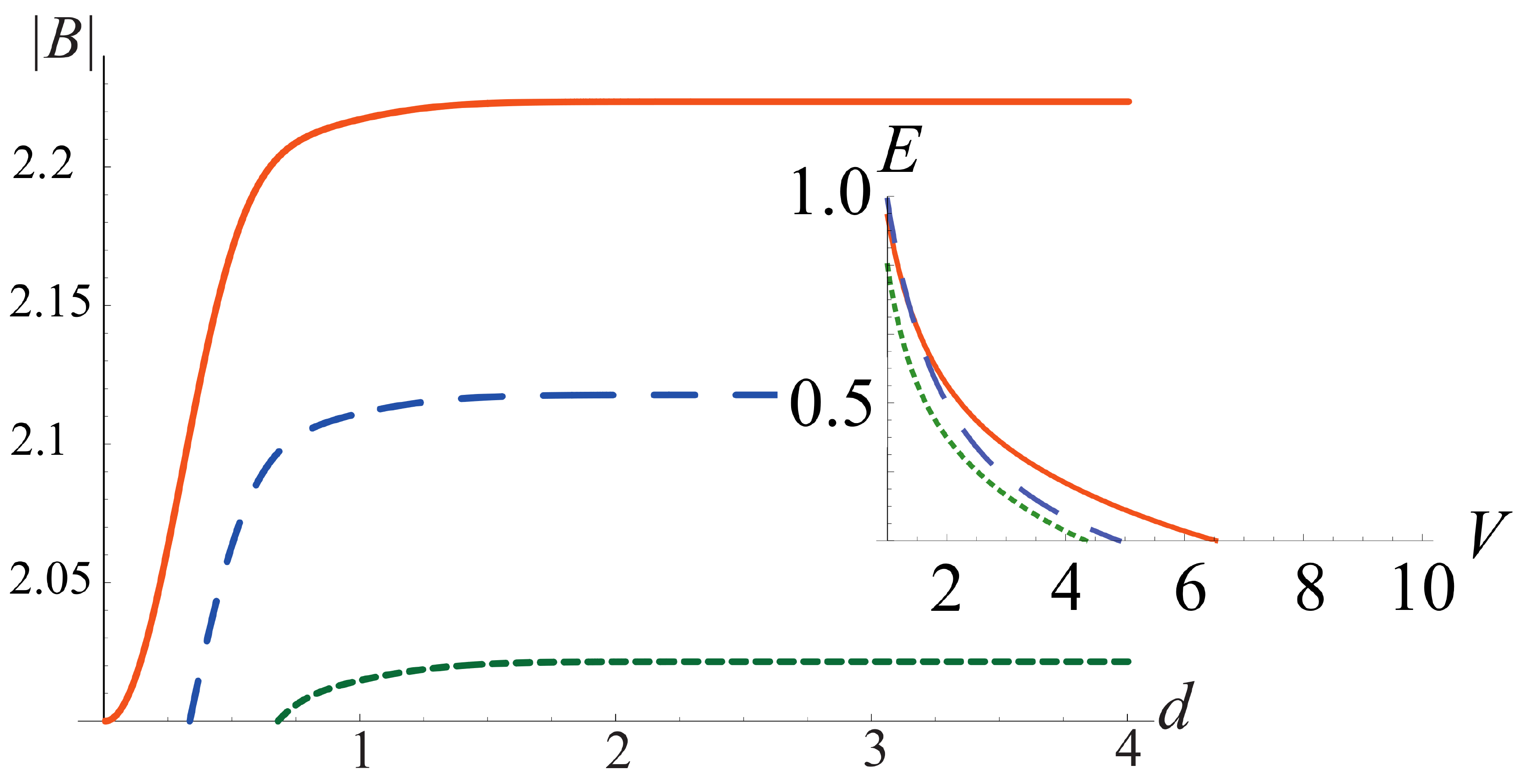}
\caption{{\bf (a)} Scheme of the system. {\bf (b)} Energy levels of the atom driven by an off-resonant two-photon Raman transition. {\bf (c)} Maximum violation of the Bell-CHSH inequality against the displacement $d$. From top to bottom, the curves correspond to $V=1,3,5$ with $\Upsilon t=2d$ and $\theta_1\simeq{3}\pi/2$ and are optimized with respect to $\theta$. The inset shows, from top to bottom, the logarithmic negativity $E$ against $V$ for projected states with $p=0,1$ and $2$, for $d=2$.} \label{CHSHoneMirr}
\end{figure*}

The system that we consider involves a three-level atom in a $\Lambda$ configuration, coupled to a single-mode optical cavity pumped by a laser field at frequency $\omega_L$ and with a movable mirror. The atom is driven by a second external field at frequency $\omega_i$ that enters the cavity radially [see Fig.~\ref{CHSHoneMirr} {\bf (a)} and {\bf (b)}]. We label $\{\ket{0},\ket{1}\}_a$ the states belonging to the fundamental atomic doublet and $\ket{e}_a$ the excited state. The atomic transition $|0\rangle_a\leftrightarrow|e\rangle_a$  is guided, at rate $\Omega,$  by the external  field at frequency $\omega_i$. On the other hand, the transition $|1\rangle_a\leftrightarrow|e\rangle_a$ is coupled to the cavity field at frequency  $\omega_C$ with coupling constant $g$. We call $\delta$ the detuning between each transition and the respective driving field, while $\Delta_{cp}=\omega_C-\omega_p$ is the cavity-pump detuning. The movable mirror with frequency $\omega_m$ is coupled to the cavity field through radiation-pressure. We assume large single-photon Raman detuning and negligible decay rate $\gamma_e$ from the atomic excited state, so that $\delta\gg{\Omega,g}\gg\gamma_e$ and an off-resonant two-photon Raman transition is realized. Moving to an interaction picture defined by the operator $\omega_p\hat{a}^\dag\hat{a}+\omega_i\ket{e}\!\bra{e}_a+\omega_{10}\ket{1}\!\bra{1}_a,$ the Hamiltonian of the overall system reads 
\begin{equation}
\hat{\cal H}_{\rm sys}=\hat{\cal H}_{a}+\hat{\cal H}_{R}+\hat{\cal H}_{m}+\hat{\cal H}_{c}+\hat{\cal H}_{mc}+\hat{\cal H}_{cp}
\label{HTot} 
\end{equation}
with
\begin{equation}
\begin{aligned}\label{HTot}
&\hat{\cal H}_{a}\!=\!\hbar{\delta}\ket{e}\bra{e}_a,\hat{\cal H}_M=\hbar\omega_m\hat{b}^\dag\hat{b},\hat{\cal H}_C=-\hbar\Delta\hat{a}^\dag\hat{a},\\
&\hat{\cal H}_{R}\!=\! \hbar\Omega\ket{e}\!\bra{0}_a
+\hbar ge^{i\Delta_{cp}{t}}\hat{a}^\dag\ket{1}\!\bra{e}_a+h.c.,\\
& \hat{\cal H}_{MC}=-\hbar\chi\hat{a}^\dag\hat{a}(\hat{b}+\hat{b}^\dag).
\end{aligned}
\end{equation}
Here, $\hat{\cal H}_{a}$ is the atomic energy, $\hat{\cal H}_{R}$ is the Raman coupling, $\hat{\cal H}_M$ ($\hat{\cal H}_C$) is the mirror (cavity) free Hamiltonian. Finally, $\hat{\cal H}_{MC}$ and $\hat{\cal H}_{cp}$ are the usual radiation-pressure and cavity-pump terms used throughout this review. 
The pumping field ensures that a few photons are always present in the cavity, allowing a mediated interaction between the atom and the mirror. On the other hand, the purpose of the external field with rate $\Omega$ is to trigger the passages between the excited level $|e\rangle_a$ and the ground level $|0\rangle_a.$

If we further assume suitable working conditions, both the atomic excited state and the cavity field are virtually populated and they can be eliminated from the dynamics of the system. We start from the elimination of the excited state of the atom $|e\rangle_a$ and the electromagnetic field inside the cavity. In order to do so, we assume $\Delta_{cp}\gg\Omega, g$ and $\delta\gg\Omega, g$.  We notice that the  only  terms in the Hamiltonian involving the atomic degrees of freedom are $\hat{\cal H}_a$ and  $\hat{\cal H}_R$. Hence, we perform first the adiabatic elimination of the exited level $|e\rangle_a$ of the atom. The Hamiltonian $\hat{\cal H}_a + \hat{\cal H}_R$ can be formally written as a $3\times3$ matrix with respect to the basis $\{|0\rangle,|1\rangle,|e\rangle\}_a$
	 \begin{equation}
\hat{\cal H}_a + \hat{\cal H}_R=\hbar \begin{pmatrix} \label{atomH}
0 & 0& \Omega\\ 
0&0& g e^{i \Delta_{cp} t} \hat{a}^\dag \\ 
\Omega & g e^{-i \Delta_{cp} t} \hat{a} & \delta \end{pmatrix}.
	 \end{equation}
By writing a generic state of the atom as $|\lambda\rangle_a = c_0 |0\rangle_a + c_1 |1\rangle_a + c_e |e\rangle_a$ and by setting to zero $\dot{c}_e$ in the corresponding Schr\"odinger equation $i\hbar \partial_t |\lambda\rangle_a = (\hat{\cal H}_a + \hat{\cal H}_R) |\lambda\rangle_a$, we find the effective Hamiltonian 
	\begin{equation}
		\begin{aligned}
\frac{\hat{\cal H}_{eff}}{\hbar}= &-\frac{\Omega^2}{\delta} |0\rangle \langle 0| - \frac{\Omega g e^{-i \Delta_{cp} t} }{\delta} \hat{a} |0\rangle \langle 1|\\
&-  \frac{\Omega g e^{i \Delta_{cp} t}}{\delta}  \hat{a}^\dag |1\rangle \langle 0| - \frac{g^2}{\delta} \hat{a}^\dag \hat{a} |1\rangle\langle 1|.
		\end{aligned}
	\end{equation}
After the adiabatic elimination, we replace the terms $\hat{\cal H}_a +\hat {\cal H}_R$ in Eq. (\ref{HTot}) with the expression above and the total Hamiltonian of the system reads now $\hat{\cal H}_{\rm sys} = \hat{\cal H}_{eff}+ \hat{\cal H}_C + \hat{\cal H}_M +  \hat{\cal H}_{MC} + \hat{\cal H}_{cp}$.

The next step is the elimination of the cavity field operators $\hat{a}$ and $\hat{a}^\dag$. In order to do so,  we consider the equations $\dot{\hat{a}} = -\frac{i}{\hbar}[\hat{\cal H}_{\rm sys},\hat{a}]$ and $\dot{\hat{a}}^\dag = -\frac{i}{\hbar}[\hat{\cal H}_{\rm sys},\hat{a}^\dag]$. 
We find that
	\begin{equation}
		\begin{aligned}
\frac{1}{\hbar}{[}\hat{\cal H}_{\rm sys}, \hat{a}] &= - \hat{a} \big[\Delta_{cp}  + \chi(\hat{b}^\dag + \hat{b})- \frac{g^2}{\delta}  |1\rangle\langle 1|_a  \big]\\
& + \frac{\Omega g }{\delta} e^{i \Delta_{cp} t} |1\rangle \langle 0|_a.
		\end{aligned}
	\end{equation} 
Taking $\Delta_{cp} \gg \chi, g^2/\delta$ we can set the derivative to zero and find
	\begin{equation}
\hat{a} = \frac{\Omega g}{\delta \Delta_{cp}} e^{i \Delta_{cp} t} |1\rangle \langle 0|,~~\hat{a}^\dag = \frac{\Omega g}{\delta \Delta_{cp}} e^{-i \Delta_{cp} t} |0\rangle \langle 1|.
	\end{equation}
By substituting these equation in the expression for $\hat{\cal H}_{MC}$ we find the effective atom-mirror interaction 
\begin{equation}
\hat{\cal H}_{eff} = \hbar\Upsilon\ket{0}\bra{0}_a(\hat{b}^\dag+\hat{b})
\label{HEffOneMirr}
\end{equation}
with $\Upsilon={\chi{g}^2\Omega^2}/{\delta^2\Delta^2_{cp}}.$ The form of the effective coupling rate $\Upsilon$ shows that all the considered coupling mechanisms are necessary in order to achieve the atom-mirror coupling. Through the two-photon Raman transition, the virtual quanta resulting from the atom-cavity field interaction are transferred (by the bus embodied by the cavity field) to the mechanical system. As a consequence, the state of the latter experiences a displacement (in phase space) conditioned on the state of the effective two-level atomic system resulting from the elimination of the excited state. $\hat{\cal H}_{\rm eff}$ involves the position quadrature operator $\hat{q}\propto\hat{b}+\hat{b}^\dag$ of the movable mirror. It is worth noticing that, if the cavity is driven by a bichromatic pump with frequencies $\omega_{p}$ and $\omega_{p}+\omega_m$ and a relative phase $\phi$, the effective coupling between the atom and the movable mirror can be made {\it flexible} in the sense that $\hat{q}$ is replaced by $\hat{b}e^{i\phi}+\hat{b}^\dag{e}^{-i\phi}$, making possible the displacement in any direction of the phase space of the movable mirror~\cite{cam,noiGP,phase,Leibfried}. It is important to stress the underlying assumption of tight confinement of the atom within the cavity, which allows us to neglect the effects of micro-motion in our analysis. In principle, micro-motion would imply the incorporation of an additional degree of freedom (the motion of the atom) that could embody a decoherence channel for the electronic atomic states, and thus an important physical mechanism to consider. However, a detailed analysis of the effects of micro-motion goes beyond the scopes of this review.

\subsection{Atom-Mirror Entanglement}
\label{subsecOMEnt}
We now focus on the quantification of  microscopic-macroscopic correlations between the atom and the mirror. First, we assume that the initial state of the movable mirror is a coherent state $\ket{\alpha}_M$ with amplitude $\alpha\in\mathbb{C},$ while the atom is assumed intially in $\ket{+}_a=(\ket{0}+\ket{1})_{a}/\sqrt{2}$. Under the action of the effective  Hamiltonian in Eq. (\ref{HEffOneMirr}), the initial state evolves into $\ket{\psi(t)}=\hat{\cal U}_t\ket{+,\alpha}_{aM}$, where
\begin{equation}
\ket{\psi(t)}=\frac{1}{\sqrt{2}}(\ket{1,\alpha}+e^{-i\varPhi(t)}\ket{0,\alpha-i\Upsilon t e^{-i\phi}})_{aM}
\label{catstate}
\end{equation}
with $\varPhi(t)=\Upsilon t\text{Re}[\alpha{e}^{i\phi}]$ and 
\begin{equation}
\hat{\cal U}_t\equiv{e}^{-i\hat{\cal H}_{eff} t}=\ket{1}\!\bra{1}_a\otimes\openone_M+\ket{0}\!\bra{0}_a\otimes \hat{D}_M(-i\Upsilon t e^{i\phi}),
\end{equation}
where $\hat{D}_M(\zeta)$ 
is the single-mode displacement operator already introduced in Sec.~\ref{Sottraggopaper}. Eq.~(\ref{catstate}) is, in general, an entangled state of a microscopic and a mesoscopic system: its Von Neumann entropy depends on the value of $\Upsilon t$ only. Intuitively, the larger the phase-space distance between $\ket{\alpha}_a$ and $\ket{\alpha-i\Upsilon t}_a$, the closer the evolved state to a balanced superposition of bipartite orthogonal states, thus maximizing the entanglement. To give a figure of merit, for $\Upsilon t=0.82$  the entropy is~$\sim0.8,$ while for $\Upsilon t>1.7$ the entropy is  $>0.996$. Interestingly, the kind of control over the mirror state reminds of the ``quantum switch'' protocol for microwave cavities~\cite{davidovicharoche}, although here it is achieved over a truly mesoscopic device.

Although impressive progresses have recently been accomplished in active and passive cooling of micro- and nano-mechanical oscillators~\cite{chan}, it is realistic to expect the  mirror to be affected by thermal randomness due to its exposure to the driving field and/or to a phononic background at temperature $T$.  Exploiting the handiness of Eq.~(\ref{catstate}), we write the initial state of the mirror at thermal equilibrium (temperature $T$) and displaced by $d$ (due to the external pump) as 
\begin{equation}
\varrho_{M}^{\rm th}=\int{d}^2\alpha{\cal P}(\alpha,V)\ket{\alpha}\!\bra{\alpha}_M
\label{thermalstate}
\end{equation}
with
${\cal P}(\alpha,V)=\frac{{2}e^{-\frac{2|\alpha-d|^2}{V-1}}}{{\pi(V-1)}}$, $V=\coth (\omega_m/2K_BT)$. Under $\hat{\cal U}_t$, the state $\ket{+}\bra{+}_a\otimes\varrho_{M}^{\rm th}$ evolves into
\begin{equation}
\label{final1}
\hat{\cal U}_t(\ket{+}\bra{+}_a\otimes\varrho_{M}^{\rm th}\,)\hat{\cal U}^\dag_t\!=\!\int\!{d}^2\alpha{\cal P}(\alpha,V)\ket{\psi(t)}\!\bra{\psi(t)},
\end{equation}
which reduces to the pure case of Eq.~(\ref{catstate}) for $T=0$. We proceed to show that the coupling mechanism described above is characterized by special features, at the core of current experimental and theoretical interests~\cite{demartini,jeong1,jeongralph}. Let us consider the case of $\phi=\pi/2$, $V=1$ (i.e. $T=0$) and $\alpha\in\mathbb{R}$, which gives $\ket{\psi(\tau)}(\ket{1,\alpha}+\ket{0,\alpha-\Upsilon t})_{aM}/\sqrt2$. This entangled state represents a mesoscopic instance of a pure Schr\"odinger-cat state. Interestingly, it has been discussed that a faithful implementation of the Schr\"o{d}inger's cat paradox would use a mesoscopic subsystem initially prepared in a thermal state, rather than a pure one~\cite{demartini,jeong1,jeongralph}. The state in Eq.~(\ref{final1}) is a significant example of such case. Unravelling the entanglement properties of this state is demanding due to the difficulty of finding an analytical tool for its undisputed revelation. In order to gain insight, here we propose to follow two paths. 

The first relies on the nonlocality properties of this class of states, induced by the strong entanglement between the subsystems. Following Ref.~\cite{banaszek,banwod}, the microscopic part is projected along the direction ${\bm n}=(\sin\theta,0,\cos\theta)$ of the single-qubit Bloch sphere while the mesoscopic one is probed by using the displaced parity observable $\hat{\Pi}(\beta)=\hat{D}^\dag(\beta)(-1)^{\hat{b}^\dag\hat{b}}\hat{D}(\beta)$, with $\beta=\beta_r+i\beta_i$. This approach has been used recently to address the micro-macro non-locality in an all-optical setting~\cite{spagnolo2011}. The correlation function for a joint measurement is thus 
\begin{equation}
{\cal C}(\beta,\theta)=\int{d}^2\alpha{\cal P}(\alpha,V)\bra{\psi(t)}({\bm n}\cdot\hat{{\bm \sigma}})\otimes\hat{\Pi}(\beta)\ket{\psi (t)}
\end{equation}
 and a Bell-Clauser-Horne-Shimony-Holt (Bell-CHSH) inequality is formulated as $|{\cal C}(0,\theta_1)+{\cal C}(0,\theta)+{\cal C}(\beta,\theta_1)-{\cal C}(\beta,\theta)|\le{2}$~\cite{chsh}. Any state satisfying this constraint can be described by a local-realistic theory. Let us first discuss the pure case of $V=1$, which gives  
\begin{equation} 
\begin{aligned}
{\cal C}(\beta,\theta)&=\frac{1}{2}e^{-2(d^2+\Upsilon^2 t^2+|\beta|^2+\beta_r\Upsilon t-2\beta_rd)}\\ 
&\times[\cos\theta(e^{4d\Upsilon t-2\Upsilon t \beta_r}\!-\!e^{2\Upsilon^2 t^2+2\Upsilon t \beta_r})\!\\ &+\!2e^{\Upsilon t(2{d}+\frac{3}{2}\Upsilon t)}\cos({2\Upsilon t \beta_i})\sin\theta].
\end{aligned}
\end{equation}
At $\Upsilon t=0$, the microscopic and mesoscopic subsystems are uncorrelated and ${\cal C}(\beta,\theta)$ can indeed be factorized. For a set value of $d$ and a non-zero value of $\Upsilon t$, we observe violation of the Bell-CHSH inequality as illustrated in Fig.~\ref{CHSHoneMirr}. Moreover,  there is a range of values of $\theta$ ($\sim\pi/2$) where, for $d\neq{0}$, the local-realistic bound is violated, symmetrically with respect to $d=0$. When the thermal character of the mesoscopic part is considered, the expression for the correlation function becomes cumbersome and we omit it. However, {\it the strong entanglement between microscopic and mesoscopic subsystems allows violation of Bell-CHSH inequality also in the mixed-state case}: the dotted curve in Fig.~\ref{CHSHoneMirr} corresponds to $V\simeq{5}$.  Beyond this value, the inequality is no longer violated. 

The second path we follow uses the technique put forward in Ref.~\cite{bosekim} and later reprised by Ferreira {\it et al.} in Ref.~\cite{prima3}. In this approach, Eq.~(\ref{final1}) is projected onto a bidimensional subspace spanned by the microscopic states $\{\ket{0},\ket{1}\}_a$ and the phononic ones $\{\ket{p},\ket{p+1}\}_M$ ($p\in\mathbb{Z}$). The entanglement within Eq.~(\ref{final1}) cannot be increased by this projection, which is just a local operation. Thus, by quantifying the entanglement for fixed $p$, we provide a lower bound to the 
overall quantum correlations in the state of the system. As a measure for entanglement in each $2\times{2}$ subspace we use the {logarithmic negativity}, formulated for spin-like systems~\cite{peres,horo3,plenio05}. An example of the results achieved with this method is given in the inset of Fig.~\ref{CHSHoneMirr}, where we show the case of $d=2$ and $p=0,1,2$. Entanglement is found in each subspace with fixed $p$, up to values of $V\sim{5}$, strengthening our findings about the resilience of non-classical correlations set by the coupling being studied.

\subsection{Non-classicality of the mirror}
We now consider the effects of the microscopic-mesoscopic interaction over the state of the movable mirror. This is a hot topic in the current research of opto and electro-mechanical systems. The grounding of opto/electro-mechanical devices as potential candidates for quantum information processing requires the design of protocols for the preparation of non-classical states of massive mechanical systems. Various attempts have been performed in this direction, mainly at the nano-scale level, where a cantilever can be capacitively coupled to a superconducting two-level system~\cite{armour,rabl,tian,rodrigues}. 
\begin{figure*}[t] 
{\bf (a)}\hskip3.5cm{\bf (b)}\hskip3.5cm{\bf (c)}\hskip3.5cm{\bf (d)}
\scalebox{0.33}{\includegraphics{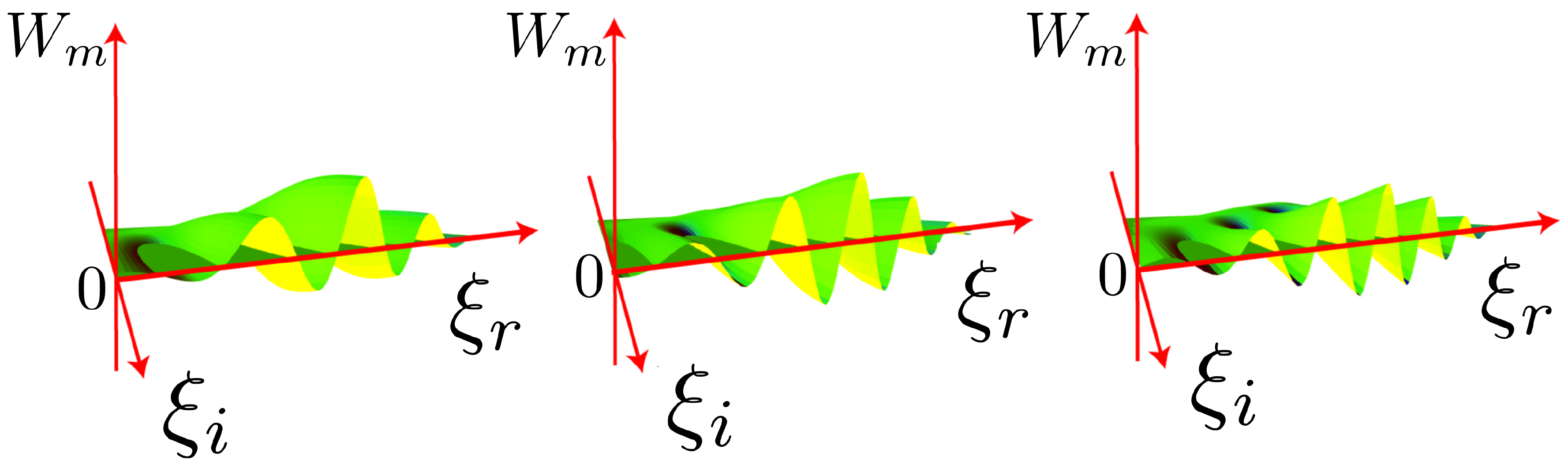}}\includegraphics[width=.49\columnwidth]{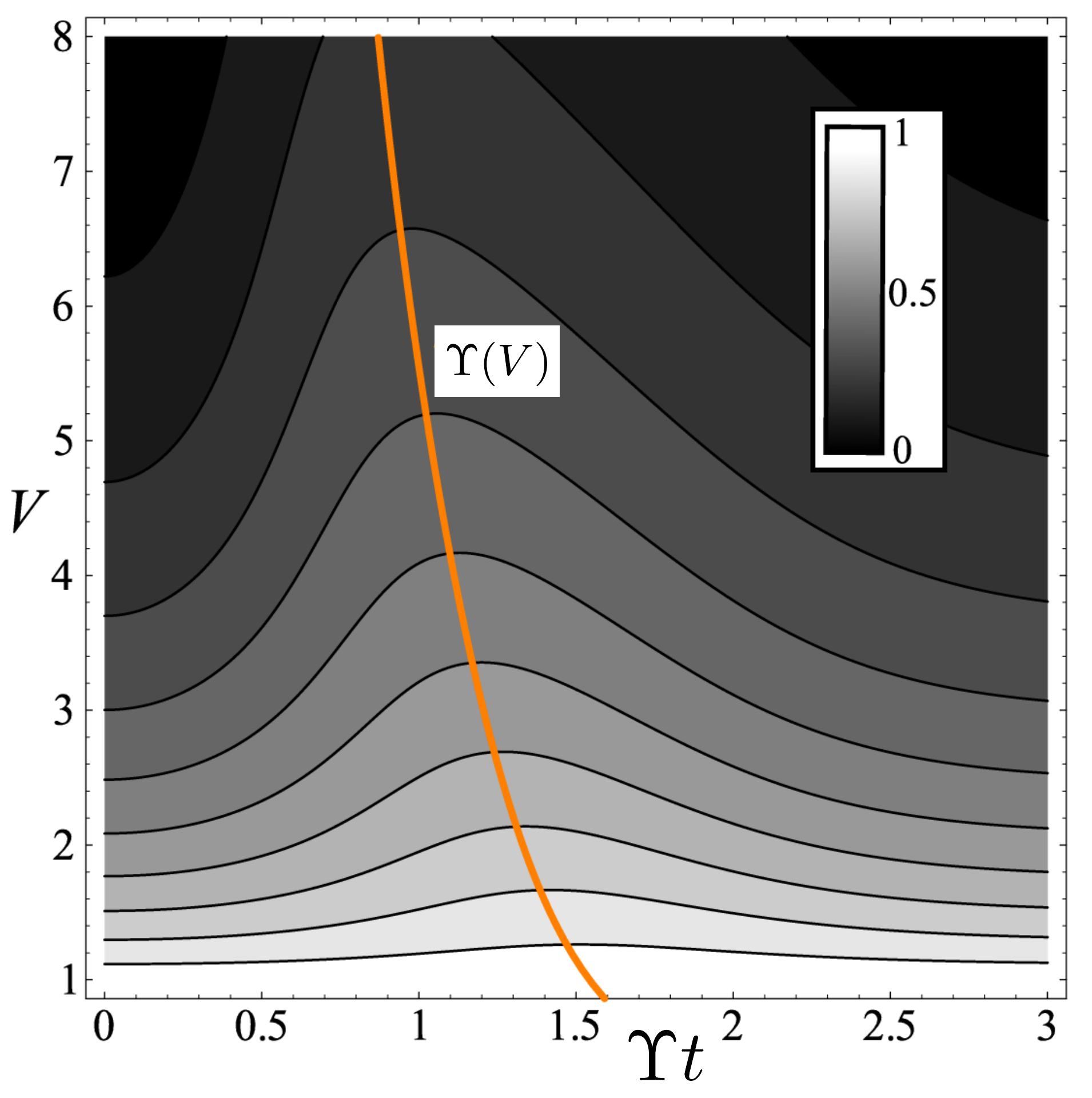}
\caption{{\bf (a)}-{\bf (c)} Wigner function of the conditional mirror state against $\xi_r=\text{Re}(\xi)$ and $\xi_i=\text{Im}(\xi)$, for $V=3$ and $d=0$. Panels {\bf (a)}, {\bf (b)}, {\bf (c)} correspond to $\Upsilon t=2,3,4$ respectively. {\bf (d)} Density plot of fidelity against $V$ and $\Upsilon$. Darker regions correspond to smaller values of $F_{W}$. {\bf (b)} Wigner function of the mirror under dissipation, for $\gamma\sim{0.1}\Upsilon$ and $V=5$.}
\label{Wignerevolution}
\end{figure*}

Let us consider the case of $\phi=0$. The optomechanical evolution encompassed by $\hat{\cal U}_t$ alone is unable to give rise to any non-classicality in the state of the mirror. This is easy to check simply by tracing out the state of the atom in Eq.~(\ref{catstate}), which would leave us with a statistical mixture of two displaced mirror's states. On the other hand, a conditional process is able to project the coherence of a quantum mechanical superposition and simultaneously get rid of the atomic degree of freedom \cite{KochPRL,RitterNature,SpechtNature,MonteiroNJP,Kiesel,MauroPRA}. In order to illustrate our claim, we consider an initial state of the system having the form $\rho(0)=|\varphi\rangle \langle \varphi |\otimes\rho_M(0)$ where $|\varphi\rangle_a=c_0 |0\rangle_a+c_1 |1\rangle_a$  is a pure state of the atom and $\rho_M(0)$ is an arbitrary state of the mechanical mode. We then project the atomic part of the evolved state $\hat{\cal U}_t \rho(0)\hat{\cal U}_t^\dag$ onto $|\varphi\rangle \langle \varphi |$, thus post-selecting the mechanical state $\rho_M(t) =  \langle \varphi | \hat{\cal U}_t |\varphi \rangle \rho_M(0)\langle \varphi | \hat{\cal U}_t^\dag |\varphi\rangle$. Therefore, the state of the mirror undergoes an effective evolution driven by the operator 
	\begin{equation}
\langle \varphi | \hat{\cal U}_t | \varphi \rangle = |c_1|^2 \hat\openone + |c_0|^2\hat{D}(-i \Upsilon t). 
\label{pUpOneMirr}
	\end{equation}
In the remainder of this paper, we consider again the case where $|{\varphi}\rangle_a=|{+}\rangle_a\equiv(|0\rangle + |1\rangle)_a/\sqrt{2}$, which optimizes the performance of our scheme  terms of the degree of non-classicality enforced in the mechanical subsystem. For an initial coherent state of the mirror, i.e. $\rho_M(0)=|\alpha\rangle\langle\alpha|_M,$ applying the conditional time evolution operator in Eq.~(\ref{pUpOneMirr}) leads to $\ket{\mu_+}_M={\cal N}_+(\ket{\alpha}+{e}^{-i\varPhi(t)}\ket{\alpha-i\Upsilon t})_M$, where ${\cal N}_+$ is the normalization factor. Depending on the value of $\Upsilon t$, such states exhibit quantum coherences. Obviously, the thermal convolution inherent in the preparation of mirror's state $\varrho_M$ may {\it blur} them. In what follows we prove that this is not the case for quite a wide range of values of $V$. 

The figure of merit that we use to estimate non-classicality is the negativity in the Wigner function $W_M(\mu)~[\mu=\mu_r+i\mu_i]$ associated with the mirror state resulting from the measurement performed over the atomic part of the system.
Considering an initial thermal state of the mirror and applying the conditional unitary evolution operator given in Eq.~(\ref{pUpOneMirr}), the Wigner function of the mirror after the post-selection process is 
\begin{equation}
\begin{split}
W_M(\mu)=&\frac{2e^{-\frac{2|\mu|^2+2\Upsilon t \mu_i+\Upsilon^2 t^2}{V}}}{(1+e^{-\frac{V\Upsilon^2}{2}})\pi{V}}\\ \times&\left[\cosh\left(\frac{\Upsilon^2 t^2+2\Upsilon t\mu_i}{V}\right)+e^{\frac{\Upsilon^2 t^2}{2V}}\cos(2\Upsilon t\mu_r)\right].
\end{split}
\end{equation}
The behaviour of $W_M(\mu)$ in the phase space is shown in Fig.~\ref{Wignerevolution}, where we clearly see the appearance of regions of negativity, witnessing non-classicality of the corresponding state as induced by our microscopic-to-mesoscopic coupling. Interference fringes are created between two positive Gaussian peaks (not shown in the figure) corresponding to the position, in the phase space, of mutually displaced coherent states. This reminds of the Wigner function of a pure Schr\"odinger cat state although, as we see later, the analogy cannot be pushed further. Remarkably, in contrast with the fragility of the nonlocality properties of the microscopic-mesoscopic state, $W_M(\mu)$ has a negative peak of $-0.01$ up to $V\sim100$, which implies strong thermal nature of the mirror state. For a mechanical system embodying one of the mirrors of a cavity, $\omega_m/2\pi\sim5$MHz is realistic. 
For $V=10$ ($100$), this corresponds to an effective temperature of $1$mK ($10$mK), {\it i.e.} energies $10$ ($100$) times larger than the ground-state energy of the mirror.

It is interesting to compare the mixed state resulting from the thermal convolution to a pure state in Eq.~(\ref{catstate}) (with $\phi=0$). As a measure of the closeness of two states, we use quantum fidelity between a mixed and a pure state written as the overlap between the corresponding Wigner functions $F_W=\pi\int{d}^2\mu{W}_{Pure}(\mu)W_{Mix}(\mu)$, where $W_{Pure}(\mu)$ [$W_{Mix}(\mu)$] is the Wigner function of the pure [mixed] state. $F_W$ is shown in Fig.~\ref{Wignerevolution} {\bf (d)} against $\Upsilon\tau$ and $V$. While the thermal effect reduces the value of the fidelity as $V$ grows, the behaviour of $F_W$ against $\Upsilon t$ is, surprisingly, non-monotonic. At a given $V$, there is always a finite value of $\Upsilon t$ associated with a maximum of $F_W$. Remarkably, the values of $\Upsilon t$ maximizing $F_W$ differ from those at which the Wigner function achieves its most negative value. 

\begin{figure}[b]
\includegraphics[width=.9\columnwidth]{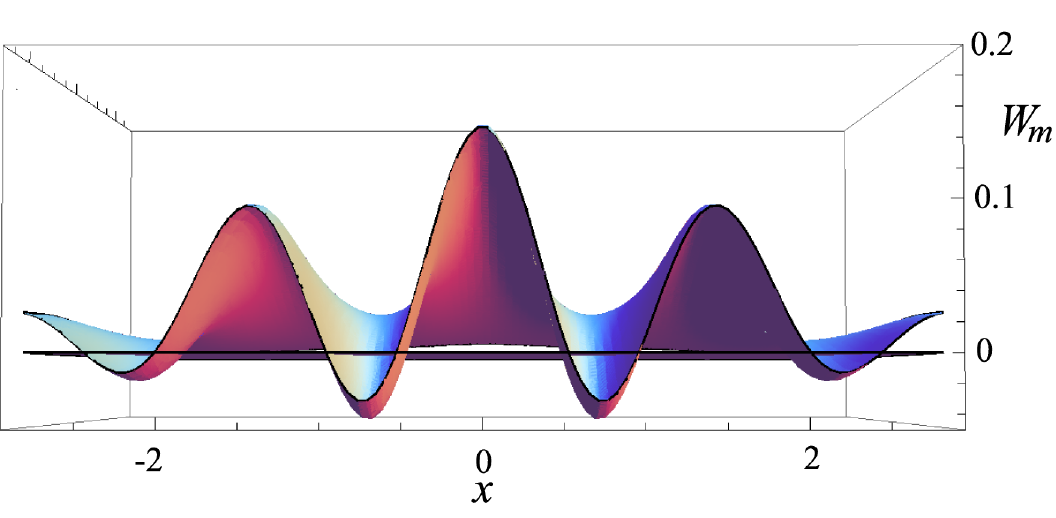}
\caption{Wigner function of the mirror under dissipation, for $\gamma\sim{0.1}\Upsilon$ and $V=5$.}
\label{WdissOneMirr}
\end{figure}
\subsection{Finite temperature dissipative dynamics}
\label{subsectionOMDissipative}
So far, we have assumed a movable mirror of large mechanical quality factor. The progresses recently accomplished in fabrication processes guarantee very small mechanical dissipation rates. However, they are not yet negligible and their effect should be considered in any proposal for quantumness in optomechanical devices. We thus include mechanical losses in our analysis, looking for their effects onto the non-classicality induced in the movable mirror. We concentrate on the finite-temperature dissipative mechanism described by 
\begin{equation}
{\cal L}^{V}(\rho)=\frac{\gamma}{2}\big[(2\hat{b}\rho\hat{b}^\dag-\{\hat{b}^\dag\hat{b},\rho\})+(V-1)(\hat{b}\rho-\rho\hat{b},\hat{b}^\dag)\big], 
\end{equation}
which is the weak-damping limit of the Brownian-motion master equation~\cite{Walls}. The density matrix $\rho$ describes the state of the atom-mirror system. The full master equation, including the unitary part $-i[\hat{\cal H}_{eff},\rho]$, is easily translated into a set of equations of motion for the mirror reduced density matrix obtained by considering the projections onto the relevant atomic states $\rho_{ij}={}_a\langle{i}|\rho|j\rangle_a~(i,j=0,1)$. These can then be recast as Fokker-Planck equations for the Wigner functions $W_{ij}$ of such mirror's state components. These read 
\begin{equation}
\partial_t{\bf W}(x,p, t)={\bf M}{\bf W}(x,p, t)+\tilde{\cal L}_d{\bf W}(x,p, t),
\label{FokkerPlanck}
\end{equation}
where  
\begin{equation}
\begin{split}
&{\bf W}(x,p, t)=\left(\begin{matrix}W_{00}(x,p,t)\\W_{01}(x,p,t)\\W_{10}(x,p,t)\\W_{11}(x,p,t)\end{matrix}\right),\\
&{\bf M}=\sqrt 2\Upsilon\text{Diag}\left[\partial_p,-\frac{ix+\partial_p}{2},\frac{ix+\partial_p}{2},0\right],\\
&\tilde{\cal {L}}_d=\Big[\frac{\gamma}{2}(x\partial_x+p\partial_p)+\frac{\gamma}{4}V(\partial^2_{p^2}+\partial^2_{x^2})+\gamma\Big]\openone,
\end{split}
\end{equation}
where we have introduced the quadrature variables $x=\sqrt 2\text{Re}(\mu), p=\sqrt 2\text{Im}(\mu)$. Each of these equations preserves the Gaussian nature of the corresponding Wigner function's component, whose time-evolved form is taken from the ansatz 
\begin{equation}
W_{ij}(x,p,t)\propto[{\text{det}}({\bf D}_{ij})]^{-1/2}{e^{-\frac{1}{2}{{\bf q}^T_{ij}{\bf D}^{-1}_{ij}{\bf q}_{ij}}+i\Theta_{ij}(t)}}
\label{ansatz}
\end{equation}
with 
\begin{equation}
{\bf q}_{ij}=\left(\begin{matrix}x-\overline{x}_{ij}\\p-\overline{p}_{ij}\end{matrix}\right),~{\bf D}_{ij}=\left(\begin{matrix}\sigma^x_{ij}&\sigma^{xp}_{ij}\\\sigma^{xp}_{ij}&\sigma^{p}_{ij}\end{matrix}\right)
\label{qandcovariance}
\end{equation}
 parameterized by the time-dependent mean values $\overline{x}_{ij},\overline{p}_{ij}$ and variances $\sigma^{x,p,xp}_{ij}$ of the variables $x,p$ and $xp$. We have also introduced the time-dependent phases $\Theta_{ij}$'s which account for the contributions from $\varPhi(t)$ in Eq.~(\ref{catstate}). The solution is readily found to be $\sum_{i,j=0,1}W_{ij}(x,p,t)$ (apart from the normalization factor), which gives back the non-Gaussian character of the mirror's state. The negativity of the Wigner function can be studied at set values of $\gamma$ and $T$ and chosing the time at which the ideal case achieves the most negative value. The results are shown in Fig.~\ref{WdissOneMirr}, where we see that non-classicality is found even for quite a large value of $\gamma/\Upsilon$. Clearly, this results from a subtle trade off between temperature and mechanical quality factor. Although small $\gamma$ and $T$ guarantee non-classicality, such a behaviour is still present at $\gamma/\Upsilon\sim{0.1}$ and for $T$ well above the ground-state one.

\subsection{Feasibility analysis}

In the proposal above, the dissipative part of the dynamics induced by damping processes in the mechanical oscillator plays an important role, and the achievement of the condition $\Upsilon\sim\gamma$ is crucial. A comment about the possibility of reaching this regime is thus in order. For state-of-the-art mechanical systems, typical values of $\gamma$ are in the range of a few Hz, as we have seen in the previous Sections of this review. On the other hand, the effective coupling rate $\Upsilon$ depends directly on the strength of the radiation pressure interaction constant  $\chi = (\omega_c /L) \sqrt{\hbar/2 m \omega_m}.$ Let us consider a mechanical modes having $\omega_m/(2\pi)=300$KHz and $m \sim 50$ng placed in a cavity of ${\cal L} = 10$mm~\cite{Groblacher1,Groblacher2,Groeblacher}: assuming $g^2 \Omega^2 / \delta^2 \Delta^2 \sim 0.1$ and $\omega_c \sim 10^{15}$Hz, we can easily get $\Upsilon \sim 1$Hz. This  value is indeed comparable to $\gamma$, thus demonstrating the achievability of the conditions required by our proposal. 

\section{Conclusions and Outlook}
\label{conclusions}

This review aimed at providing an overview of some of the possibilities for quantum-empowered tasks that can be made possible by  the adoption of a hybrid approach to the dynamics of optically driven quantum mechanical oscillators. We have illustrated specific examples of hybridisation, ranging from the merging of a diluted BEC into an optomechanical cavity to the magnetic interaction between spinor atomic gases and mechanical cantilevers, passing through the use of a single few-level atom as an enforcer of mechanical non-classicality. Each of the instances addressed in this paper aims at addressing an important class of problems in the area of quantum technologies: state preparation, the manipulation and control of the inherently quantum features of an optomechanical device, and the detection of such properties. Needless to say, our analysis is not exhaustive nor definitive: many are the problems that have yet to find an appropriate addressing in the context of hybrid quantum optomechanics, and we have provided a set of suggestions that, we hope, would raise the interest of the community working in related areas and, ultimately, will contribute to the full development of such a promising architecture for quantum devices.

\section*{Acknowledgments} 
We thank Th. Busch, T. Donner, T. Esslinger, A. Farace, R. Fazio, V. Giovannetti, P. Massignan, J. F. McCann, M. G. A. Paris, A. Sanpera, G. Vacanti, V. Vedral, and A. Xuereb for many discussions on the topic of this review article over the past few years. We acknowledge financial support from the Alexander von Humboldt Foundation, MIUR (FIRB 2012 RBFR12NLNA), MIUR (PRIN 2010-2011), the UK EPSRC (EP/G004579/1 and EP/L005026/1), the John Templeton Foundation (grant ID 43467), and the EU Collaborative Project TherMiQ (Grant Agreement 618074).


\begin{thebibliography}{99}

\bibitem{dwave} {\it Quantum quantified}, accessible from {\textcolor{blue}{http://www.economist.com/blogs/babbage/2014/01/quantum-computing}}

\bibitem{insane} http://royalsociety.org/news/2013/chancellor-investment-quantum-technology/

\bibitem{NielsenChuang} M. A. Nielsen, I. L. Chuang, {\it Quantum Computation and Quantum Information} (Cambridge University Press, 2010). 

\bibitem{Wallquist} M. Wallquist, K. Hammerer, P. Rabl, M. Lukin, and P. Zoller, Phys. Sc. T {\bf 137}, 014001 (2009); C. Sias, and M. K\"ohl, in {\it Quantum gas experiments-exploring many-body states}, P. T\"orma and K. Sengstock eds. (Inperial College Press, 2014); A. M. Stephens, J. Huang, K. Nemoto, and W. J. Munro, Phys. Rev. A {\bf 87}, 052333 (2013); P. Treutlein, C. Genes, K. Hammerer, M. Poggio, and P. Rabl, in {\it Cavity Optomechanics}, M. Aspelmeyer, T. Kippenberg, and F. Marquardt eds. (Springer, to appear); P. van Loock, T. D. Ladd, K. Sanaka, F. Yamaguchi, K. Nemoto, W. J. Munro, and Y. Yamamoto, Phys. Rev. Lett. {\bf 96}, 240501 (2006).

\bibitem{reviews}  F. Marquardt and S. M. Girvin, Physics {\bf 2}, 40 (1993); M. Aspelmeyer, S. Gr\"oblacher, K. Hammerer, and N. Kiesel, J. Opt. Soc. Am. B {\bf 27}, A189 (2010). 

\bibitem{AspelmeyerRMP} M. Aspelmeyer, T. J. Kippenberg, and F. Marquardt, arXiv:1303.0733 (2013).

\bibitem{Bassi} A. Bassi, K. Lochan, S. Satin, T. P. Singh, and H. Ulbricht, Rev. Mod. Phys. {\bf 85}, 471 (2013).

\bibitem{arcizetNV} O. Arcizet, V. Jacques, A. Siria, P. Poncharal, P. Vincent, and S. Seidelin, Nature Phys. {\bf 7}, 879 (2011).

\bibitem{tip} P. Treutlein, D. Hunger, S. Camerer, T. W. H\"{a}nsch, and J. Reichel, \prl {\bf 99}, 140403 (2007).

\bibitem{hunger} D. Hunger, S. Camerer, T. W. H\"ansch, D. K\"onig, J. P. Kotthaus, J. Reichel, and P. Treutlein, Phys. Rev. Lett. {\bf 104}, 143002 (2010). 

\bibitem{sillanpaa} J.-M. Pirkkalainen, S. U. Cho, J. Li, G. S. Paraoanu, P. J. Hakonen, and M. A. Sillanp\"a\"a, Nature {\bf 494}, 211 (2013).

\bibitem{grangier} J. Wenger, R. Tualle-Brouri, and P. Grangier, Phys. Rev. Lett. {\bf 92}, 153601 (2004).

\bibitem{kim2} M. S. Kim, J. Phys. B {\bf 41}, 133001 (2008).

\bibitem{Vitali2007}
D. Vitali, S. Gigan, A. Ferreira, H. R. B\"ohm, P. Tombesi, A. Guerreiro, V. Vedral, A. Zeilinger, and M. Aspelmeyer, {Phys. Rev. Lett.} {\bf 98}, 030405 (2007).

\bibitem{mauro2} M. Paternostro, Phys. Rev. Lett. {\bf 106}, 183601 (2011).

\bibitem{logneg}
G. Vidal and R. F. Werner, {Phys. Rev. A} {\bf 65}, 032314 (2002).





\bibitem{dakna} M. Dakna, {\it et al.}, Phys. Rev. A {\bf 55}, 3184 (1997). 


\bibitem{glauber} K. E. Cahill and R. J. Glauber, Phys. Rev. {\bf 177}, 1857 (1969).

\bibitem{kenfack} A. Kenfack and K. Zyczkowski, J. Opt. B {\bf 6}, 396 (2004).

\bibitem{hudson} R. L. Hudson, Rep. Math. Phys. {\bf 6}, 249 (1974).

\bibitem{mariWigner} A. Mari, {\it et al.}, Phys. Rev. Lett. {\bf 106}, 010403 (2011).

\bibitem{commentsq} As the block matrix ${\bm M}$ is always diagonal, squeezing of the state of $M$ occurs along one of the quadratures in phase-space.

\bibitem{MauroPRL} M. Paternostro, D. Vitali, S. Gigan, M. S. Kim, C. Brukner, J. Eisert, and M. Aspelmeyer, {Phys. Rev. Lett.} {\bf 99}, 250401 (2007).

\bibitem{mari}
A. Mari and J. Eisert, Phys. Rev. Lett. {\bf 103}, 213603 (2009); New J. Phys. {\bf 14}, 075014 (2012).

\bibitem{fabre} J. Laurat, {\it et al.}, J. Opt. B {\bf 7}, S577 (2005) achieve uncertainties on the entries of a covariance matrix of a few percents of the nominal values. 

\bibitem{MauroJPB} M. Paternostro, J. Phys. B {\bf 41}, 155503 (2008).

\bibitem{grangier1} A. Ourjoumtsev, {\it et al.}, \prl {\bf 98}, 030502 (2007).

\bibitem{grad} I. S. Gradshteyn and I. M. Ryzhik, {\it Tables of Integrals, Series, and Products} (Academic Press, New York, 1965).




\bibitem{esteve2008} J. Est\`eve, {\it et al.}, {Nature} (London) {\bf 455}, 1216 (2008); J. F. Sherson, {\it et al.}, {\it ibid.} {\bf 443}, 557 (2006);
M. Greiner, {\it et al.}, {\it ibid.} {\bf 415}, 39 (2002); F. Brennecke, {\it et al.}, {\it ibid.} {\bf 450}, 268 (2007); Y. Colombe, {\it et al.}, {\it ibid.} {\bf 450}, 272 (2007); R. J\"ordens, {\it et al.}, {\it ibid.} {\bf 455}, 204 (2008).

\bibitem{Esslinger2008} F. Brennecke, S. Ritter, T. Donner, and T. Esslinger, {Science} {\bf 322}, 235 (2008).

\bibitem{Stamper2008} K. W. Murch, {\it et al.}, {Nature Phys.} {\bf 4}, 561 (2008); T. Botter, {\it et al.}, {Procs. Int. Conf. Atomic Phys. (ICAP) 2008}, arXiv:0810.3841.

\bibitem{StringariPitaevskii}
L. Pitaevskii and S. Stringari, {\it Bose-Einstein condensation} (Oxford University Press, Oxford, 2003).

\bibitem{Larson} J. Larson, B. Damski, G. Morigi, and M. Lewenstein, {Phys. Rev. Lett.} {\bf 100}, 050401 (2008).

\bibitem{Meystre2009} K. Zhang, {\it et al.}, Phys. Rev. A {\bf 81}, 013802 (2010). 

\bibitem{Giovannetti2001} V.~Giovannetti~and~D.~Vitali,~{Phys.~Rev.~A}~{\bf 63}, 023812 (2001).

\bibitem{gro2009A} S. Gr\"oblacher, {\it et al.}, {Nature} (London) {\bf 460}, 724 (2009); S. Gr\"oblacher, {\it et al.}, {Nature Phys.} {\bf 5}, 485 (2009); A. Schliesser, O. Arcizet, R. Rivi\`ere, and T. J. Kippenberg, {\it ibid.} {\bf 5}, 509 (2009).

\bibitem{MauroNJP}
M. Paternostro, S. Gigan, M. S. Kim, F. Blaser, H. R. B\"ohm and M. Aspelmeyer, {New J. Phys.} {\bf 8}, 107 (2006). 

\bibitem{natures2006A} S. Gigan, H. R. B\"ohm,  M. Paternostro, F. Blaser,  G. Langer,  J. B. Hertzberg,  K. C. Schwab,  D. B\"auerle,  M. Aspelmeyer and  A. Zeilinger, Nature (London) {\bf 444}, 67 (2006); O. Arcizet,  P.-F. Cohadon,  T. Briant,  M. Pinard and  A. Heidmann,  Nature (London) {\bf 444}, 71 (2006); D. Kleckner and D. Bouwmeester, Nature (London) {\bf 444}, 75 (2006).

\bibitem{Probes} K. Eckert, O. Romero-Isart, M. Rodriguez, M. Lewenstein, E. S. Polzik, and A. Sanpera, {Nature Phys.} {\bf 4}, 50 (2008); 
S. Singh and P. Meystre, {Phys. Rev. A} {\bf 81}, 041804(R) (2010); 
G. De Chiara, O. Romero-Isart, and A. Sanpera, Phys. Rev. A {\bf 83}, 021604(R) (2011).

\bibitem{genes}
C. Genes, D. Vitali and P. Tombesi, Phys. Rev. A {\bf 77}, 050307(R) (2008).

\bibitem{Esslinger2007} 
F. Brennecke T. Donner, S. Ritter, T. Bourdel, M. K\"ohl, and T. Esslinger, {\it Nature (London)} {\bf 450}, 268 (2007); 
Y. Colombe, T. Steinmetz, G. Dubois, F. Linke, D. Hunger, J. Reichel, {\it ibid.} {\bf 450}, 272 (2007).

\bibitem{giedke}
G. Giedke, B. Kraus, M. Lewenstein, and J. I. Cirac,  {Phys. Rev. Lett.} {\bf 87}, 167904 (2001); {Phys. Rev. A} {\bf 64}, 052303 (2001).

\bibitem{contangle} A. Ferraro, S. Olivares, and M. G. A. Paris, {\it Gaussian states in continuous variable quantum information} (Bibliopolis, Napoli, 2005).

\bibitem{contangle2} G. Adesso, A. Serafini, and F. Illuminati, {Phys. Rev. A} {\bf 73}, 032345 (2006).

\bibitem{RitschNatPhys}
I. B. Mekhov, C. Maschler, H. Ritsch, {Nature Phys.} {\bf 3}, 319 (2007).

\bibitem{Walls} 
D. F. Walls and G. J. Milburn, {\it Quantum Optics} (Springer, Berlin, 1994).

\bibitem{schmidt} See also M. Schmidt, M. Ludwig, and F. Marquardt, New J. Phys. {\bf 14}, 125005 (2012).



\bibitem{farace} A. Farace and V. Giovannetti, Phys. Rev. A {\bf 86}, 013820 (2012).

\bibitem{doria}
P. Doria, T. Calarco, and S. Montangero,
Phys. Rev. Lett. {\bf 106}, 190501 (2011).

\bibitem{DeChiara2011} G. De Chiara, M. Paternostro and G. M. Palma, Phys.~Rev.~A {\bf 83}, 052324 (2011).

\bibitem{thompson}
J. D. Thompson,  B. M. Zwickl,  A. M. Jayich,  Florian Marquardt,  S. M. Girvin  and  J. G. E. Harris, Nature (London) {\bf 452}, 72 (2008). 

\bibitem{sphere} D. E. Chang, {\it et al.}, Proc. Natl. Acad. Sci. USA {\bf 107}, 1005 (2010); O. Romero-Isart, M. L. Juan, R. Quindant, and J. I. Cirac, New J. Phys. {\bf 12}, 033015 (2010). 










\bibitem{SMAfail} H. Pu, C. K. Law, S. Raghavan, J. H. Eberly and N. P. Bigelow, \pra {\bf 60}, 1463 (1999).

\bibitem{rotor}R. Barnett, J. D. Sau, S. Das Sarma \pra {\bf 82}, 031602 (2010); R. Barnett, H.-Y. Hui, C.-H. Lin, J. D. Sau, S. Das Sarma, arXiv:1011.3517 (2010).


\bibitem{spinorham}  T. L. Ho, Phys. Rev. Lett. {\bf 81}, 742 (1998); C. K. Law, H. Pu and N. P. Bigelow, {\it ibid.} {\bf 81}, 5257 (1998); T. Ohmi and K. Machida, J. Phys. Soc. Jpn. {\bf 67}, 1822 (1998).

\bibitem{scatteringlengths} E. G. M. van Kempen, S. J. J. M. F. Kokkelmans, D. J. Heinzen, and B. J. Verhaar, \prl {\bf 88}, 093201 (2001).

\bibitem{spindynamics} M.-S Chang, Q. Qin, W. Zhang, L.You and M. S. Chapman, Nature Phys. {\bf 1}, 111 (2005).




\bibitem{feshbach} S. Inouye, M. R. Andrews, J. Stenger, H. J. Miesner, D. M. Stamper-Kurn, and W. Ketterle, Nature (London) {\bf 392}, 151 (1998).

\bibitem{gyro} E. Arimondo, M. Inguscio, and P. Violino, \rmp {\bf 49}, 31 (1977).

\bibitem{magntip} Y. J. Wang, M. Eardley, S. Knappe, J. Moreland, L. Holberg and J. Kitching, \prl {\bf 97}, 227602 (2006).


\bibitem{polartensor} I. H. Deutsch, P. S. Jessen, \pra {\bf 57}, 1972 (1998).

\bibitem{faradayQNM} Y. Takahashi, K. Honda, N. Tanaka, K. Toyoda, K. Ishikawa, and T. Yabuzaki \pra {\bf 60}, 4974 (1999).

\bibitem{faradayback} G. A. Smith, S. Chaudhury, and P. S. Jessen J. Opt. B: Quantum Semiclass. Opt. {\bf 5}, 323 (2003).

\bibitem{faradayNa} Y. Liu, E. Gomez, S. E. Maxwell, L. D. Turner, E. Tiesinga, and P. D. Lett, \prl {\bf 102}, 125301 (2009); {\it ibid.} {\bf 102}, 225301 (2009).








\bibitem{armour} A. D. Armour, M. P. Blencowe, and K. C. Schwab, \prl {\bf 88}, 148301 (2002).

\bibitem{rabl}
P. Rabl, A. Shnirman, and P. Zoller, \prb {\bf 70}, 205304 (2004). 

\bibitem{tian} 
L. Tian, \prb {\bf 72}, 195411 (2005). 

\bibitem{rodrigues}
D. A. Rodrigues, J. Imbers, and A. D. Armour, \prl {\bf 98}, 067204 (2007).

\bibitem{cam} S. Mancini, D. Vitali, and P. Tombesi, \prl {\bf 90}, 137901 (2003).

\bibitem{noiGP}
 G. Vacanti, R. Fazio, M. S. Kim, G. M. Palma, M. Paternostro and V. Vedral, Phys. Rev. A {\bf 85}, 022129 (2012).


\bibitem{phase} 
G. J. Milburn, S. Schneider, and D. F. V. James, Fortschr. Phys. {\bf 48}, 801 (2000).

\bibitem{Leibfried}
D. Leibfried, B. DeMarco,  V. Meyer,  D. Lucas,  M. Barrett,  J. Britton,  W. M. Itano,  B. Jelenkovic,  C. Langer,  T. Rosenband, and  D. J. Wineland, Nature (London) {\bf 422}, 412 (2003).

\bibitem{davidovicharoche} L. Davidovich, A. Maali, M. Brune, J. M. Raimond, and S. Haroche, \prl {\bf 71}, 2360 (1993).

\bibitem{chan} J. Chan,  T. P. Mayer Alegre,  A. H. Safavi-Naeini,  J. T. Hill,  A. Krause,  S. Gr\"oblacher,  M. Aspelmeyer, and O. Painter, Nature (London) {\bf 478}, 89 (2011). 

\bibitem{demartini} F. De Martini, F. Sciarrino, and C. Vitelli, \prl {\bf 100}, 253601 (2008). 

\bibitem{jeong1} H. Jeong and T. C. Ralph, {Phys. Rev. Lett.} {\bf 97}, 100401 (2006).

\bibitem{jeongralph} H. Jeong, M. Paternostro, and T. C. Ralph,  \prl {\bf 102}, 060403 (2009).

\bibitem{banaszek} K. Banaszek and K. W\'odkiewicz, {Phys. Rev. A} {\bf 58}, 4345  (1998).

\bibitem{banwod} K. W\'odkiewicz, New. J. Phys. {\bf 2}, 21 (2000).

\bibitem{spagnolo2011} N. Spagnolo, C. Vitelli, M. Paternostro, F. De Martini, and F. Sciarrino, Phys. Rev. A {\bf 84}, 032102 (2011).

\bibitem{chsh} J. F. Clauser, M. A. Horne, A. Shimony, and R. A. Holt, \prl {\bf 23}, 880 (1969).

\bibitem{bosekim} S. Bose,  I. Fuentes-Guridi, P. L. Knight and V. Vedral, \prl {\bf 87}, 050401 (2001).

\bibitem{prima3} A. Ferreira, A. Guerreiro, and V. Vedral, \prl {\bf 96}, 060407 (2006).

\bibitem{peres} A. Peres, \prl {\bf 77}, 1413 (1996).

\bibitem{horo3}
M. Horodecki, P. Horodecki, and R. Horodecki, Phys. Lett. A {\bf 223}, 1 (1996).

\bibitem{plenio05}
M. B. Plenio, Phys. Rev. Lett. {\bf 95}, 090503 (2005).


\bibitem{KochPRL}
M. Koch, C. Sames, M. Balbach, H. Chibani, A. Kubanek, K. Murr, T. Wilk, and G. Rempe, \prl {\bf 107}, 023601 (2011).

\bibitem{RitterNature}
S. Ritter, C. N\"{o}lleke, C. Hahn, A.  Reiserer, A. Neuzner, M. Uphoff, M. M\"{u}cke, E. Figueroa, J. Bochmann, and G. Rempe, Nature {\bf 484}, 195 (2012).

\bibitem{SpechtNature}
H. P. Specht, C. N\"{o}lleke, A. Reiserer, M. Uphoff, E. Figueroa, S. Ritter, and G. Rempe, Nature { \bf 473}, 190 (2011).

\bibitem{MonteiroNJP}
T. S. Monteiro, J. Millen, G. A. T. Pender, F. Marquardt, D. Chang, and P. F. Barker, New J. Phys. {\bf 15}, 015001, (2012).

\bibitem{Kiesel}
N. Kiesel, F. Blaser, U. Delic, D. Grass, R. Kaltenbaek, and M. Aspelmeyer, arXiv:1304.6679 (2013).

\bibitem{MauroPRA}
A. Xuereb, and M. Paternostro, \pra {\bf 87}, 023830 (2013).

\bibitem{Groblacher1}
S. Gr\"oblacher,  K. Hammerer,  M. R. Vanner, and  M. Aspelmeyer,  Nature (London) {\bf 460}, 724 (2009).

\bibitem{Groblacher2}
S. Gr\"oblacher,  J. B. Hertzberg,  M. R. Vanner,  G. D. Cole,  S. Gigan,  K. C. Schwab, and  M. Aspelmeyer, Nat. Phys. {\bf 5}, 485 (2009). 

\bibitem{Groeblacher} S. Gr\"oblacher (Private communication).







\end{thebibliography}
\end{document}